\documentclass[prd,aps,showpacs,nofootinbib,preprint,eqsecnum]{revtex4}
\usepackage{graphicx,color,amsmath,amsxtra}
\usepackage{epsf}
\usepackage{amssymb}
\usepackage{enumerate}
\usepackage{hhline}
\usepackage{array}
\usepackage{tabularx}

\newcommand{\be}{\begin{equation}}
\newcommand{\ee}{\end{equation}}
\newcommand{\bea}{\begin{eqnarray}}
\newcommand{\eea}{\end{eqnarray}}
\newcommand{\beaa}{\begin{eqnarray*}}
\newcommand{\eeaa}{\end{eqnarray*}}

\newcommand{\e}{\mathrm{e}}

\newcommand{\Eqn}[1]{&\hspace{-0.2em}#1\hspace{-0.2em}&}

\def\Vec#1{\mbox{\boldmath $#1$}}
\def\Vecs#1{\mbox{\boldmath\tiny $#1$}}

\def\be{\begin{equation}}
\def\ee{\end{equation}}
\def\bea{\begin{eqnarray}}
\def\eea{\end{eqnarray}}

\def\e{\mathrm{e}}
\begin{document}

\title{
Generation of large-scale magnetic fields, non-Gaussianity, and 
primordial gravitational waves in inflationary cosmology
}

\author{Kazuharu Bamba$^{1, 2, 3, }$\footnote{E-mail address: 
bamba.kazuharu@ocha.ac.jp}
}
\affiliation{
$^1$Leading Graduate School Promotion Center, 
Ochanomizu University, 2-1-1 Ohtsuka, Bunkyo-ku, Tokyo 112-8610, Japan\\ 
$^2$Department of Physics, Graduate School of Humanities and Sciences, Ochanomizu University, Tokyo 112-8610, Japan\\ 
$^3$Kobayashi-Maskawa Institute for the Origin of Particles and the
Universe, Nagoya University, Nagoya 464-8602, Japan\footnote{The author's previous affiliation}
}

%\date{\today}

%%%%%%%%%%%%%%%%%%%%%
%  Abstract
%%%%%%%%%%%%%%%%%%%%%
\begin{abstract}
The generation of large-scale magnetic fields in inflationary cosmology 
is explored, in particular, in a kind of moduli inflation motivated by racetrack inflation in the context of the Type IIB string theory. 
In this model, the conformal invariance of the hypercharge electromagnetic fields is broken thanks to the coupling of both the scalar and pseudoscalar fields to the hypercharge electromagnetic fields. 
The following three cosmological observable quantities are first evaluated: 
The current magnetic field strength on the Hubble horizon scale, 
which is much smaller than the upper limit from the back reaction problem, 
local non-Gaussianity of the curvature perturbations due to the existence of the massive gauge fields, and the tensor-to-scalar ratio. 
It is explicitly demonstrated that the resultant values of local non-Gaussianity and the tensor-to-scalar ratio are consistent with the Planck data. 
\end{abstract}
%%%%%%%%%%%%%%%%%%%%%

%----------------------------
%\pacs{Keywords:}
\hspace{13.0cm} OCHA-PP-330
%----------------------------

\maketitle
%==============================================================================

%%%%%%%%%%%%%%%%%%%%%%%%%%%
%%%  Sec. I
%%%%%%%%%%%%%%%%%%%%%%%%%%%
\section{Introduction}

It is observationally confirmed that 
there are galactic magnetic fields on $1$--$10$kpc scale with 
the strength of $\sim 10^{-6}$G, 
and that also in clusters of galaxies, 
there exist the magnetic fields on $10$~kpc--$1$Mpc scale 
with their amplitude of $10^{-7}$--$10^{-6}$G. 
The origins of cosmic magnetic fields, particularly, such 
large-scale magnetic fields in clusters of galaxies
have not yet been established 
(for reviews, see, e.g.,~\cite{M-F-Reviews}). 
There have been proposed various generation mechanisms  
such as the plasma instability~\cite{Biermann1, PI}, 
cosmological electroweak and quark-hadron phase transitions~\cite{PT}, 
cosmic string~\cite{M-CS}, primordial density perturbations~\cite{DP}, 
and the secondary dynamo amplification mechanism~\cite{EParker}. 
However, it is difficult for these mechanism 
to produce the large-scale magnetic fields. 

It is known that electromagnetic quantum fluctuations generated during 
inflation are the most natural origin of large-scale magnetic 
fields~\cite{Inflation}, because the coherent scale of 
magnetic fields can be extended larger 
than the Hubble horizon at the inflationary stage~\cite{Turner:1987bw}. 
The Maxwell theory has its conformal invariance. 
Moreover, the Friedmann-Lema\^{i}tre-Robertson-Walker (FLRW) metric, 
which describes the homogeneous and isotropic universe consistent with 
observations, is conformally flat\footnote{
For the breaking mechanisms of the conformal flatness, see, 
for example,~\cite{Campanelli:2013mea, B-C-F, ANS-ANSL}.}. 
Hence, at the inflationary stage, 
the conformal invariance of the electromagnetic fields 
has to be broken so that 
the quantum fluctuations of the electromagnetic fields can be 
generated~\cite{Parker:1968mv} and eventually 
result in the large-scale magnetic fields at the present 
time~\cite{Turner:1987bw, N-G-C}. 
There are several well-known ideas of the breaking mechanism: e.g., 
($i$) A non-minimal coupling between 
the scalar curvature and the electromagnetic fields 
produced by a one-loop vacuum-polarization effect 
in quantum electrodynamics in the curved space-time~\cite{Drummond:1979pp}; 
($ii$) A coupling of a scalar field to the electromagnetic 
fields~\cite{Ratra:1991bn, MF-Scalar, B-Y, Garretson:1992vt}; 
($iii$) The trace anomaly~\cite{Dolgov:1993vg}. 

In this paper, we investigate the generation of large-scale magnetic fields 
from a kind of moduli inflation inspired by racetrack 
inflation~\cite{BlancoPillado:2004ns} 
in the framework of the Type IIB string theory with 
the so-called Kachru-Kallosh-Linde-Trivedi volume stabilization 
mechanism~\cite{Kachru:2003aw}. 
In this model, the conformal invariance of the hypercharge 
electromagnetic fields is broken through their coupling 
to both a scalar field and an axion-like pseudoscalar one. 
%%%%% 
It should be noted that 
our model is still a toy model motivated by 
racetrack inflation or so-called axion inflation, 
where the axion plays a role of the inflaton. 
The main purpose of this work is that by using a simple model, 
we reveal cosmological consequences in racetrack (or axion) 
inflation\footnote{Various cosmological results in axion inflation~\cite{Barnaby:2011qe, Barnaby:2012xt, Ferreira:2014zia, 
Cheng:2014kga} including the 
generation of large-scale magnetic fields~\cite{Garretson:1992vt, Anber:2006xt, Anber:2009ua} or primordial black holes~\cite{Bugaev:2013fya} 
and observational constraints on axion inflation~\cite{Meerburg:2012id} have also been explored (for a recent review on inflation driven by axion, 
see~\cite{Pajer:2013fsa}).}. 
In Refs.~\cite{Barnaby:2010vf, Barnaby:2011vw}, 
it has been indicated that a coupling of the pseudoscalar inflaton field 
to the electromagnetic fields can generate 
non-Gaussianity~\cite{Komatsu:2001rj, Tanaka:2010km} 
of power spectrum of the curvature perturbations 
coming from the quantum fluctuations of the inflaton field. 
%%%%%
Thus, we analyze non-Gaussianity 
of the curvature perturbations in the present scenario by following 
the procedure in Refs.~\cite{Meerburg:2012id, Linde:2012bt}\footnote{ 
For non-Gaussianity from magnetic fields, see~\cite{Brown:2005kr}}. 
Moreover, we study the so-called tensor-to-scalar ratio defined by 
the ratio of scalar modes of the curvature perturbations to 
their tensor modes (namely, the primordial gravitational waves)~\cite{Barnaby:2011qe, Barnaby:2012xt}. 
We show that if the magnetic fields on the Hubble horizon scale 
with their current strength compatible with the back reaction problem 
are generated, 
local non-Gaussianity and the tensor-to-scalar ratio in the cosmic microwave background (CMB) radiation with those values smaller than 
the limits from the Planck satellite~\cite{Ade:2013ydc} 
can be produced\footnote{The recent BICEP2 result~\cite{Ade:2014xna} on the tensor-to-scalar ratio is also mentioned in Sec.~IV C.}. 
%%%%%%%% 
The most important result of this work is that 
the explicit values of three cosmological observable quantities, 
i.e., the large-scale magnetic fields, local non-Gaussianity, and the tensor-to-scalar ratio are first derived. 
%%%%%%%%
Furthermore, we should emphasize the novelty of our present model 
in comparison with the other recent works on non-Gaussianity of the curvature perturbations and the tensor-to-scalar ratio in a kind of axion inflation~\cite{Meerburg:2012id, Linde:2012bt, Bugaev:2013fya, Barnaby:2010vf, Barnaby:2011vw}. 
In our model, a scalar field as well as the axion-like 
pseudoscalar field couple to the hypercharge electromagnetic field, 
whereas in the other past models, only the pseudoscalar field couples to 
the hypercharge electromagnetic field. 
The existence of such a scalar field coupling to 
the (hypercharge) electromagnetic field is suggested by 
the Kaluza-Klein (KK) compactification mechanism~\cite{K-K} for 
the fundamental higher-dimensional space-time theories including 
string theories. 
In fact, both couplings appears in the framework of racetrack inflation. 
Thus, the setting of our model is closer to the realistic one than that in the past related works, although it is a toy model. 
In addition, there is one more significant advantage that 
thanks to the coupling of the scalar filed to the hypercharge 
electromagnetic field, in principle, the large-scale magnetic fields with 
the current strength enough to explain the observations 
without any secondary amplification mechanism like the galactic dynamo. 
This point cannot be realized in the past models. 
%%%%%

The observational test of this model is the severest, 
therefore it is very difficult for the model to be viable, 
because we use the three independent observations of 
the large-scale magnetic fields, local non-Gaussianity, and 
tensor-to-scalar ratio. 
Furthermore, this model is the most general within 
the fundamental theories which we are considering. 
Thus, we develop the generic discussions in order 
not only to extend the theoretical possibility 
but also to strictly constrain the freedom of the theory. 
%%%%% Units %%%%%
We use the units $k_{\mathrm{B}} = c = \hbar = 1$ and describe the 
Newton's constant by 
$G=1/M_\mathrm{P}^2$, where $M_\mathrm{P} =2.43 \times 10^{18}$ GeV 
is the reduced Planck mass. 
In terms of electromagnetism, we adopt Heaviside-Lorentz units. 
%%%%%%%%%%%%%%%%%

%%%%%%%% Construction %%%%%%%%%
The paper is organized as follows. 
In Sec.\ II, we explain our model action and derive the basic equations. 
In Sec.\ III, we investigate the evolution of each field and estimate 
the current strength of the large-scale magnetic fields. 
In Sec.\ IV, we explore the power spectrum of the curvature perturbations, non-Gaussianity, and the tensor-to-scalar ratio. 
In Sec.\ V, conclusions are presented. 
%%%
In Appendix A, 
we examine the large-scale magnetic fields, 
non-Gaussianity, and tensor-to-scalar ratio for the 
axion (monodromy) inflation, and comparison these results with 
the ones for a kind of moduli inflation motivated by racetrack 
inflation in the previous sections. 
%%%
In Appendix B, the issues of the backreaction and the strong coupling 
are stated. 
In Appendix C, the observational constraints on the field strength of 
magnetic fields are summarized. 
Cosmological implications related to this work are also stated in 
Appendix D. 
%%%%%%%%%%%%%%%%%%%%%%%%%%%%%%%%

%%%%%%%%%%%%%%%%%%%%%%%%%%%
%%%  Sec. II
%%%%%%%%%%%%%%%%%%%%%%%%%%%
\section{Model}

Our model Lagrangian is given by\footnote{Such a kind of the action in Eq.~(\ref{eq:2.1}) has also been studied for a baryogenesis scenario due to 
the anomaly~\cite{Bamba:2006km, Bamba:2007hf}.}
\begin{eqnarray}
{\mathcal L} \Eqn{=}  \frac{M_\mathrm{P}^2}{2} R 
-\frac{1}{4} X F_{\mu\nu}F^{\mu\nu}
-\frac{1}{4} g_{\mathrm{ps}} \frac{Y}{M} F_{\mu\nu}\tilde{F}^{\mu\nu} 
 \nonumber \\
&& {}-\frac{1}{2}g^{\mu\nu}{\partial}_{\mu}{\Phi}{\partial}_{\nu}{\Phi} 
- U(\Phi) 
-\frac{1}{2}g^{\mu\nu}{\partial}_{\mu}{Y}{\partial}_{\nu}{Y} 
- V(Y) \,,
\label{eq:2.1} \\
X \Eqn{\equiv} \exp \left(-\lambda \frac{\Phi}{M_P} \right)\,, 
\label{eq:2.2} 
\end{eqnarray} 
\begin{equation} 
V(Y) \approx \bar{V} - \frac{1}{2} m^2 Y^2\,, 
\label{eq:2.4} 
\end{equation} 
where $R$ is the Ricci scalar, 
%%%
$g_{\mathrm{ps}}$ is a dimensionless coupling constant, 
%%%
$\Phi$ is the canonically normalized 
field of the scalar field $X$ with the normalization constant
$\lambda$, $Y$ is a canonical pseudoscalar field, 
and $M$ is a constant with the dimension of mass 
corresponding to the decay constant of $Y$. 
Furthermore, $F_{\mu\nu} = \nabla_{\mu} F_{\nu} - \nabla_{\nu} F_{\mu}$ 
is the field strength of the $\mathrm{U}(1)_Y$ hypercharge gauge field $F_{\mu}$, where $\nabla_{\mu}$ is the covariant derivative, and $\tilde{F}^{\mu\nu}$ are the dual field strength of $F_{\mu}$. While we do not specify the exact form of scalar potentials $U(X = X(\Phi))$, $Y$ would be expected to have a potential, given by Eq.~(\ref{eq:2.4}) with 
a normalization factor $\bar{V}$ and the mass $m$ of the pseudoscalar $Y$. 
%%%
The pseudoscalar field $Y$ couples to the dual of the field strength, and 
hence it acts as an axion. 
Throughout our analysis, we assume that inflation is driven by the potential energy of $Y$ as in the so-called natural inflation or axion inflation~\cite{Anber:2009ua, N-A-I, CN-inflation}. 
%%%%%
We take the flat FLRW space-time 
\begin{equation}
{ds}^2 = -{dt}^2 + a^2(t)d{\Vec{x}}^2 \,, 
\label{eq:2.8}
\end{equation}
with $a$ the scale factor. In this background, 
the field equations of $\Phi$ (i.e., $X$) and $Y$ read\footnote{Here, 
we have used the fact that the contribution of 
the hypercharge electromagnetic field is negligible 
because it exists as a quantum fluctuation during inflation and 
the amplitude is so small that its squared 
can be neglected.}
\begin{equation}
\ddot{\Phi} + 3H\dot{\Phi} + \frac{dU(\Phi)}{d\Phi} = 0\,, 
\quad 
\ddot{Y} + 3H\dot{Y} + \frac{dV(Y)}{dY} = 0\,,
\label{eq:2.10}  
\end{equation}
where $H \equiv \dot{a}/a$ is the Hubble parameter and 
the dot denotes the derivative with respect to the cosmic time $t$. 
Using the Coulomb gauge 
$F_0(t,\Vec{x}) = 0$ and ${\partial}_j F^j (t,\Vec{x}) =0$, 
we find that the field equation of $F_{\mu}$ is described as 
\begin{equation}
\ddot{F_i}(t,\Vec{x}) 
+ \left( H + \frac{\dot{X}}{X} 
\right) \dot{F_i}(t,\Vec{x}) 
- \frac{1}{a^2}{\partial}_j {\partial}_j F_i(t,\Vec{x}) 
- \frac{g_{\mathrm{ps}}}{M} \frac{1}{aX} \dot{Y} 
\epsilon^{ijk} {\partial}_j F_k(t,\Vec{x}) = 0\,, 
\label{eq:2.11}
\end{equation}   
where the second term within the round bracket $(\,)$ and 
the fourth term originate from the breaking of the 
conformal invariance of the hypercharge electromagnetic fields.

%%%%%%%%%%%%%%%%%%%
%%%  Sec. III
%%%%%%%%%%%%%%%%%%%
\section{Current strength of large-scale magnetic fields}

In this section, we explore the evolutions of the 
$\mathrm{U}(1)_Y$ gauge field, 
the scalar field $X$, and the pseudoscalar field $Y$, 
and estimate the strength of large-scale magnetic fields 
at the present time. 

%%%%%%%%%%%%%%%%%%%
%%%  Sec. III A
%%%%%%%%%%%%%%%%%%%
\subsection{Scalar and pseudoscalar fields}

We suppose that inflation is basically driven by the potential of $Y$.  
In the FLRW background~(\ref{eq:2.8}), the Friedmann equation becomes 
$3 M_\mathrm{P} H^2 = \left[ \left(1/2\right) 
{\dot{Y}}^2 + V(Y) \right]$. 
If the so-called slow-roll approximation $\dot{Y}^2/2 \ll V(Y)$ 
is satisfied, we have $H \approx H_\mathrm{inf} = \mathrm{constant}$ 
with $H_\mathrm{inf}$ the Hubble parameter during inflation, so that 
the exponential inflation can be realized. 
In this case, the scale factor $a(t)$ can be expressed as 
$a(t) = a_k \exp \left[ H_{\mathrm{inf}} \left(t-t_k\right) \right]$ 
with $a_k=a(t_k)$, 
where $t_k$ is the time when a comoving wavelength 
$2\pi/k$ of the $\mathrm{U}(1)_Y$ gauge field first crosses the horizon at the inflationary stage, and thus $k/(a_k H_{\mathrm{inf}}) = 1$ is met. 
The analytic solution of Eq.~(\ref{eq:2.10}) 
is given by~\cite{Garretson:1992vt} 
\begin{equation} 
Y = Y_k \exp \left\{ 
\frac{3}{2} \left[-1 \pm \sqrt{1 + \left( \frac{2m}{3H_{\mathrm{inf}}} 
\right)^2 } \right] H_{\mathrm{inf}} \left( t-t_k \right)
\right\}\,, 
\label{eq:3.4} 
\end{equation}
with $Y_k= Y(t_k)$. 
In the following, we use this solution. 
In particular, without generality, we take 
the ``$+$'' sign on the right-hand side of this solution. 
On the other hand, regarding $X$, we study the case that the 
concrete dynamics of $X$ during inflation does not 
influence on the results and only the difference between 
the initial and final values during inflation is important.

%%%%%%%%%%%%%%%%%%%
%%%  Sec. III B
%%%%%%%%%%%%%%%%%%%
\subsection{$\mathrm{U}(1)_Y$ gauge field}

%%%%%%%%%%%%%%%%%%%
%%%  Sec. III B 1
%%%%%%%%%%%%%%%%%%%
\subsubsection{Quantization}

First, we quantize the $\mathrm{U}(1)_Y$ gauge field $F_{\mu}(t,\Vec{x})$. 
It follows from the hypercharge electromagnetic part of 
the action constructed by the Lagrangian~(\ref{eq:2.1}), 
we find that the canonical momenta conjugate to 
$F_{\mu}(t,\Vec{x})$ read 
${\pi}_0 = 0$ and ${\pi}_{i} = X a \dot{F_i}(t,\Vec{x})$. 
The canonical commutation relation between $F_i(t,\Vec{x})$ and ${\pi}_{j}(t,\Vec{x})$ is imposed as 
\begin{equation}
\left[ F_i(t,\Vec{x}), {\pi}_{j}(t,\Vec{y}) \right] = i 
\int \frac{d^3 k}{(2\pi )^{3/2} }
\e^{i \Vecs{k} \cdot \left( \Vecs{x} - \Vecs{y} \right)}
\left[ {\delta}_{ij} - \left(k_i k_j / k^2 \right) \right] \,. 
\end{equation}
Here, $\Vec{k}$ is the comoving wave number and its amplitude is 
expressed as $k = |\Vec{k}|$. 
This relation leads to the description of $F_i(t,\Vec{x})$ as 
\begin{equation} 
F_i(t,\Vec{x}) = \int \frac{d^3 k}{(2\pi )^{3/2} }
\left[ \hat{b}(\Vec{k}) 
F_i(t,\Vec{k}) \e^{i \Vecs{k} \cdot \Vecs{x} }
       + {\hat{b}}^{\dagger}(\Vec{k})
{F_i^*}(t,\Vec{k}) \e^{-i \Vecs{k} \cdot \Vecs{x}} \right] \,, 
\end{equation}
where $\hat{b}(\Vec{k})$ and ${\hat{b}}^{\dagger}(\Vec{k})$ 
are the annihilation and creation operators, respectively. 
These operators obey the relations
\begin{equation} 
\left[ \hat{b}(\Vec{k}), {\hat{b}}^{\dagger}({\Vec{k}}^{\prime}) \right] = 
{\delta}^3 (\Vec{k}-{\Vec{k}}^{\prime})\,, 
\quad 
\left[ \hat{b}(\Vec{k}), \hat{b}({\Vec{k}}^{\prime}) \right] = 
\left[ {\hat{b}}^{\dagger}(\Vec{k}), {\hat{b}}^{\dagger}({\Vec{k}}^{\prime})
 \right] = 0\,. 
\end{equation}
We also have the normalization condition as 
\begin{equation} 
F_i(k,t){\dot{F}}_j^{*}(k,t) - {\dot{F}}_j(k,t){F_i^{*}}(k,t)
= \frac{i}{ X a } 
\left[ {\delta}_{ij} - \left( k_i k_j / k^2 \right) \right]\,. 
\end{equation}
%

%%%%%%%%%%%%%%%%%%%
%%%  Sec. III B 2
%%%%%%%%%%%%%%%%%%%
\subsubsection{Set up}

We set the $x^3$ axis to lie along the direction of the spatial momentum 
\Vec{k} and express the transverse directions as 
$x^{1}$ and $x^{2}$. 
By using Eq.~(\ref{eq:2.11})
and defining the circular polarizations
$F_{\pm}(k,t) \equiv F_1(k,t) \pm i F_2(k,t)$ with 
the Fourier modes 
$F_1(k,t)$ and $F_2(k,t)$ of the $\mathrm{U}(1)_Y$ gauge field, 
we acquire 
\begin{equation}
\ddot{F}_{\pm}(k,t) 
+ \left( H_{\mathrm{inf}} + \frac{\dot{X}}{X} 
\right) \dot{F}_{\pm}(k,t) 
+ \left[ 1 \pm \frac{g_{\mathrm{ps}}}{M} \frac{\dot{Y}}{X} 
\left( \frac{k}{a} \right)^{-1} 
\right] \left( \frac{k}{a} \right)^2 F_{\pm}(k,t) = 0\,, 
\label{eq:3.12}
\end{equation}
During inflation, 
we numerically solve this equation 
by following the procedure in Ref.~\cite{Bamba:2006km}, because 
it is very hard to acquire the analytic solution of Eq.~(\ref{eq:3.12}). 
For the sub-horizon scale $k/\left(aH\right) \gg 1$, 
the $F_-(k,t)$ corresponds to the decaying mode, and therefore 
we only examine the evolution of $F_+(k,t)$. 

An approximate amplitude $F_+(k,t=t_k)$ at the horizon crossing, 
where $k/\left( a H_{\mathrm{inf}} \right) = 1$, 
is represented as~\cite{Barnaby:2010vf, Barnaby:2011vw, Anber:2009ua} 
$F_+(k,t_k) \simeq \left(1/\sqrt{2k}\right) 
\left(1/\sqrt{X(t_k)}\right) \left( 2 \xi_k \right)^{-1/4} 
\exp \left(\pi \xi_k -2\sqrt{2 \xi_k} \right)$ 
with $\xi_k = \xi (t = t_k)$. Here, 
\begin{equation} 
\xi \equiv \frac{1}{2} \frac{g_{\mathrm{ps}}}{M} \frac{1}{X} 
\frac{\dot{Y}}{H_{\mathrm{inf}}}\,.
\label{eq:4.2}
\end{equation}
This amplification comes from the tachyonic instability. 
Thus, when we numerically calculate Eq.~(\ref{eq:3.12}), 
we take into account the above amplification factor in the initial 
conditions. 
We define the following amplification factor as 
\begin{equation} 
C_{+}(k,t) \equiv \frac{F_{+}(k,t)}{F_{+}(k,t_k)}\,. 
\label{eq:3.13}
\end{equation}   
We estimate the initial amplitude of $F_{+}(k,t)$, i.e., $F_{+}(k,t_k)$, 
by matching with the solution for sub-horizon scales $k/(aH) \gg 1$
 at the horizon exit~\cite{Bamba:2006km}. 
Here, we assume that 
in the short-wavelength limit of $k \rightarrow \infty$, 
the amplitude of $F_{+}(k,t)$ is described by 
$\left| F_{+}^{(\mathrm{in})} (k,t) \right| = 
\left(1/\sqrt{2k}\right) \left(1/\sqrt{X(t)}\right)$, 
where the coefficients of modes have been chosen 
so that the vacuum can be reduced to the one in the Minkowski space-time in the short-wavelength limit (the so-called Bunch-Davies vacuum~\cite{B-D}).

%%%%%%%%%%%%%%%%%%%
%%%  Sec. III B 3
%%%%%%%%%%%%%%%%%%%
\subsubsection{Numerical analysis} 

We derive the strength of large-scale magnetic fields, 
provided that during inflation, $X$ can approximately be regarded 
as a constant. 
This means that a dynamical quantity to the hypercharge 
electromagnetic fields is only the pseudoscalar field $Y$. 
Such a case has been explored in 
Refs.~\cite{Garretson:1992vt, Barnaby:2010vf, Barnaby:2011vw, Barnaby:2011qe, Pajer:2013fsa}. 
Indeed, the field strength of the large-scale magnetic fields 
can be amplified in our model, where the hypercharge electromagnetic fields 
couple to the scalar field $X$. 
In other words, the important quantity to characterize the amplification of 
the magnetic fields is the ratio of the final value of $X$ to the initial one 
at inflationary stage.  
The ordinary theory of the electromagnetic fields, where $X=1$, 
has to be recovered by the epoch of the Big Bang Nucleosynthesis (BBN). 
Accordingly, we suppose that $X$ stays almost constant 
during inflation, and after inflation it quickly 
reaches $X (t = t_\mathrm{R}) = 1$ at the reheating stage $t_R$ 
owing to an appropriate form of $V(\Phi)$. 

%%%%%%%%%%%%%%
% Figure 1
%%%%%%%%%%%%%%
\begin{figure}[tbp]
\begin{center}
%\resizebox{!}{8cm}{
   \includegraphics{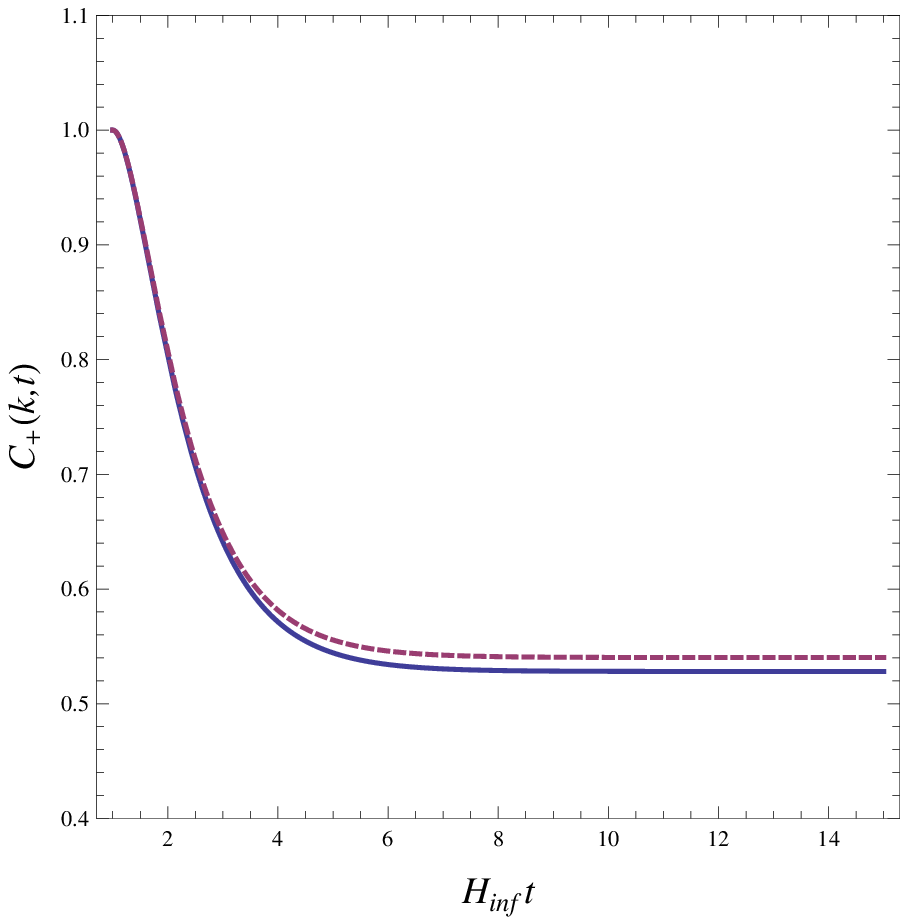}
%                  }
\caption{$C_{+}(k, t)$ as a function of $H_{\mathrm{inf}} t$ 
for $X(t_k) \equiv \exp \left( \chi_k \right)$ with $\chi_k = -0.940$, 
$H_{\mathrm{inf}} = 1.0 \times 10^{10}$GeV, $m = 2.44 \times 10^{9}$GeV, 
$Y_k = 7.70 \times 10^{-2} M_{\mathrm{P}} = 1.87 \times 10^{17} \, \mathrm{GeV}$, $M =1.0 \times 10^{-1} M_\mathrm{P} = 2.43 \times 10^{17} \, 
\mathrm{GeV}$, 
$\bar{V} = 5.07 \times 10^{-17} M_\mathrm{P}^4$, $\xi_k = 2.5590616$, 
and $g_{\mathrm{ps}} = 1.0$ (the case (b) in Table~\ref{table-1}). 
The solid line shows the case including the dynamics of $Y$, 
whereas the dotted line depicts that without it, namely, 
$\dot{Y} = 0$ in Eq.~(\ref{eq:3.12}).}
\label{fig-1}
\end{center}
\end{figure}
%%%%%%%%%%%%%%

In Fig.~\ref{fig-1}, we depict the evolution of $C_{+}(k, t)$ 
during inflation with the solid line for 
$X(t_k) \equiv \exp \left( \chi_k \right)$ with $\chi_k = -0.940$,  
$H_{\mathrm{inf}} = 1.0 \times 10^{10}$GeV, $m = 2.44 \times 10^{9}$GeV,  
$M =1.0 \times 10^{-1} M_\mathrm{P} = 2.43 \times 10^{17} \, 
\mathrm{GeV}$, 
$\bar{V} = 5.07 \times 10^{-17} M_\mathrm{P}^4$, 
$\xi_k = 2.5590616$, 
and $g_{\mathrm{ps}} = 1.0$.
This is the case (b) in Table~\ref{table-1} shown later. 
We have numerically solved Eq.~(\ref{eq:3.12}) for $k = a_k H_{\mathrm{inf}}$ 
mode for the exponential inflation from the initial time at 
$t=t_k= H_{\mathrm{inf}}^{-1}$, when we set $C_{+}(k, t_k) = 1$. 
We define the values of these parameters by the Cosmic Background Explorer (COBE)~\cite{Smoot:1992td} 
normalization and Planck data~\cite{Ade:2013uln} on the CMB radiation. 
For comparison, we have also plotted the numerical results 
for the case that $\dot{Y} = 0$ in Eq.~(\ref{eq:3.12}) with the dotted line. 
Here, the behavior for $\dot{Y} \neq 0$ is quite similar to that 
for $\dot{Y} = 0$, 
because the pseudoscalar field $Y$ rolls down its potential very slowly. 

{}From Fig.~\ref{fig-1}, we see that 
$C_{+}(k, t)$ asymptotically approaches a constant 
within about 10 Hubble expansion time 
after the horizon crossing during inflation. 
This is an important feature of evolution of $C_{+}(k, t)$, 
that is, the amplitude becomes a finite value and does not decay. 
It contributes to the resultant strength of the large-scale magnetic fields. 
Such a behavior of $C_{+}(k, t)$ does not depend on the model parameters. 
This result is also consistent with that in Ref.~\cite{Barnaby:2010vf}. 
The way of determining the values of $m$ and $Y_k$ are explained 
in the last paragraph of Sec.~IV A.

%%%%%%%%%%%%%%%%%%%
%%%  Sec. III C
%%%%%%%%%%%%%%%%%%%
\subsection{Current magnetic field strength}

Next, we evaluate the magnetic field strength at the present time. 
The proper hypermagnetic and hyperelectric fields are represented 
with the comoving hypermagnetic fields ${B_Y}_i(t,\Vec{x})$ 
and hyperelectric ones ${E_Y}_i(t,\Vec{x})$, respectively, 
as~\cite{Ratra:1991bn} 
\begin{eqnarray}
{B_Y}_i^{\mathrm{proper}}(t,\Vec{x}) \Eqn{=} 
\frac{1}{a^2} {B_Y}_i(t,\Vec{x}) = \frac{1}{a^2} 
{\epsilon}_{ijk}{\partial}_j F_k(t,\Vec{x}) \,, 
\label{eq:3.24} \\ 
{E_Y}_i^{\mathrm{proper}}(t,\Vec{x}) \Eqn{=} 
a^{-1}{E_Y}_i(t,\Vec{x}) = -a^{-1}\dot{F}_i(t,\Vec{x}) \,, 
\label{eq:MR1-33-IIIC-ADD-01} 
\end{eqnarray}
where ${\epsilon}_{ijk}$ is the totally antisymmetric tensor 
(${\epsilon}_{123}=1$). 
Multiplying the energy density of the proper hypermagnetic field 
in the Fourier space ${\rho}_{B_Y}(k,t)$ by the phase-space density 
$4\pi k^3/(2\pi)^3$, we obtain the energy density of the proper hypermagnetic field in the physical space 
\begin{equation}
{\rho}_{B_Y}(L,t) = 
 \frac{k^3}{4{\pi}^2}
\left[  
\left| {B_Y}_+^{\mathrm{proper}}(k,t)
\right|^2 
+
\left| {B_Y}_-^{\mathrm{proper}}(k,t)
\right|^2 \right] X\,. 
\label{eq:3.31}
\end{equation}
Here, 
$\left| {B_Y}_\pm^{\mathrm{proper}}(k,t)
\right|^2 = 
\left(1/a^2\right) \left( k/a \right)^2 
|F_\pm(k,t)|^2
$, which follows from Eq.~(\ref{eq:3.24}), 
and $L=2\pi/k$ is a comoving scale.

The instantaneous reheating at $t=t_\mathrm{R}$ after 
inflation occurs much earlier than the electroweak phase transition 
(EWPT) at $T_\mathrm{EW} \sim 100$GeV. 
The conductivity of the universe ${\sigma}_\mathrm{c}$ should be very small at the inflationary stage, because few particle present. 
In the reheating process, charged particles are created, and 
therefore ${\sigma}_\mathrm{c}$ increases and 
would become large enough as $(t \geq t_\mathrm{R})$. 
Hence, when ${\sigma}_\mathrm{c} \gg H$, 
the hyperelectric fields dissipate by accelerating the charged particles. 
In the following radiation- and matter-dominated stages 
$(t \geq t_\mathrm{R})$, we have 
$B_Y \propto a^{-2}$~\cite{Ratra:1991bn, B-Y}. 
Thus, at a later time after the EWPT when $X$ reached 
the true minimum of $X=1$, the energy density of the hypermagnetic fields 
$\rho_{B_Y} (L,t)$ reduces to that of the magnetic fields $\rho_B (L,t)$. 
The expression of $\rho_B (L,t)$ is given by~\cite{Bamba:2006km}
\begin{equation}    
\rho_B (L,t) \simeq
\frac{1}{8\pi^2} 
\frac{1}{X(t_k)} \frac{1}{\sqrt{2 \xi_k}} 
\exp \left[ 2\left(\pi \xi_k -2\sqrt{2 \xi_k}\right) \right] 
\left( \frac{k}{a} \right)^4 
|C_+(k, t_\mathrm{R})|^2\,, 
\label{eq:3.35}
\end{equation}
where we have imposed $X(t_{\rm R})=1$ 
and neglected the different coefficient factor 
between the magnetic field of $\mathrm{U}(1)_Y$ and that of $\mathrm{U}(1)_{\rm em}$ because it is order of unity.

%
%%%%%%%%%%%%%%%%%%%%%%%%%%%%
%%%%%%%%  Table I  %%%%%%%%%
%%%%%%%%%%%%%%%%%%%%%%%%%%%%
\begin{table}[t]
\caption{Current strength of magnetic fields on the Hubble horizon scale 
and $1$Mpc scale 
for $X(t_k) = \exp \left( \chi_k \right)$ with $\chi_k = -0.940$, 
$M =1.0 \times 10^{-1} M_\mathrm{P} = 2.43 \times 10^{17} \, 
\mathrm{GeV}$, 
$g_{\mathrm{ps}} = 1.0$, 
$\xi_k = 2.5590616$, 
and 
$k = 2\pi/\left( 2997.9 h^{-1} \right) \, {\mathrm{Mpc}}^{-1}$ with $h=0.673$. 
For the cases (i) (i=a, b, c, d, e, f), we have 
$T_\mathrm{R} \, [\mathrm{GeV}] = (1.02 \times 10^{14},\, 3.22 \times 10^{13},\, 3.22 \times 10^{12},\, 3.22 \times 10^{11},\, 3.22 \times 10^{10},\, 3.22 \times 10^{9})$ and 
$\bar{V}/M_\mathrm{P}^4 = (5.07 \times 10^{-15},\, 5.07 \times 10^{-17},\, 5.07 \times 10^{-21},\, 5.07 \times 10^{-25},\, 5.07 \times 10^{-29},\, 5.07 \times 10^{-33})$.}
\begin{center}
\tabcolsep = 2mm 
\begin{tabular}
{ccccccc}
%{ccccccc}
\hline
\hline
& $B(H_0^{-1},t_0) \hspace{1mm} [\mathrm{G}]$  
& $B(1\mathrm{Mpc},t_0) \hspace{1mm} [\mathrm{G}]$  
& $H_{\mathrm{inf}} \hspace{1mm} [\mathrm{GeV}]$ 
& $m \hspace{1mm} [\mathrm{GeV}]$  
& $Y_k/M_\mathrm{P}$
& $C_+(k, t_{\mathrm{R}})$ 
\\[0mm]
\hline
(a)  
& $7.15 \times 10^{-64}$
& $1.42 \times 10^{-56}$
& $1.0 \times 10^{11}$ 
& $2.44 \times 10^{10}$ 
& $7.70 \times 10^{-2}$ 
& $0.528$ 
\\[0mm]
(b)  
& $7.15 \times 10^{-64}$
& $1.42 \times 10^{-56}$
& $1.0 \times 10^{10}$ 
& $2.44 \times 10^{9}$ 
& $7.70 \times 10^{-2}$ 
& $0.528$ 
\\[0mm]
(c)  
& $2.33 \times 10^{-64}$
& $4.62 \times 10^{-57}$
& $1.0 \times 10^{8}$
& $1.0 \times 10^{7}$ 
& $1.62 \times 10^{1}$ 
& $0.172$ 
\\[0mm]
(d)  
& $2.33 \times 10^{-64}$
& $4.62 \times 10^{-57}$
& $1.0 \times 10^{6}$ 
& $1.0 \times 10^{5}$ 
& $1.62 \times 10^{1}$ 
& $0.172$ 
\\[0mm] 
(e)  
& $2.85 \times 10^{-64}$
& $5.66 \times 10^{-57}$
& $1.0 \times 10^{4}$ 
& $8.0 \times 10^{2}$  
& $2.23 \times 10^{1}$ 
& $0.211$
\\[0mm]
(f)  
& $2.85 \times 10^{-64}$
& $5.66 \times 10^{-57}$
& $1.0 \times 10^{2}$ 
& $8.0$  
& $2.23 \times 10^{1}$ 
& $0.211$  
\\[1mm]
\hline
\hline
\end{tabular}
\end{center}
\label{table-1}
\end{table}
%%%%%%%%%%%%%%%%%%%%%%%%%%%%%%%%%%%%%%%%%%%%%%%%%%%%%%%%%%%%%%%%%%
%

We estimate the current strength of the large-scale magnetic fields. 
We identify a $k$-mode as the present horizon scale $H_0^{-1}$ by 
setting $k = 2\pi/\left( 2997.9 h^{-1} \right) \, 
{\mathrm{Mpc}}^{-1}$ with $h=0.673$~\cite{Ade:2013lta}. 
In this case, the Hubble parameter at the inflationary stage 
is written as 
\begin{equation}
H_{\rm inf} \left( t_\mathrm{R} - t_k \right) = 45 + \ln \left( \frac{L_k}{[\mathrm{Mpc}]} \right) + 
\ln \left\{ \frac{\left[ 30/\left( \pi^2 g_\mathrm{R} \right) \right]^{1/12} 
            \left({\rho}^{(Y)} \left( t_{\mathrm{R}} \right) \right)^{1/4}}{
          10^{38/3} \hspace{1mm} \mathrm{[GeV]} } \right\}\,,
\label{eq:MR1-29-IIIC-01}
\end{equation} 
under the assumption of instantaneous reheating after inflation~\cite{Kolb and Turner}.
In Table~\ref{table-1}, we list the parameter sets to generate the current strength of magnetic fields of $B(H_0^{-1}, t_0) = {\cal O}(10^{-64})$ G  
at the Hubble horizon scale, 
for $X(t_k) = \exp \left( \chi_k \right)$ with $\chi_k = -0.940$ 
and $g_{\mathrm{ps}} = 1.0$. 
We find that for the wide range of $H_{\mathrm{inf}}$ and $m$, 
$C_+(k, t_{\mathrm{R}})$ is ${\cal O}(0.1)$. 
For the clear comparison with the results in the literature, 
we also calculate the current field strength of the magnetic fields at 
$1$Mpc scale. 
We note that the most important parameter to determine the magnetic field strength is $\chi_k$. 
The essence is that the amplitude of quantum fluctuations of the $\mathrm{U}(1)_Y$ fields generated inside the Hubble horizon can be a factor of $1/\sqrt{X(t_k)}$ larger than that in the ordinary Maxwell theory. 
Thus, the energy density of the (hypercharge) magnetic fields 
can be amplified by the factor of ratio of the final value of 
$X(t_{\mathrm{R}}) = 1$ at the inflationary stage to 
the initial value of $X(t_k)$. 

One of the important properties in this model is that 
the smaller $X(t_k)$ is, the larger 
the strength of the current magnetic fields 
$B(H_0^{-1}, t_0)$ on the Hubble horizon scale becomes. 
For all the cases (a)--(f) in Table~\ref{table-1}, 
the results are compatible with 
the observational constraints on non-Gaussianity~\cite{Ade:2013ydc} 
and the tensor-to-scalar ratio~\cite{Ade:2013uln} 
obtained from the Planck satellite, which are explained in the next section. 

We discuss the case of the non-instantaneous reheating 
and consider the sensitivity of the results on the duration 
of the reheating stage and the dependence of the results on 
the final reheating temperature. 
For the non-instantaneous reheating, 
the stage of oscillation of the inflaton should be taken into 
account, in which the energy density of the inflaton field 
evolves as being proportional to $a^{-3}$, namely, 
it behaves as matter. 
According to Ref.~\cite{Bassett:2000aw}, in which 
the evolution of the magnetic fields during preheating has 
been examined, 
if the conductivity of the universe $\sigma_\mathrm{c}$ is 
much larger than the Hubble expansion rate at the reheating stage, 
the amplification of the resultant magnetic fields does not occur. 
Thus, in our scenario, provided that $\sigma_\mathrm{c} \ll H$ 
at the reheating stage, the quantitative results could not 
differ very much from those for the instantaneous reheating stage. 
Moreover, when the final reheating temperature is lower, 
the value of the Hubble parameter at the end of the 
reheating stage is also smaller, and 
therefore, from Table I, it is seen that the 
the current strength of the magnetic fields becomes weaker.

%%%%%%%%%%%%%%%%%%%
%%%  Sec. IV
%%%%%%%%%%%%%%%%%%%
\section{Power spectrum, non-Gaussianity, and tensor-to-scalar ratio of the curvature perturbations}

In this section, we study the power spectrum of the curvature perturbations 
and estimate non-Gaussianity and the tensor-to-scalar ratio, provided that 
the curvature perturbations generated during inflation originate from only the 
quantum fluctuations of $Y$, the inflaton field, and the contribution of the 
scalar field $X$ is negligible because we consider the case in which 
the energy density of the potential of $Y$ is much larger than that 
of $X$ at the inflationary stage.

%%%%%%%%%%%%%%%%%%%
%%%  Sec. IV A
%%%%%%%%%%%%%%%%%%%
\subsection{Power spectrum of the curvature perturbations} 

First, we explore the power spectrum of the curvature perturbations 
originating from the quantum fluctuations of $Y$ corresponding to 
the inflaton field. 
It is known that the coupling term between $Y$ and 
$F_{\mu\nu}\tilde{F}^{\mu\nu}$ can lead to 
the quantum fluctuations $\delta Y (t, \Vec{x})$ in terms of $Y$. 
These fluctuations satisfy the following equation~\cite{Anber:2009ua, Barnaby:2010vf, Barnaby:2011vw, BHKP-B} 
\begin{equation}
\frac{\partial^2 \delta Y (t, \Vec{x})}{\partial t^2} 
+ 3H \frac{\partial \delta Y (t, \Vec{x})}{\partial t} 
-\frac{\nabla^2 \delta Y (t, \Vec{x})}{a^2} 
= \frac{g_{\mathrm{ps}}}{M}
F_{\mu\nu}\tilde{F}^{\mu\nu}\,.
\end{equation}
The generic solution consists of two parts. One is the solution of the 
homogeneous equation, namely, 
the ordinary vacuum fluctuations at the inflationary stage. 
The other is the particular solution coming from the source term. 
The origin of the latter is considered to be the inverse decay of 
two quanta of the gauge field to the quantum fluctuation of $Y$. 
These two terms are independent each other. 
The power spectrum of scalar modes of the curvature perturbations 
on hypersurfaces of the uniform density 
$\mathcal{R} = -\left(H/\dot{Y}\right) \delta Y$
is defined by the two-point correlation function 
in the Fourier space~\cite{Barnaby:2011vw} as 
$<\mathcal{R}_{\Vec{k}} \mathcal{R}_{\Vec{k}'}> \equiv 
\left( 2\pi^2/k^3 \right) P_{\mathcal{R}} (k) \delta^{(3)} 
\left(\Vec{k} + \Vec{k}' \right)$. 
Thus, the resultant power spectrum becomes~\cite{Barnaby:2010vf, Barnaby:2011vw, Meerburg:2012id} 
\begin{eqnarray}
P_{\mathcal{R}} (k) \Eqn{\simeq} \Delta_{\mathcal{R}}^2 \left( \frac{k}{k_*} 
\right)^{n_\mathrm{s} -1} 
\left( 1 + \Delta_{\mathcal{R}}^2 f_\mathrm{S} (\xi) \exp\left(4\pi\xi \right) \right)\,, 
\label{eq:MR1-22-IVA-1} \\
\Delta_{\mathcal{R}}^2 \Eqn{=} \left( \frac{H_{\mathrm{inf}}}{2\pi} \right)^2 
\frac{H_{\mathrm{inf}}^2}{|\dot{Y}|^2}\,,
\label{eq:MR1-22-IVA-2} \\
f_\mathrm{S} (\xi) 
\Eqn{\cong} 
\begin{cases}
7.5 \times 10^{-5} \, \xi^{-6} \hspace{6.3mm} \mbox{for \, $\xi \gg 1$\,,}\\
3.0 \times 10^{-5} \, \xi^{-5.4} \quad \mbox{for \, $2 \leq \xi \leq 3$\,.}
\end{cases}
\label{eq:fS}
\end{eqnarray}
Here, $k_* = 0.002 \, \mathrm{Mpc}^{-1}$. 
In addition, we have 
\begin{equation}
\dot{Y} (t_\mathrm{R}) = \frac{3}{2} 
\left[-1 + \sqrt{1 + \left( \frac{2m}{3H_{\mathrm{inf}}} 
\right)^2 } \right] H_{\mathrm{inf}} Y_k \exp \left\{ 
\frac{3}{2} \left[-1 + 
\sqrt{1 + \left( \frac{2m}{3H_{\mathrm{inf}}} 
\right)^2 } \right] \left(N - 1\right) \right\}\,,
\label{eq:4.3}
\end{equation}
with $N$ the number of $e$-folds, where 
in deriving Eq.~(\ref{eq:4.3}), we have used Eq.~(\ref{eq:3.4}). 
Moreover, the spectral index $n_\mathrm{s}$ of scalar modes of 
the curvature perturbations is given by~\cite{Meerburg:2012id, Hinshaw:2012aka}
\begin{eqnarray}  
n_\mathrm{s} \Eqn{\simeq} 1 -6 \epsilon +2 \eta \,,
\label{eq:MR1-22-IVA-6} \\
\epsilon \Eqn{\equiv} \frac{M_\mathrm{P}^2}{2} 
\left( \frac{V' (Y)}{V(Y)} \right)^2 \,,
\label{eq:MR1-22-IVA-7} \\
\eta \Eqn{\equiv} M_\mathrm{P}^2 \frac{V'' (Y)}{V(Y)}\,,
\label{eq:MR1-22-IVA-8}
\end{eqnarray}
where the prime denotes the derivative with respect to $Y$ of 
$\partial/\partial Y$, and $\epsilon$ and $\eta$ are 
the so-called slow-roll parameters in terms of 
the potential $V(Y)$. 
According to the Planck result~\cite{Ade:2013uln}, 
by using the Planck and Wilkinson Microwave Anisotropy Probe (WMAP) 
data, the value of the spectral index is estimated as 
$n_\mathrm{s} = 0.9603 \pm 0.0073\, (95 \% \, \mathrm{CL})$. 
With the COBE~\cite{Smoot:1992td} 
normalization for the power spectrum 
of the curvature perturbation 
$\Delta_{\mathcal{R}}^2 (k) = 2.4 \times 10^{-9}$ 
at $k=k_* = 0.002 \, \mathrm{Mpc}^{-1}$, 
which is consistent with the Nine-Year WMAP result~\cite{Hinshaw:2012aka}, 
and the Planck result of $n_\mathrm{s} = 0.9603$, 
for $H_{\mathrm{inf}} = 1.0 \times 10^{13}$GeV 
and $\bar{V} = 5.07 \times 10^{-11} M_\mathrm{P}^4$ 
in Eq.~(\ref{eq:2.4}), from Eq.~(\ref{eq:MR1-22-IVA-1}) for 
$k = 2\pi/\left( 2997.9 h^{-1} \right) \, {\mathrm{Mpc}}^{-1}$ 
with $h=0.673$ and Eq.~(\ref{eq:MR1-22-IVA-6}), 
we acquire $m = 2.44 \times 10^{12} \, \mathrm{GeV}$ and 
$Y_k = 7.70 \times 10^{-2} M_{\mathrm{P}} = 
1.87 \times 10^{17} \, \mathrm{GeV}$. 
By using Eqs.~(\ref{eq:MR1-31-IVA-01}) and (\ref{eq:MR1-31-IVA-02}), 
the values of $m$ and $Y_k$ can be derived 
for other various values of $H_{\mathrm{inf}}$, e.g., 
those in Table~\ref{table-1}.

%%%%%%%%%%%%%%
% Figure 2
%%%%%%%%%%%%%%
\begin{figure}[tbp]
\begin{center}
%\resizebox{!}{8cm}{
   \includegraphics{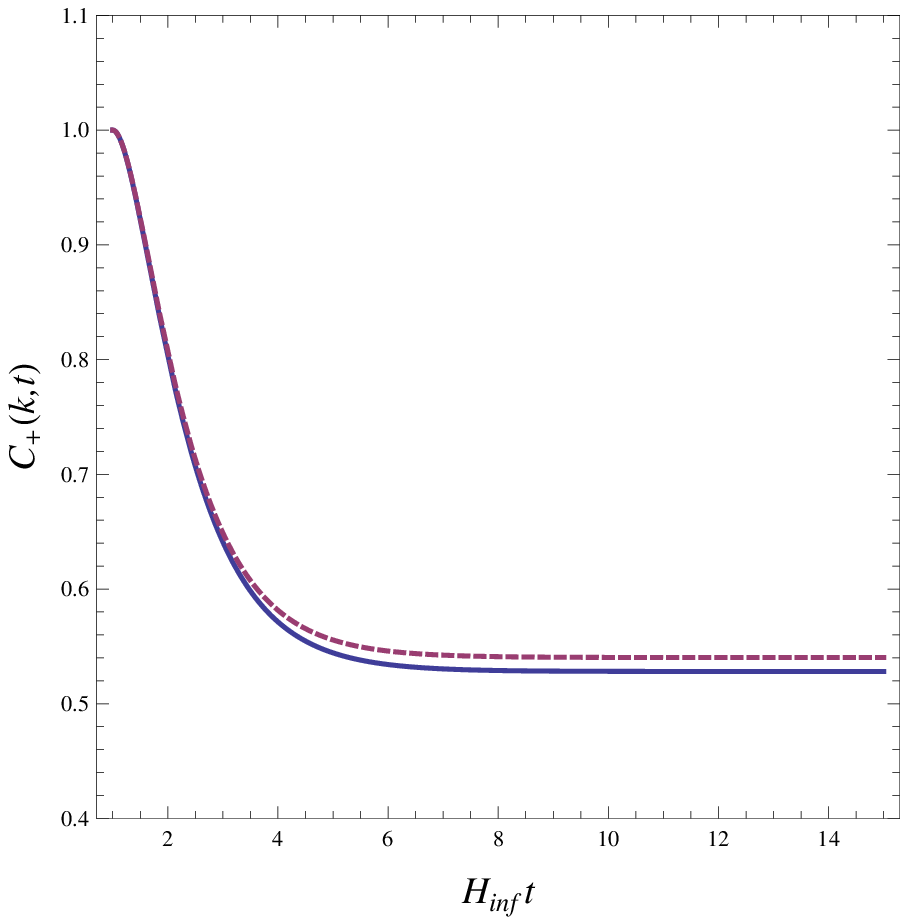}
%                  }
\caption{$C_{+}(k, t)$ as a function of $H_{\mathrm{inf}} t$. 
The legend is the same as in Fig.~\ref{fig-1} 
except $H_{\mathrm{inf}} = 1.0 \times 10^{13}$GeV and 
$m = 2.44 \times 10^{12}$GeV. 
(the case (A) in Tables~\ref{table-2} and~\ref{table-3}). 
}
\label{fig-2}
\end{center}
\end{figure}
%%%%%%%%%%%%%%

In Fig.~\ref{fig-2}, 
we display the evolution of $C_{+}(k, t)$ 
during inflation with the solid line 
for $X(t_k) \equiv \exp \left( \chi_k \right)$ with 
$\chi_k = -0.940$, 
$H_{\mathrm{inf}} = 1.0 \times 10^{13}$GeV, 
$M =1.0 \times 10^{-1} M_\mathrm{P} = 2.43 \times 10^{17} \, 
\mathrm{GeV}$, 
$m = 2.44 \times 10^{12}$GeV, 
$\bar{V} = 5.07 \times 10^{-11} M_\mathrm{P}^4$, 
$Y_k = 7.70 \times 10^{-2} M_{\mathrm{P}} = 
1.87 \times 10^{17} \, \mathrm{GeV}$, 
and $g_{\mathrm{ps}} = 1.0$. 
This is the case (A) in Tables~\ref{table-2} and~\ref{table-3} 
presented later. 
The procedure of the numerical calculation 
is the same as the one used to derive the results in Fig.~\ref{fig-1}. 
The qualitative features of evolution of $C_{+}(k, t)$ 
is equivalent to those shown in Fig.~\ref{fig-1}, 
namely, $C_{+}(k, t)$ becomes a constant around the 10 Hubble 
expansion time after the first horizon crossing during inflation. 
Even for different values of $H_{\mathrm{inf}}$, 
the evolution of $C_{+}(k, t)$ is the same as 
that in the case described above. 
Namely, the value of $C_{+}(k, t)$ asymptotically approaches 
a constant whose value is ${\cal O}(0.1)$. 

It follows from the values of the COBE normalization and Planck data 
that 
\begin{eqnarray} 
&&
f_\mathrm{S} (\xi) \exp\left(4\pi\xi \right) 
= \frac{25}{144} \times 10^8\,,  
\label{eq:MR1-31-IVA-01} \\
&&
Y_k = \pm \frac{M_{\mathrm{P}}}{\sqrt{2}\beta} 
\left[ 4\frac{\bar{V}}{M_{\mathrm{P}}^2 m^2 \beta^{-2}}+ 1 
\pm \sqrt{ 12\frac{\bar{V}}{M_{\mathrm{P}}^2 m^2 \beta^{-2}} +1} 
\right]^{1/2}\,,  
\label{eq:MR1-31-IVA-02} 
\end{eqnarray} 
with 
\begin{equation}
\beta \equiv
\sqrt{\frac{-\left(n_{\mathrm{s}} -1 \right)}{8}} \,.
\label{eq:MR1-31-IVA-03} 
\end{equation} 
Since $Y$ slowly rolls during inflation, $\xi$ can be considered to 
be a constant at the inflationary stage. 
Therefore, we use 
$\xi \simeq \xi_k = \left(g_{\mathrm{ps}} \dot{Y}(t_k) \right)/\left( 2 M X(t_k) H_{\mathrm{inf}}\right) 
= \left[3Y_k/\left(4MX(t_k)\right)\right] 
\left\{-1 + 
\sqrt{1 + \left[2m/\left(3H_{\mathrm{inf}}\right) 
\right]^2 } \right\}
\approx \left[Y_k/\left(6MX(t_k)\right)\right] 
\left(m^2/H_{\mathrm{inf}}^2\right)$, 
where the last approximate equality can be met for $m/H_{\mathrm{inf}} \ll 1$. 
Hence, if the values of $n_{\mathrm{s}}$, $\bar{V}$, and $H_{\mathrm{inf}}$ are given, we can determine those of $m$ and $Y_k$. 
Here, $\bar{V}$ and $M$ can be regarded as free parameters. 
We take the value of $\bar{V}$ derived from 
the relation $\bar{V} = 3 H_{\mathrm{inf}}^2 M_{\mathrm{P}}^2$, 
which corresponds to the Friedmann equation with $\dot{Y} = 0$ at $Y=0$. 
In this case,  
in Eq.~(\ref{eq:MR1-31-IVA-02}), we find 
$\bar{V}/\left(M_{\mathrm{P}}^2 m^2 \beta^{-2}\right) 
= 3H_{\mathrm{inf}}^2/\left(m^2 \beta^{-2}\right)$.
We also get the values of $m$ and $Y_k$ with Eqs.~(\ref{eq:MR1-31-IVA-01}) and (\ref{eq:MR1-31-IVA-02}). 
In addition, since the values of $m$ and $Y_k$ are real numbers, 
the values within the square root in Eqs.~(\ref{eq:MR1-31-IVA-01}) and (\ref{eq:MR1-31-IVA-02}) have to be larger than or equal to zero. Thus, we obtain 
the constraint on $\bar{V}$ as $\bar{V} > 2 \gamma H_{\mathrm{inf}}^4$. 
In what follows, we take the ``$+$'' sign in front of the right-hand side of $Y_k$ in Eq.~(\ref{eq:MR1-31-IVA-02}). 
Consequently, for $m/H_{\mathrm{inf}} \ll 1$, such cases are 
reasonable during inflation, we have 
\begin{eqnarray} 
m \Eqn{\approx} \sqrt{6} \xi_k \frac{M}{M_{\mathrm{P}}} X(t_k) 
H_{\mathrm{inf}}\,,
\label{eq:m-00} \\
Y_k \Eqn{\approx} \sqrt{6} M_{\mathrm{P}} \frac{H_{\mathrm{inf}}}{m} 
= \frac{M_{\mathrm{P}}^2}{\xi_k X(t_k) M}\,,
\label{eq:Yk-00} 
\end{eqnarray} 
where in deriving the last equality in Eq.~(\ref{eq:Yk-00}), 
we have used Eq.~(\ref{eq:m-00}). 
As a result, with the value of $\xi_k$ from Eq.~(\ref{eq:MR1-31-IVA-01}) and substituting it into Eqs.~(\ref{eq:m-00}) and (\ref{eq:Yk-00}), 
we obtain the approximate values of $m$ and $Y_k$. 
Indeed, from the lower relation for $2 \leq \xi \leq 3$ in Eq.~(\ref{eq:fS}), 
we numerically find that a solution of Eq.~(\ref{eq:MR1-31-IVA-01}) 
is $\xi_k = 2.5590616$. 
In the following, we evaluate the value of $m$ with Eq.~(\ref{eq:m-00}) 
and that of $Y_k$ with Eq.~(\ref{eq:MR1-31-IVA-02}).

%%%%%%%%%%%%%%%%%%%
%%%  Sec. IV B
%%%%%%%%%%%%%%%%%%%
\subsection{Non-Gaussianity} 

We suppose that the $\mathrm{U}(1)_Y$ gauge field couples 
to another scalar field, e.g., the Higgs-like field $\varphi$. 
In this case, the covariant derivative for $\varphi$ is defined by 
$D_\mu \equiv \partial_\mu + i g' F_\mu$, 
where $g'$ is the gauge coupling, and thus 
the kinetic term of $\varphi$ becomes $\left| D \varphi 
\right|^2$~\cite{Meerburg:2012id}. 
Provided that the gauge field obtains its mass through 
the Higgs mechanism in terms of $\varphi$. 
The quantum fluctuations of the gauge field mass 
are produced by the quantum fluctuations of $\varphi$. 
Eventually, the quantum fluctuations yield 
in the amount of quanta of the generated gauge field. 
As a result, the generation of the gauge field leads to 
the perturbations of number of $e$-folds of inflation 
$\delta N$. This produces the local type non-Gaussinanity in the 
anisotropy of the CMB radiation. 
Non-Gaussianity can be calculated 
by using the $\delta N$ formalism~\cite{Sasaki:1995aw, Starobinsky:1986fxa, Lyth:2005fi} and deriving the curvature perturbations originating from 
the quantum fluctuations of $\varphi$. 
When we consider the inflationary model in Ref.~\cite{Linde:2012bt}\footnote{For the model in Ref.~\cite{Barnaby:2010vf, Barnaby:2011vw}, 
the equilateral-type non-Gaussianity appears. 
Since the constraints on the local-type non-Gaussianity from the Planck data~\cite{Ade:2013ydc} are stronger 
than those on the equilateral-type on the local-type non-Gaussianity, in this work we examine the local-type on the local-type non-Gaussianity.}, 
by using the COBE~\cite{Smoot:1992td} 
normalization for the power spectrum of the curvature perturbations 
$\Delta_{\mathcal{R}}^2 (k) = 2.4 \times 10^{-9}$ 
at $k=k_* = 0.002 \, \mathrm{Mpc}^{-1}$, 
the local type non-Gaussianity $f_\mathrm{NL}^\mathrm{local}$ 
is expressed as~\cite{Meerburg:2012id}
\begin{equation}
f_\mathrm{NL}^\mathrm{local} \approx 1.0 \times 10^{14} \Delta 
N_\mathrm{max}^3 \frac{g'{}^4}{\xi^{6}} 
\frac{m^2}{H_{\mathrm{inf}}^2}\,.
\label{eq:4.1}
\end{equation}
Here, $\Delta N_\mathrm{max}$ is the maximum value of an extra numbers of 
$e$-folds, and $\xi$ is defined by Eq.~(\ref{eq:4.2}) with 
$\dot{Y}$ in Eq.~(\ref{eq:4.3}). 

The reason why in the previous sections, the coupling between 
$F_\mu$ and $\varphi$ through the covariant derivative of 
$D_\mu$ is as follows. 
Such a coupling might lead to the amplification 
of the $\mathrm{U}(1)_Y$ hypercharge gauge field $F_{\mu}$ 
during the reheating stage 
because the conformal invariance of the hypercharge electromagnetic 
fields is broken through this coupling. 
However, it has been indicated in Ref.~\cite{Bassett:2000aw} that 
if the conductivity of the universe 
is much larger than the Hubble parameter during the reheating stage, 
such a amplification cannot be realized. 
Therefore, when we estimate the resultant field strength of the 
large-scale magnetic fields, it is not necessary to 
take into consideration this coupling. 
On the other hand, the physical motivation why we consider 
the existence of the additional scalar field $\varphi$ 
and introduce it is the following. 
It is known that in string theories, the gauge symmetry 
is broken spontaneously, and the gauge fields obtain 
their mass. Hence, by introducing the coupling of $F_{\mu}$ 
to $\varphi$, which evolves to its vacuum expectation value 
like a Higgs field, we investigate the cosmological consequence 
of the spontaneous symmetry breaking. 
In such a case, the number of $e$-folds $N$ during inflation 
could be changed by the perturbations of $\varphi$, 
so that the curvature perturbations can be generated 
through the perturbations of $\varphi$~\cite{Meerburg:2012id}. 
As a result, the local-type non-Gaussianity in terms of the 
curvature perturbations is produced.

%%%%%%%%%%%%%%%%%%%%%%%%%%%%
%%%%%%%%  Table II  %%%%%%%%
%%%%%%%%%%%%%%%%%%%%%%%%%%%%
\begin{table}[t]
\caption{
Local type non-Gaussianity of the curvature perturbations. 
Legend is the same as in Table~\ref{table-1} with 
$\Delta N_\mathrm{max} = 1.0$. 
The value of $\bar{V}$ is determined by using 
the relation $\bar{V} = 3 H_{\mathrm{inf}}^2 M_\mathrm{P}^2$ 
as 
$\bar{V} = 5.07 \times 10^{-11} M_\mathrm{P}^4$ for the case (A) 
and 
$\bar{V} = 5.07 \times 10^{-13} M_\mathrm{P}^4$ for the case (B). 
The value of $C_+(k, t_{\mathrm{R}})$ is 
(the case (A), the case (B)) = $(0.528, 0.528)$. 
Moreover, the current field strength of the magnetic fields on $1$Mpc 
scale $B(1\mathrm{Mpc}, t_0) \hspace{1mm} [\mathrm{G}]$ 
is (the case (A), the case (B)) = 
$(1.42 \times 10^{-56}, 1.42 \times 10^{-56})$. 
}
\begin{center}
\tabcolsep = 2mm 
\begin{tabular}
{ccccccc}
\hline
\hline
& 
$f_\mathrm{NL}^\mathrm{local}$
& $g'{}^2$
& $H_{\mathrm{inf}} \hspace{1mm} [\mathrm{GeV}]$
& $m \hspace{1mm} [\mathrm{GeV}]$
& $Y_k/M_\mathrm{P}$ 
& $B(H_0^{-1},t_0) \hspace{1mm} [\mathrm{G}]$  
\\[0mm]
\hline
(A) 　
& 
$2.70$　
& $1.13 \times 10^{-5}$ 
& $1.0 \times 10^{13}$
& $2.44 \times 10^{12}$　
& $7.70 \times 10^{-2}$
& $7.15 \times 10^{-64}$ 
\\[0mm] 
(B) 　
& 
$2.12 \times 10^{8}$　
& $1.0 \times 10^{-1}$ 
& $1.0 \times 10^{12}$
& $2.44 \times 10^{11}$　
& $7.70 \times 10^{-2}$
& $7.15 \times 10^{-64}$ 
\\[1mm]
\hline
\hline
\end{tabular}
\end{center}
\label{table-2}
\end{table}
%%%%%%%%%%%%%%%%%%%%%%%%%%%%%%%%%%%%%%%%%%%%%%%%%%%%%%%%%%%%%%%%%
%

%%%%%%%%%%%%%%%%%%%%%%%%%%%%
%%%%%%%%  Table IIb  %%%%%%%%
%%%%%%%%%%%%%%%%%%%%%%%%%%%%
\begin{table}[t]
\caption{
Local type non-Gaussianity of the curvature perturbations. 
Legend for the case (C) is the same as 
the case (B) in Table~\ref{table-2} except  
$M =1.0 \times 10^{-2} M_\mathrm{P} = 2.43 \times 10^{16} \, 
\mathrm{GeV}$. 
In the case (C), we obtain $C_+(k, t_{\mathrm{R}}) = 0.423$. 
Furthermore, the current field strength of the magnetic fields on $1$Mpc scale 
is $B(1\mathrm{Mpc}, t_0) \hspace{1mm} [\mathrm{G}] = 3.59 \times 10^{-57}$. 
}
\begin{center}
\tabcolsep = 2mm 
\begin{tabular}
{cccccccc}
\hline
\hline
& $f_\mathrm{NL}^\mathrm{local}$
& $g'{}^2$
& $H_{\mathrm{inf}} \hspace{1mm} [\mathrm{GeV}]$
& $m \hspace{1mm} [\mathrm{GeV}]$
& $Y_k/M_\mathrm{P}$ 
& $\chi_k$
& $B(H_0^{-1},t_0) \hspace{1mm} [\mathrm{G}]$
\\[0mm]
\hline
(C)  
& $2.12 \times 10^{8}$ 
& $1.0 \times 10^{-1}$ 
& $1.0 \times 10^{12}$
& $2.44 \times 10^{11}$ 
& $7.70 \times 10^{-2}$
& $1.36$
& $1.81 \times 10^{-64}$ 
\\[1mm]
\hline
\hline
\end{tabular}
\end{center}
\label{table-2b}
\end{table}
%%%%%%%%%%%%%%%%%%%%%%%%%%%%%%%%%%%%%%%%%%%%%%%%%%%%%%%%%%%%%%%%%
%

In Table~\ref{table-2}, we display the numerical results of 
local non-Gaussianity $f_\mathrm{NL}^\mathrm{local}$ of 
the curvature perturbations by taking 
$\Delta N_\mathrm{max} = 1.0$, 
$M =1.0 \times 10^{-1} M_\mathrm{P} = 2.43 \times 10^{17} \, 
\mathrm{GeV}$, 
$\bar{V} = 5.07 \times 10^{-11} M_\mathrm{P}^4$ 
($5.07 \times 10^{-13} M_\mathrm{P}^4$) 
for the case (A) (the case (B)), 
$g_{\mathrm{ps}} = 1.0$, and 
$k = 2\pi/\left( 2997.9 h^{-1} \right) \, {\mathrm{Mpc}}^{-1}$ with $h=0.673$. 
Here, we have used the absolute value of $C_+(k, t_{\mathrm{R}})$
to estimate the resultant strength of magnetic fields as 
in Eq.~(\ref{eq:3.35}). 
According to the Planck satellite~\cite{Ade:2013ydc}, 
the constraint on $f_\mathrm{NL}^\mathrm{local}$ is given by 
$f_\mathrm{NL}^\mathrm{local} = 2.7 \pm 5.8 \, (68 \% \, \mathrm{CL})$. 
This has been improved very much in comparison with 
the Seven-Year WMAP analysis 
$-10 < f_\mathrm{NL}^\mathrm{local} < 74 \, (95 \% \, \mathrm{CL})$~\cite{Komatsu:2010fb}. {}From Table~\ref{table-2}, 
we find that for the case (A), the values of $f_\mathrm{NL}^\mathrm{local}$ can be compatible with the Planck data, 
whereas for the case (B), 
that of $f_\mathrm{NL}^\mathrm{local}$ is much larger. 
The upper limit on $f_\mathrm{NL}^\mathrm{local}$ of 
less than or equal to ${\cal O}(1)$ makes 
the space for our model parameters very small.  
However, there exists a viable room for 
the parameters such as the case (A) displayed in Table~\ref{table-2}. 
The constraint on $f_\mathrm{NL}^\mathrm{local}$ 
can be met by other close values of the parameters. 

We also demonstrate the case (C) of Table~\ref{table-2b}, 
in which $\Delta N_\mathrm{max}$, $\bar{V}$, $g'{}^2$, $g_{\mathrm{ps}}$, and
$k$ are the same as those in the case (B) of Table~\ref{table-2}, 
while the value of $M$ is smaller than that in Table~\ref{table-2}. 
Even though the value of $M$ is larger, 
the value of $f_\mathrm{NL}^\mathrm{local}$ is not changed. 
Since the upper limit of $f_\mathrm{NL}^\mathrm{local}$ 
is less than or equal to ${\cal O}(1)$, we see that 
the case (C) is not consistent with the observations. 
Thus, for a region of our model parameters, 
non-Gaussianity for the spectrum of the curvature perturbations 
can be compatible with the constraint from the Planck result. 

We emphasize that 
the main feature of our model is 
the presence of term of $X(t_k)$, which can make the 
large-scale magnetic field stronger. 
The contribution of this factor to non-Gaussianity $f_\mathrm{NL}^\mathrm{local}$ in Eq.~(\ref{eq:4.1}) is included through $\xi$ in Eq.~(\ref{eq:4.2}), 
$m$ in Eq.~(\ref{eq:m-00}), and $Y_k$ in Eq.~(\ref{eq:Yk-00}).

%%%%%%%%%%%%%%%%%%%
%%%  Sec. IV C
%%%%%%%%%%%%%%%%%%%
\subsection{Tensor-to-scalar ratio} 

In addition to the scalar modes of the curvature perturbations, 
the tensor modes, namely, gravitational waves, can be generated. 
The tensor-to-scalar ratio $r$ is defined by 
the ratio of amplitude of the tensor modes 
to that of the scalar modes. 
In the context of the present scenario, $r$ reads~\cite{Barnaby:2011vw} 
\begin{eqnarray}
&&
r= 
\begin{cases}
16 \epsilon (t_k) & \mbox{for $\xi \lesssim 3$\,,}\\
7.2 \epsilon^2 (t_k) & \mbox{for $\xi \to \infty$\,,}
\end{cases}
\label{eq:4.4} \\
&&
\epsilon (t_k) =
\frac{2M_\mathrm{P}^2 m^4 Y_k^2}{\left(2\bar{V} - m^2 Y_k \right)^2}\,, 
\label{eq:140318-1}
\end{eqnarray}
where $\epsilon (t_k) = \epsilon (t=t_k)$ 
in Eq.~(\ref{eq:MR1-22-IVA-7}), and 
we have used Eqs.~(\ref{eq:2.4}) and (\ref{eq:3.4}). 

We show the estimations of the tensor-to-scalar ratio $r$ 
in Tables~\ref{table-3} and \ref{table-3b}. 
The cases (A) and (B) are the same as those in Table~\ref{table-2}, 
that is, the values of $H_{\mathrm{inf}}$, 
$M$, $m$, $Y_k$, and $\chi_k$ are the same. 
Similarly, the case (C) is equivalent to that in Table~\ref{table-2b}.
We remark that since the values of $H_{\mathrm{inf}}$ and the ratio of $m$ to $H_{\mathrm{inf}}$ in the case (C) 
are the same as those in the case (B), the value of $r$ 
in the case (C) is also equal to that in the case (B). 
The upper limit from the Planck data is estimated as 
$r < 0.11 (95 \% \, \mathrm{CL})$~\cite{Ade:2013uln}. 
It is expected that future/current experiments for the polarization of the 
CMB radiation such as POLARBEAR~\cite{POLARBEAR} and LiteBIRD~\cite{LiteBIRD} 
can detect $r < 0.01$, and 
the future plan of LiteBIRD can observe $r < 0.001$~\cite{LiteBIRD}. 
As a result, when the magnetic fields on the Hubble horizon scale without 
the back reaction problem are generated at the present time, 
both the local non-Gaussianity and tensor-to-scalar ratio of the CMB 
radiation meeting the constraints from the Planck satellite 
can be produced in a region of the parameters. 

In order to check the effect of the dynamics of the $X$ field, 
we have also investigated a toy model with 
the dynamical $X$ field, in which the potential of 
$X = \exp \left(-\lambda \Phi/M_{\mathrm{P}} \right)$ is given by 
$U(X) = U(\Phi) =
\bar{U} \exp \left( -\tilde{\lambda} \Phi/M_{\mathrm{P}} \right)
$ with $\lambda$ a dimensionless constant and  
$\bar{U}$ a constant. 
As a consequence, we have acquired qualitatively 
similar results on the current field strength of the large-scale 
magnetic fields, non-Gaussianity $f_\mathrm{NL}^\mathrm{local}$ 
in Eq.~(\ref{eq:4.1}), and the tensor-to-scalar ratio $r$ 
in Eq.~(\ref{eq:4.4}).

%%%%%%%%%%%%%%%%%%%%%%%%%%%%
%%%%%%%%  Table III  %%%%%%%%
%%%%%%%%%%%%%%%%%%%%%%%%%%%%
\begin{table}[t]
\caption{
Tensor-to-scalar ratio of the curvature perturbations 
for the cases (A) and (B). 
Legend is the same as Table~\ref{table-2}. 
}
\begin{center}
\tabcolsep = 2mm 
\begin{tabular}
{ccccc} 
\hline
\hline
& $r$
& $H_{\mathrm{inf}} \hspace{1mm} [\mathrm{GeV}]$
& $m \hspace{1mm} [\mathrm{GeV}]$
& $Y_k/M_\mathrm{P}$ 
\\[0mm]
\hline
(A)  
& $1.87 \times 10^{-5}$
& $1.0 \times 10^{13}$
& $2.44 \times 10^{12}$
& $7.70 \times 10^{-2}$
\\[0mm]
(B)  
& $1.87 \times 10^{-5}$ 
& $1.0 \times 10^{12}$
& $2.44 \times 10^{11}$
& $7.70 \times 10^{-2}$
\\[1mm]
\hline
\hline
\end{tabular}
\end{center}
\label{table-3}
\end{table}
%%%%%%%%%%%%%%%%%%%%%%%%%%%%%%%%%%%%%%%%%%%%%%%%%%%%%%%%%%%%%%%%%
%
%%%%%%%%%%%%%%%%%%%%%%%%%%%%
%%%%%%%%  Table IIIb  %%%%%%%%
%%%%%%%%%%%%%%%%%%%%%%%%%%%%
\begin{table}[t]
\caption{
Tensor-to-scalar ratio of the curvature perturbations for the case (C). 
Legend is the same as Table~\ref{table-2b}. 
}
\begin{center}
\tabcolsep = 2mm 
\begin{tabular}
{ccccc} 
\hline
\hline
& $r$
& $H_{\mathrm{inf}} \hspace{1mm} [\mathrm{GeV}]$
& $m \hspace{1mm} [\mathrm{GeV}]$
& $Y_k/M_\mathrm{P}$ 
\\[0mm]
\hline
(C)  
& $1.87 \times 10^{-5}$
& $1.0 \times 10^{12}$
& $2.44 \times 10^{11}$
& $7.70 \times 10^{-2}$ 
\\[1mm]
\hline
\hline
\end{tabular}
\end{center}
\label{table-3b}
\end{table}
%%%%%%%%%%%%%%%%%%%%%%%%%%%%%%%%%%%%%%%%%%%%%%%%%%%%%%%%%%%%%%%%%
%

In addition, we mention that 
the BICEP2 experiment has recently observed the $B$-mode polarization of the CMB radiation with $r=0.20_{-0.05}^{+0.07}\, (68 \% \, \mathrm{CL})$~\cite{Ade:2014xna}. There are discussions on the way of subtracting the foreground data~\cite{A-A, MS-KK}. Our investigations related to the BICEP2 result on 
$r$ is described in Appendix A. 

In comparison with the past works, the important property of our model 
is that there exists the term of 
$X(t_k)$ leading to the strong magnetic fields. 
This term contributes to the tensor-to-scalar ratio $r$ in Eq.~(\ref{eq:4.4}) 
with $\epsilon (t_k)$ in Eq.~(\ref{eq:140318-1}) via 
$m$ in Eq.~(\ref{eq:m-00}) and $Y_k$ in Eq.~(\ref{eq:Yk-00}). 
In our model, in principle, thanks to the factor of $X(t_k)$, 
the large-scale magnetic fields with 
its strong amplitude to account for the observational values 
only through the adiabatic compression without dynamo mechanism. 
The reason why we only have small values of 
the magnetic field strength in Tables I--III is 
that in this work, we attempt to simultaneously explain three 
observational quantities, namely, 
large-scale magnetic fields, 
non-Gaussianity of the curvature perturbations, 
and the tensor-to-scalar ratio. 
This point is the crucial advantage of our model.

%%%%%%%%%%%%%%%%%%%
%%%  Sec. V
%%%%%%%%%%%%%%%%%%%
\section{Conclusions}

In the present paper, we have explored 
the generation of large-scale magnetic fields in 
a toy model of the so-called moduli inflation. 
In this model, the conformal invariance of the hypercharge 
electromagnetic fields are broken 
due to their coupling to both the scalar and pseudoscalar fields appearing in the framework of string theories. 
We have studied the current strength of the magnetic fields on 
the Hubble horizon scale, 
local non-Gaussianity of the curvature perturbations 
originating from the existence of the massive gauge fields, 
and the tensor-to-scalar ratio. 
As a consequence, it has been shown that 
in addition to the magnetic fields on the Hubble horizon scale, whose 
current field strength is compatible with the back reaction problem, 
local non-Gaussianity and the tensor-to-scalar ratio 
of the power spectrum of the CMB radiation 
can be generated, 
the values of which are consistent with the constraints observed 
by the Planck satellite, i.e., $f_\mathrm{NL}^\mathrm{local} = \mathcal{O}$(1) 
and $r < 0.11 (95 \% \, \mathrm{CL})$~\cite{Ade:2013uln}. 

It should be remarked that one of the most important achievement of this work is to derive the explicit values of three cosmological observables such as 
the large-scale magnetic fields, local non-Gaussianity, and the tensor-to-scalar ratio for the first time.

%%%%%%%%%%%%%%%%%%%%%%%%
%%%  Acknowledgments
%%%%%%%%%%%%%%%%%%%%%%%%
\section*{Acknowledgments}

%%%
The author would like to sincerely appreciate the discussions on 
the initial stage of this work  
with Professor Tatsuo Kobayashi and Professor Osamu Seto. 
This work was not able to be executed 
without their very important and helpful suggestions. 
Moreover, he would like to express his gratitude especially to 
Professor Akio Sugamoto, who kindly read the draft of this 
manuscript and presented me a number of quite sincere suggestions 
and comments, 
and also to Professor Masahiro Morikawa for his very hearty discussions 
and advice. 
In addition, he really appreciate the very warm hospitality of the Kobayashi-Maskawa Institute for the Origin of Particles and the Universe (KMI) at Nagoya University very much, in which he has completed almost all the parts of this work. 
%%%
Furthermore, he acknowledges 
KEK Theory Center Cosmophysics Group who organized the Workshop: 
Accelerators In the Universe 2012 (KEK-CPWS-AIU2012)``Axion Cosmophysics''
and organizers of ``the 3rd International Workshop on 
Dark Matter, Dark Energy and Matter-Antimatter Asymmetry'' 
for their warm hospitality as well as financial support. 
In these workshops, this work was initiated. 
In addition, he is grateful to Professor Gi-Chol Cho, Professor Shin'ichi Nojiri, Professor Sergei D. Odintsov, Professor Mohammad Sami, Professor Misao Sasaki, Professor Tatsu Takeuchi, 
Professor Koichi Yamawaki, and Professor Jun'ichi Yokoyama 
for their really hearty continuous encouragements. 
He also thanks Professor Shinji Mukhoyama, Dr. Ryo Namba, 
and Professor Shuichiro Yokoyama for their useful comments.
%%%
This work was partially supported by 
the JSPS Grant-in-Aid for Young Scientists (B) \# 25800136 (K.B.).

%%%%%%%%%%%%%%%%%%%
%%%  Appendix 
%%%%%%%%%%%%%%%%%%%
\appendix

%%%%%%%%%%%%%%%%%%%
%%%  A
%%%%%%%%%%%%%%%%%%%
\section{Axion monodromy inflation}
 
The tensor-to-scalar ratio $r$ in moduli inflation is much smaller than the BICEP2 result\footnote{There have been proposed scalar field models of inflation to realize the BICEP2 result on $r$, e.g., 
in Refs.~\cite{S-B, KS, KSY-HKSY}.}, although it is still consistent with the Planck data. In this Appendix, we explore 
axion monodromy inflation and derive the value of $r$ in order to compare it with that in moduli inflation.  
We explore the following potential~\cite{KSY-HKSY}
\begin{equation}
V(Y) = A Y^q \,, \quad 
q=1\,,
\label{eq:5.1}
\end{equation}
with $A$ a constant. 
In axion monodromy inflation, there is only the pseudoscalar field $Y$, 
and therefore the scalar field $\Phi$, i.e., the scalar quantity $X = 1$ in Eq.~(\ref{eq:2.2}) does not exist. 
Hence, the total Lagrangian 
becomes $\mathcal{L}$ in Eq.~(\ref{eq:2.1}) with $X=1$ (namely, $\Phi = 0$) and $V(Y)$ in Eq.~(\ref{eq:5.1}) instead of that in Eq.~(\ref{eq:2.4}). 
We note that as the other form of the potential, we can consider  
$V(Y) = \left(1/2\right) m^2 Y^2$, 
which follows from the limit $ Y/f \ll 1$ of the potential 
$V(Y) = \lambda^4 \left( 1-\cos \left(Y/f\right) \right)$ 
analyzed in Refs.~\cite{Barnaby:2010vf, Barnaby:2011vw}. 

For the potential $V(Y)$ in Eq.~(\ref{eq:5.1}), 
the slow-roll inflation is supposed to be realized, the solution 
of Eq.~(\ref{eq:2.10}) is given by 
\begin{equation} 
Y=\bar{Y} t\,, \quad \bar{Y} = -\frac{A}{3 H_{\mathrm{inf}}}\,.
\label{eq:5.3}
\end{equation}
The field equation of $F_\mu$ in Eq.~(\ref{eq:2.11}) becomes 
\begin{equation}
\ddot{F_i}(t,\Vec{x}) + 
H\dot{F_i}(t,\Vec{x}) 
- \frac{1}{a^2}{\partial}_j {\partial}_j F_i(t,\Vec{x}) 
+ \frac{g_{\mathrm{ps}}}{M} \frac{1}{a} \frac{A}{3 H_{\mathrm{inf}}} 
\epsilon^{ijk} {\partial}_j F_k(t,\Vec{x}) = 0\,, 
\label{eq:5.4}
\end{equation}   
where we have used Eq.~(\ref{eq:5.3}). 
Moreover, with Eq.~(\ref{eq:5.3}), $\xi$ in Eq.~(\ref{eq:4.2}) reads
\begin{equation} 
\xi = -\frac{1}{6} \frac{g_{\mathrm{ps}}}{M}
\frac{A}{H_{\mathrm{inf}}^2}\,. 
\label{eq:5.5}
\end{equation}
Clearly, this is not a dynamical quantity but a constant. 

By using Eqs.~(\ref{eq:MR1-22-IVA-2}) with the COBE normalization $\Delta_{\mathcal{R}}^2 (k) = 2.4 \times 10^{-9}$  and (\ref{eq:MR1-22-IVA-6})--(\ref{eq:MR1-22-IVA-8}) with the Planck data $n_\mathrm{s} = 0.9603$ and assuming that $t \approx H_{\mathrm{inf}}^{-1}$ during inflation, we find $\epsilon = 6.62 \times 10^{-3}$ and 
$A = \left[3/\left(4\sqrt{6} \pi \right) \right] \times 10^5  H_{\mathrm{inf}}^3$. This value of $\epsilon$ is realized if $H_{\mathrm{inf}} = 6.51 \times 10^{15}$GeV. 
Moreover, it follows from Eq.~(\ref{eq:5.5}) that if $M = 3.55 \times 10^{18}$GeV, $g_{\mathrm{ps}} = 1.0$, and $H_{\mathrm{inf}} = 6.51 \times 10^{15}$GeV, 
we get $|\xi| = 2.98$. {}From Eq.~(\ref{eq:5.5}), we obtain $r = 16 \epsilon = 0.106$. This is the same order of the BICEP2 result. 
Thus, in axion monodromy inflation, the tensor-to-scalar ratio compatible with the BICEP2 result can be produced.

%%%%%%%%%%%%%%%%%%%
%%%  B
%%%%%%%%%%%%%%%%%%%
\section{Issues of the backreaction and the strong coupling}

In this Appendix, we explain the issues of the backreaction and the strong coupling. 
The back reaction problem by the generation of electromagnetic fields 
during inflation has been found~\cite{Campanelli:2013mea, Demozzi:2009fu, Kanno:2009ei, Suyama:2012wh, Fujita:2012rb, DMR-C} (for more recent related works on the relation between the generated gauge fields and inflation, 
see~\cite{Maleknejad:2012fw, Cembranos:2012ng, Bartolo:2013msa, 
Namba:2013kia, Nurmi:2013gpa, Fujita:2014sna}). 
It has been pointed out~\cite{Demozzi:2009fu} that 
the amplitude of the current magnetic fields on ${\cal O}(1)$ Mpc scale 
should be less than $10^{-32}$ G. 
In such a case, the dynamics of inflation is not disturbed by 
the back reaction originating from the generation of electromagnetic fields. 
This means that the strength of the magnetic fields on the Hubble horizon 
scale should be less than $10^{-35}$G, which can be derived by 
$B(k, t) \propto 
\left(k/H_{\mathrm{inf}}^{-1}\right)^{0.8}$~\cite{Demozzi:2009fu}. 
Throughout this paper, we take parameter sets (for a given coherence scale 
$\propto k^{-1}$) in which the current magnetic field strength can 
satisfy this constraint. 

In addition, the strong coupling problem that 
the very strong gauge coupling is necessary to amplify the 
gauge fields during inflation has been indicated 
in Ref.~\cite{Demozzi:2009fu}. 
Recently, as a solution for this problem, 
the so-called sawtooth model for the coupling between a scalar 
field and the $\mathrm{U}(1)_Y$ fields has been proposed 
in Ref.~\cite{Ferreira:2013sqa}. 
In this scenario, the behavior of the scalar field is a sawtooth path.  
As a result, the magnetic field strength about $10^{-16}$ G 
on $1$ Mpc scale at the present time can be generated without facing the strong coupling problem as well as the back reaction problem. 
Furthermore, according to the updated analysis in Ref.~\cite{Ferreira:2014hma} 
with the recent data from the BICEP2 experiment~\cite{Ade:2014xna}, the magnetic field strength on $1$ Mpc scale should be less than $10^{-30}$G. 
It is quite interesting to apply our analysis on 
the generation of large-scale magnetic fields and 
the estimation of power spectrum, non-Gaussianity and 
tensor-to-scalar ratio of the curvature perturbations 
to more realistic moduli inflation models such as 
the racetrack inflation model. 

The strong coupling problem could be solved 
in the sawtooth scenario~\cite{Ferreira:2013sqa, Ferreira:2014hma}. 
Moreover, in Ref.~\cite{Caprini:2014mja}, it has been pointed out that 
thanks to the inverse cascade mechanism, the constraints obtained in 
Ref.~\cite{Demozzi:2009fu} can be evaded. 
Thus, it is important to study 
whether the sawtooth-like evolution of the dilaton field
leading to the large-scale magnetic fields with their sufficient 
strength can be realized in moduli inflation or not. 
It may be useful to investigate the racetrack inflation model 
with positive exponent potential terms, 
because they induce a quite high potential wall 
for a large value of the dilaton field~\cite{AHK-AHKS}.

%%%%%%%%%%%%%%%%%%%
%%%  C
%%%%%%%%%%%%%%%%%%%
\section{Constraints on the strength of cosmic magnetic fields}

In this Appendix, we present 
the upper bounds of the magnetic field strength.  
The observations of the CMB radiation imply that 
the upper limit on the magnetic field strength on $1$Mpc scale is 
$\sim 10^{-9}$G~\cite{CMB-Limit, Giovannini:2008df}
and that on the magnetic field strength on the scale 
larger than the present Hubble horizon 
is $4.8 \times 10^{-9} \mathrm{G}$~\cite{Barrow:1997mj}. 
In Ref.~\cite{Pogosian:2013dya}, 
by using the data of the polarized radiation imaging 
and spectroscopy mission (PRISM)~\cite{Andre:2013afa}, 
it has been indicated that 
the magnetic fields with $\sim 10^{-9}$G can be detected. 

Moreover, there are other methods, such as 
the 21cm fluctuations of the neutral hydrogen~\cite{Tashiro:2006uv}, 
the parameter $\sigma_8$ for the density perturbation of matter~\cite{YIKM}, 
the correlation of the curvature perturbations with the magnetic fields~\cite{Ganc:2014wia}, 
the data of the fifth science (S5) run from 
the Laser Interferometer Gravitational-wave Observatory 
(LIGO)~\cite{Wang:2008vp}, the X-ray galaxy cluster survey by 
Chandra, the Sunyaev-Zel'divich (S-Z) survey~\cite{TSLS-TS-TTI}, and 
primordial gravitational waves, namely, 
the tensor modes of the curvature perturbations, 
generated during inflation~\cite{Kuroyanagi:2009ez}. 
The upper limits from these observations are compatible with 
or weaker than those estimated by using the CMB radiation data. 
Generic investigations on the spectrum of 
the large-scale magnetic fields from inflation 
have been executed in Refs.~\cite{Bamba:2007hm, Bonvin:2013tba}. 
With the observations of a blazar, 
the lower bounds on the cosmic magnetic fields in void regions 
have also been estimated in Ref.~\cite{Takahashi:2013uoa}. 

On the other hand, for the magnetic fields on smaller scales, 
there are the upper bounds from the BBN. 
The upper limit of the magnetic field strength on the Hubble horizon scale 
at the BBN epoch $\sim 9.8 \times 10^{-5} h^{-1}\mathrm{Mpc}$ 
with $h=0.673$~\cite{Ade:2013lta}, 
is less than $10^{-6}$G~\cite{BBN}. 

Incidentally, various issues related to the cosmic magnetic fields 
have been discussed: Intergalactic magnetic fields~\cite{Nikiel-Wroczynski:2013zwa}, the relation between cosmological magnetic fields and 
blazars~\cite{Tashiro:2013bxa}, the influence of decay of 
the cosmic magnetic fields on the CMB radiation~\cite{Miyamoto:2013oua}, 
and the secondary anisotropies of the CMB radiation originating from 
stochastic magnetic fields~\cite{Kunze:2013iwa}. 
Moreover, constraints on the primordial magnetic fields have been proposed 
from the conversion between the CMB photon and graviton~\cite{Chen:2013gva}, 
the interaction of the CMB radiation with an axion~\cite{Tashiro:2013yea} in the context of the axiverse~\cite{Axiverse}, 
the trispectrum of the CMB radiation~\cite{Trivedi:2013wqa}, 
and the measurement of the Faraday rotation~\cite{BS-HF}.

%%%%%%%%%%%%%%%%%%%
%%%  D
%%%%%%%%%%%%%%%%%%%
\section{Cosmological implications}

In this Appendix, we state cosmological implications obtained from this work. 
There exists the possibility of baryogenesis coming from 
the large-scale magnetic fields generated from inflation. 
These magnetic fields can yield gravitational waves 
because the space-time is distorted by the existence of the 
magnetic fields, and eventually 
the magnetic helicity can be produced~\cite{Caprini:2003vc}. 
Moreover, the relation between the magnetic helicity and the cosmic chiral asymmetry has been investigated in detail~\cite{Tashiro:2012mf}. 
If the magnetic helicity exists before the EWPT, 
baryon numbers can be produced through the effect of the quantum anomaly~\cite{Joyce:1997uy, GS-GS}. The coupling of the electromagnetic fields 
to the pseudoscalar field can lead to the magnetic helicity, and 
thus moduli inflation driven by an axion-like pseudoscalar field 
can generate not only the large-scale magnetic fields but also 
the baryon asymmetry of the universe (for trial scenarios, see, e.g.,~\cite{Bamba:2006km, Bamba:2007hf}). 
It is meaningful to build a concrete inflationary model, 
in which both cosmic magnetic fields and baryons can be generated 
in the framework of fundamental theories such as string theories 
describing the physics in the early universe. 
In addition, a leptogenesis scenario due to the existence of the primordial magnetic fields has been proposed in Ref.~\cite{Long:2013tha}. 
In Ref.~\cite{Montiel:2014dia}, the idea that the component of dark energy 
may be non-linear electromagnetic fields has been proposed.  

We also state the detectability of cosmic magnetic fields. 
Current and/or future experiments on the polarizations of the CMB radiation, 
for example, Planck~\cite{Ade:2013uln, Ade:2013lta}, 
QUIET~\cite{QUIET-1, Samtleben:2008rb, Araujo:2012yh}, 
POLARBEAR~\cite{POLARBEAR}, B-Pol~\cite{B-Pol}, and LiteBIRD~\cite{LiteBIRD} 
can detect the large-scale magnetic fields with 
the current strength $\sim 4 \times 10^{-11} - 10^{-10}$G~\cite{Caprini:2003vc, Test}. For the magnetic fields with the left-handed magnetic helicity, 
the field strength $\sim 10^{-14}$G on $ \sim 10$Mpc scale 
can be observed~\cite{Tashiro:2013ita}. Further theoretical investigations 
on the properties of $B$-mode polarization of the CMB radiation has recently 
been examined in Ref.~\cite{Giovannini:2014oia}. 
Furthermore, there have been appeared 
various ideas to detect primordial magnetic fields 
such as future observations for low-medium redshift~\cite{Calabrese:2013lga} 
and the bias of magnification of lensing effects~\cite{Camera:2013fva}. 
Since there are a number of ways of detecting the cosmic magnetic fields, 
it is possible to examine the physics in both the 
early- and late-time universes through the detections of the primordial 
large-scale magnetic fields, especially, in the void structures or 
the inter-galactic region.

%%%%%%%%%%%%%%%%%%%%%%%%%%%%%%%%%
%% thebibliography environment
%%%%%%%%%%%%%%%%%%%%%%%%%%%%%%%%%


\begin{thebibliography}{99}

%%%%%%%%%%%%%%
%%%%%%%%%%%%%%
%%%%%%%%%%%%%%

%%%%%%%% M-F-part %%%%%%%%

\bibitem{M-F-Reviews}
%
%For reviews on cosmic magnetic fields, see 
%\cite{Kronberg:1993vk}
%\bibitem{Kronberg:1993vk}
  P.~P.~Kronberg,
  %``Extragalactic magnetic fields,''
  Rept.\ Prog.\ Phys.\  {\bf 57}, 325 (1994);\ 
  %%CITATION = RPPHA,57,325;%%
%
%\cite{Grasso:2000wj}
%\bibitem{Grasso:2000wj}
  D.~Grasso and H.~R.~Rubinstein,
  %``Magnetic fields in the early universe,''
  Phys.\ Rept.\  {\bf 348}, 163 (2001)
  [arXiv:astro-ph/0009061];\
  %%CITATION = PRPLC,348,163;%%
%
%\cite{Carilli:2001hj}
%\bibitem{Carilli:2001hj} 
  C.~L.~Carilli and G.~B.~Taylor,
  %``Cluster magnetic fields,''
  Ann.\ Rev.\ Astron.\ Astrophys.\  {\bf 40}, 319 (2002)
  [astro-ph/0110655];\
  %%CITATION = ASTRO-PH/0110655;%%
%
%\cite{Widrow:2002ud}
%\bibitem{Widrow:2002ud}
  L.~M.~Widrow,
  %``Origin of Galactic and Extragalactic Magnetic Fields,''
  Rev.\ Mod.\ Phys.\  {\bf 74}, 775 (2002)
  [arXiv:astro-ph/0207240];\ 
  %%CITATION = RMPHA,74,775;%%
%
%\cite{Giovannini:2003yn}
%\bibitem{Giovannini:2003yn}
  M.~Giovannini,
  %``The magnetized universe,''
  Int.\ J.\ Mod.\ Phys.\  D {\bf 13}, 391 (2004)
  [arXiv:astro-ph/0312614];\ 
  %%CITATION = IMPAE,D13,391;%%
%
%\cite{Giovannini:2004rj}
%\bibitem{Giovannini:2004rj} 
%  M.~Giovannini,
  %``Theoretical tools for the physics of CMB anisotropies,''
%  Int.\ J.\ Mod.\ Phys.\ D {\bf 14}, 363 (2005)
\textit{ibid}.\  {\bf 14}, 363 (2005) 
  [astro-ph/0412601];\
  %%CITATION = ASTRO-PH/0412601;%%
%
%\cite{Giovannini:2006kg}
%\bibitem{Giovannini:2006kg}
%  M.~Giovannini,
  %``Magnetic fields, strings and cosmology,''
  Lect.\ Notes Phys.\  {\bf 737}, 863 (2008)
  [arXiv:astro-ph/0612378];\
  %%CITATION = LNPHA,737,863;%%
%
%\cite{Kandus:2010nw}
%\bibitem{Kandus:2010nw}
  A.~Kandus, K.~E.~Kunze and C.~G.~Tsagas,
  %``Primordial magnetogenesis,''
  Phys.\ Rept.\  {\bf 505}, 1 (2011)
  [arXiv:1007.3891 [astro-ph.CO]];\ 
  %%CITATION = PRPLC,505,1;%%
%
%\cite{Yamazaki:2012pg}
%\bibitem{Yamazaki:2012pg} 
  D.~G.~Yamazaki, T.~Kajino, G.~J.~Mathew and K.~Ichiki,
  %``The Search for a Primordial Magnetic Field,''
%  Phys.\ Rept.\  {\bf 517}, 141 (2012)
\textit{ibid}.\ {\bf 517}, 141 (2012)
  [arXiv:1204.3669 [astro-ph.CO]];\ 
  %%CITATION = ARXIV:1204.3669;%%
  %10 citations counted in INSPIRE as of 12 Oct 2013
%
%\cite{Durrer:2013pga}
%\bibitem{Durrer:2013pga} 
  R.~Durrer and A.~Neronov,
  %``Cosmological Magnetic Fields: Their Generation, 
  %Evolution and Observation,''
  Astron.\ Astrophys.\ Rev.\  {\bf 21}, 62 (2013)
  [arXiv:1303.7121 [astro-ph.CO]].
  %%CITATION = ARXIV:1303.7121;%%
  %40 citations counted in INSPIRE as of 07 Mar 2014
%

\bibitem{Biermann1}
L.~Biermann and A.~Schl\"{u}ter,
Phys.\ Rev.\ {\bf 82}, 863 (1951). 

\bibitem{PI}
%\cite{Weibel:1959zz}
%\bibitem{Weibel:1959zz}
  E.~S.~Weibel,
  %``Spontaneously Growing Transverse Waves In A Plasma Due To An Anisotropic
  %Veloc Ity Distribution,''
  Phys.\ Rev.\ Lett.\  {\bf 2}, 83 (1959).

\bibitem{PT}
%
%\bibitem{Quashnock}
J.~M.~Quashnock, A.~Loeb, and D.~N.~Spergel,
Astrophys.\ J.\ {\bf 344}, L49 (1989);\ 
%
%\cite{Baym:1995fk}
%\bibitem{Baym:1995fk}
  G.~Baym, D.~Bodeker and L.~D.~McLerran,
  %``Magnetic fields produced by phase transition bubbles in the electroweak
  %phase transition,''
  Phys.\ Rev.\  D {\bf 53}, 662 (1996)
  [arXiv:hep-ph/9507429];\ 
  %%CITATION = PHRVA,D53,662;%%
%
%\cite{Boyanovsky:2002wa}
%\bibitem{Boyanovsky:2002wa}
  D.~Boyanovsky, H.~J.~de Vega and M.~Simionato,
  %``Large scale magnetogenesis from a non-equilibrium phase transition in the
  %radiation dominated era,''
%  Phys.\ Rev.\  D {\bf 67}, 123505 (2003)
\textit{ibid}.\ {\bf 67}, 123505 (2003)
  [arXiv:hep-ph/0211022];\ 
  %%CITATION = PHRVA,D67,123505;%%
%
%\cite{Boyanovsky:2002kq}
%\bibitem{Boyanovsky:2002kq}
  D.~Boyanovsky, M.~Simionato and H.~J.~de Vega,
  %``Magnetic field generation from non-equilibrium phase transitions,''
%  Phys.\ Rev.\  D {\bf 67}, 023502 (2003)
\textit{ibid}.\ {\bf 67}, 023502 (2003)
  [arXiv:hep-ph/0208272];\ 
  %%CITATION = PHRVA,D67,023502;%%
%
%\cite{Durrer:2003ja}
%\bibitem{Durrer:2003ja}
  R.~Durrer and C.~Caprini,
  %``Primordial Magnetic Fields and Causality,''
  JCAP {\bf 0311}, 010 (2003)
  [arXiv:astro-ph/0305059];\ 
  %%CITATION = JCAPA,0311,010;%%
%
%\cite{Kahniashvili:2009qi}
%\bibitem{Kahniashvili:2009qi}
  T.~Kahniashvili, A.~G.~Tevzadze and B.~Ratra,
  %``Phase Transition Generated Cosmological Magnetic Field at Large Scales,''
  Astrophys.\ J.\  {\bf 726}, 78 (2011)
  [arXiv:0907.0197 [astro-ph.CO]];\ 
  %%CITATION = ASJOA,726,78;%%
%
%\cite{Kahniashvili:2012uj}
%\bibitem{Kahniashvili:2012uj} 
  T.~Kahniashvili, A.~G.~Tevzadze, A.~Brandenburg and A.~Neronov,
  %``Evolution of Primordial Magnetic Fields from Phase Transitions,''
  Phys.\ Rev.\ D {\bf 87}, 083007 (2013)
  [arXiv:1212.0596 [astro-ph.CO]].
  %%CITATION = ARXIV:1212.0596;%%
  %5 citations counted in INSPIRE as of 19 Aug 2013

\bibitem{M-CS}
%
%\cite{Vilenkin:1984ib}
%\bibitem{Vilenkin:1984ib} 
  A.~Vilenkin,
  %``Cosmic Strings and Domain Walls,''
  Phys.\ Rept.\  {\bf 121}, 263 (1985);\ 
  %%CITATION = PRPLC,121,263;%%
  %1115 citations counted in INSPIRE as of 06 Jun 2013
%
%\cite{Vachaspati:1991tt}
%\bibitem{Vachaspati:1991tt} 
  T.~Vachaspati and A.~Vilenkin,
  %``Large scale structure from wiggly cosmic strings,''
  Phys.\ Rev.\ Lett.\  {\bf 67}, 1057 (1991);\ 
  %%CITATION = PRLTA,67,1057;%%
  %126 citations counted in INSPIRE as of 06 Jun 2013
%
%\cite{Brandenberger:1992wc}
%\bibitem{Brandenberger:1992wc} 
  R.~H.~Brandenberger, A.~-C.~Davis, A.~M.~Matheson and M.~Trodden,
  %``Superconducting cosmic strings and primordial magnetic fields,''
  Phys.\ Lett.\ B {\bf 293}, 287 (1992)
  [hep-ph/9206232];\ 
  %%CITATION = HEP-PH/9206232;%%
  %21 citations counted in INSPIRE as of 06 Jun 2013
%
%\cite{Dimopoulos:1997xa}
%\bibitem{Dimopoulos:1997xa} 
  K.~Dimopoulos and A.~-C.~Davis,
  %``Friction domination with superconducting strings,''
  Phys.\ Rev.\ D {\bf 57}, 692 (1998)
  [hep-ph/9705302];\ 
  %%CITATION = HEP-PH/9705302;%%
  %13 citations counted in INSPIRE as of 06 Jun 2013
%
%\cite{Dimopoulos:1997df}
%\bibitem{Dimopoulos:1997df} 
  K.~Dimopoulos,
  %``Primordial magnetic fields from superconducting cosmic strings,''
%  Phys.\ Rev.\ D {\bf 57}, 4629 (1998)
\textit{ibid}.\  {\bf 57}, 4629 (1998) 
  [hep-ph/9706513];\  
  %%CITATION = HEP-PH/9706513;%%
  %33 citations counted in INSPIRE as of 06 Jun 2013
%
%\cite{Battefeld:2007qn}
%\bibitem{Battefeld:2007qn} 
  D.~Battefeld, T.~Battefeld, D.~H.~Wesley and M.~Wyman,
  %``Magnetogenesis from Cosmic String Loops,''
  JCAP {\bf 0802}, 001 (2008)
  [arXiv:0708.2901 [astro-ph]];\ 
  %%CITATION = ARXIV:0708.2901;%%
  %15 citations counted in INSPIRE as of 06 Jun 2013
%
%\cite{Davis:2005ih}
%\bibitem{Davis:2005ih} 
  A.~-C.~Davis and K.~Dimopoulos,
  %``Cosmic superstrings and primordial magnetogenesis,''
  Phys.\ Rev.\ D {\bf 72}, 043517 (2005)
  [hep-ph/0505242];\ 
  %%CITATION = HEP-PH/0505242;%%
  %16 citations counted in INSPIRE as of 06 Jun 2013
%
%\cite{Zadorozhna:2013jwa}
%\bibitem{Zadorozhna:2013jwa} 
  L.~V.~Zadorozhna, B.~I.~Hnatyk and Yu.~A.~Sitenko,
  %``Magnetic Field of Cosmic Strings in the Early Universe,''
  UJP {\bf 58}, 398 (2013)
  [arXiv:1305.0029 [astro-ph.CO]].
  %%CITATION = ARXIV:1305.0029;%%

\bibitem{DP}
%
%\cite{Berezhiani:2003ik}
%\bibitem{Berezhiani:2003ik}
  Z.~Berezhiani and A.~D.~Dolgov,
  %``Generation of large scale magnetic fields at recombination epoch,''
  Astropart.\ Phys.\  {\bf 21}, 59 (2004)
  [arXiv:astro-ph/0305595];\ 
  %%CITATION = APHYE,21,59;%%
%
%\cite{Matarrese:2004kq}
%\bibitem{Matarrese:2004kq}
  S.~Matarrese, S.~Mollerach, A.~Notari and A.~Riotto, 
%S.~Matarrese {\it et al.}, 
  %``Large-scale magnetic fields from density perturbations,''
  Phys.\ Rev.\  D {\bf 71}, 043502 (2005)
  [arXiv:astro-ph/0410687];\ 
  %%CITATION = PHRVA,D71,043502;%%
%
%\cite{Takahashi:2005nd}
%\bibitem{Takahashi:2005nd}
  K.~Takahashi, K.~Ichiki, H.~Ohno and H.~Hanayama, 
%K.~Takahashi {\it et al.}, 
  %``Magnetic field generation from cosmological perturbations,''
  Phys.\ Rev.\ Lett.\  {\bf 95}, 121301 (2005)
  [arXiv:astro-ph/0502283];\ 
  %%CITATION = PRLTA,95,121301;%%
%
%\bibitem{Ichiki1}
%K.~Ichiki \textit{et al}., 
K.~Ichiki, K.~Takahashi, H.~Ohno, H.~Hanayama and N.~Sugiyama,
Science {\bf 311}, 827 (2006);\ 
%
%\cite{Ichiki:2007zz}
%\bibitem{Ichiki:2007zz} 
  K.~Ichiki, K.~Takahashi, N.~Sugiyama, H.~Hanayama and H.~Ohno,
  %``Generation of large-scale magnetic fields 
  %from primordial density fluctuations,''
  Mod.\ Phys.\ Lett.\ A {\bf 22}, 2091 (2007);\ 
  %%CITATION = MPLAE,A22,2091;%%
  %1 citations counted in INSPIRE as of 12 Oct 2013
%
%\cite{Takahashi:2007ds}
%\bibitem{Takahashi:2007ds} 
  K.~Takahashi, K.~Ichiki and N.~Sugiyama,
  %``Electromagnetic Properties of the Early Universe,''
  Phys.\ Rev.\ D {\bf 77}, 124028 (2008)
  [arXiv:0710.4620 [astro-ph]];\ 
  %%CITATION = ARXIV:0710.4620;%%
  %9 citations counted in INSPIRE as of 12 Oct 2013
%
%\cite{Siegel:2006px}
%\bibitem{Siegel:2006px}
  E.~R.~Siegel and J.~N.~Fry, 
  %``Cosmological Structure Formation Creates Large-Scale Magnetic Fields,''
  Astrophys.\ J.\  {\bf 651}, 627 (2006)
  [arXiv:astro-ph/0604526];\ 
  %%CITATION = ASJOA,651,627;%%
%
%\cite{Kobayashi:2007wd}
%\bibitem{Kobayashi:2007wd}
  T.~Kobayashi, R.~Maartens, T.~Shiromizu and K.~Takahashi, 
%T.~Kobayashi {\it et al.}, 
  %``Cosmological magnetic fields from nonlinear effects,''
  Phys.\ Rev.\  D {\bf 75}, 103501 (2007)
  [arXiv:astro-ph/0701596];\ 
  %%CITATION = PHRVA,D75,103501;%%
%
%%%%%
%\cite{Kunze:2007ph}
%\bibitem{Kunze:2007ph} 
  K.~E.~Kunze,
  %``Primordial magnetic fields and nonlinear electrodynamics,''
%  Phys.\ Rev.\ D {\bf 77}, 023530 (2008)
\textit{ibid}.\  {\bf 77}, 023530 (2008) 
  [arXiv:0710.2435 [astro-ph]];\ 
  %%CITATION = ARXIV:0710.2435;%%
%
%\cite{Campanelli:2007cg}
%\bibitem{Campanelli:2007cg} 
  L.~Campanelli, P.~Cea, G.~L.~Fogli and L.~Tedesco,
  %``Inflation-Produced Magnetic Fields in Nonlinear Electrodynamics,''
%  Phys.\ Rev.\ D {\bf 77}, 043001 (2008)
\textit{ibid}.\  {\bf 77}, 043001 (2008) 
  [arXiv:0710.2993 [astro-ph]];\ 
  %%CITATION = ARXIV:0710.2993;%%
%
%\cite{Maeda:2008dv}
%\bibitem{Maeda:2008dv}
  S.~Maeda, S.~Kitagawa, T.~Kobayashi and T.~Shiromizu,
  %``Primordial magnetic fields from second-order cosmological
  %perturbations:Tight coupling approximation,''
  Class.\ Quant.\ Grav.\  {\bf 26}, 135014 (2009)
  [arXiv:0805.0169 [astro-ph]];\ 
  %%CITATION = CQGRD,26,135014;%%
%
%\cite{Fenu:2010kh}
%\bibitem{Fenu:2010kh} 
  E.~Fenu, C.~Pitrou and R.~Maartens,
  %``The seed magnetic field generated during recombination,''
  Mon.\ Not.\ Roy.\ Astron.\ Soc.\  {\bf 414}, 2354 (2011)
  [arXiv:1012.2958 [astro-ph.CO]];\ 
  %%CITATION = ARXIV:1012.2958;%%
%
%\cite{Ichiki:2011ah}
%\bibitem{Ichiki:2011ah} 
  K.~Ichiki, K.~Takahashi and N.~Sugiyama,
  %``Constraint on the primordial vector mode and 
  %its magnetic field generation from seven-year 
  %Wilkinson Microwave Anisotropy Probe Observations,''
  Phys.\ Rev.\ D {\bf 85}, 043009 (2012)
  [arXiv:1112.4705 [astro-ph.CO]];\ 
  %%CITATION = ARXIV:1112.4705;%%
  %6 citations counted in INSPIRE as of 12 Oct 2013
%
%\cite{Saga:2013glg}
%\bibitem{Saga:2013glg} 
  S.~Saga, M.~Shiraishi, K.~Ichiki and N.~Sugiyama,
  %``Generation of magnetic fields in Einstein-Aether gravity,''
  %Phys.\ Rev.\ D {\bf 87}, 104025 (2013)
\textit{ibid}.\  {\bf 87}, 104025 (2013)
  [arXiv:1302.4189 [astro-ph.CO]];\ 
  %%CITATION = ARXIV:1302.4189;%%
  %2 citations counted in INSPIRE as of 12 Oct 2013
%
%\cite{Nalson:2013jya}
%\bibitem{Nalson:2013jya} 
  E.~Nalson, A.~J.~Christopherson and K.~A.~Malik,
  %``Effects of non-linearities on magnetic field generation,''
  arXiv:1312.6504 [astro-ph.CO];\ 
  %%CITATION = ARXIV:1312.6504;%%
%
%\cite{Berger:2014wta}
%\bibitem{Berger:2014wta} 
  P.~Berger, A.~Kehagias and A.~Riotto,
  %``Testing the Origin of Cosmological Magnetic Fields 
  %through the Large-Scale Structure Consistency Relations,''
  arXiv:1402.1044 [astro-ph.CO]; 
  %%CITATION = ARXIV:1402.1044;%%
%
%\cite{Kobayashi:2014sga}
%\bibitem{Kobayashi:2014sga} 
  T.~Kobayashi,
  %``Primordial Magnetic Fields from the Post-Inflationary Universe,''
  arXiv:1403.5168 [astro-ph.CO];\ 
  %%CITATION = ARXIV:1403.5168;%%
%
%\cite{Osano:2014uma}
%\bibitem{Osano:2014uma} 
  B.~Osano,
  %``Magnetic fields from second-order interactions,''
  arXiv:1403.5505 [gr-qc].
  %%CITATION = ARXIV:1403.5505;%%

%
%%%%%

\bibitem{EParker}
E.~N.~Parker,
Astrophys.\ J.\ {\bf 163}, 255 (1971);\ 
%E.~N.~Parker, 
\textit{Cosmical Magnetic Fields}
(Clarendon, Oxford, England, 1979);\ 
Ya.~B.~Zel'dovich, A.~A.~Ruzmaikin, and D.~D.~Sokoloff,
\textit{Magnetic Fields in Astrophysics}
(Gordon and Breach, New York, 1983).

\bibitem{Inflation} 
%
%\cite{Guth:1980zm}
%\bibitem{Guth:1980zm} 
  A.~H.~Guth,
  %``The Inflationary Universe: A Possible Solution to 
  %the Horizon and Flatness Problems,''
  Phys.\ Rev.\ D {\bf 23}, 347 (1981);\ 
  %%CITATION = PHRVA,D23,347;%%
%
%\cite{Sato:1980yn}
%\bibitem{Sato:1980yn} 
  K.~Sato,
  %``First Order Phase Transition of a Vacuum and Expansion of the Universe,''
  Mon.\ Not.\ Roy.\ Astron.\ Soc.\  {\bf 195}, 467 (1981);\ 
  %%CITATION = MNRAA,195,467;%%
%
%\cite{Starobinsky:1980te}
%\bibitem{Starobinsky:1980te} 
  A.~A.~Starobinsky,
  %``A New Type of Isotropic Cosmological Models Without Singularity,''
  Phys.\ Lett.\ B {\bf 91}, 99 (1980).
  %%CITATION = PHLTA,B91,99;%%

%\cite{Turner:1987bw}
\bibitem{Turner:1987bw}
  M.~S.~Turner and L.~M.~Widrow,
  %``Inflation Produced, Large Scale Magnetic Fields,''
  Phys.\ Rev.\  D {\bf 37}, 2743 (1988).
  %%CITATION = PHRVA,D37,2743;%%

%\cite{Campanelli:2013mea}
\bibitem{Campanelli:2013mea} 
  L.~Campanelli,
  %``Origin of Cosmic Magnetic Fields,''
  Phys.\  Rev.\  Lett.\  {\bf 111}, 061301 (2013)
  [arXiv:1304.6534 [astro-ph.CO]]. 
  %%CITATION = ARXIV:1304.6534;%%
  %3 citations counted in INSPIRE as of 12 Oct 2013

%%%%%%%%%
\bibitem{B-C-F}
%
%\cite{Maroto:2000zu}
%\bibitem{Maroto:2000zu}
  A.~L.~Maroto,
  %``Primordial magnetic fields from metric perturbations,''
  Phys.\ Rev.\  D {\bf 64}, 083006 (2001)
  [arXiv:hep-ph/0008288];\ 
  %%CITATION = PHRVA,D64,083006;%%
%
%\cite{Tsagas:2004kv}
%\bibitem{Tsagas:2004kv}
  C.~G.~Tsagas,
  %``Electromagnetic fields in curved spacetimes,''
  Class.\ Quant.\ Grav.\  {\bf 22}, 393 (2005)
  [arXiv:gr-qc/0407080];\ 
  %%CITATION = CQGRD,22,393;%%
%
%\cite{Tsagas:2005nn}
%\bibitem{Tsagas:2005nn}
  C.~G.~Tsagas and A.~Kandus, 
  %``Superadiabatic-type magnetic amplification in conventional cosmology,''
  Phys.\ Rev.\  D {\bf 71}, 123506 (2005)
  [arXiv:astro-ph/0504089];\ 
  %%CITATION = PHRVA,D71,123506;%%
%
%\cite{Barrow:2008jp}
%\bibitem{Barrow:2008jp}
  J.~D.~Barrow and C.~G.~Tsagas,
  %``Slow decay of magnetic fields in open Friedmann universes,''
%  Phys.\ Rev.\  D {\bf 77}, 107302 (2008)
\textit{ibid}.\ {\bf 77}, 107302 (2008)
  [Erratum-ibid.\  D {\bf 77}, 109904 (2008)]
  [arXiv:0803.0660 [astro-ph]];\ 
  %%CITATION = PHRVA,D77,107302;%%
%
%\cite{Barrow:2011ic}
%\bibitem{Barrow:2011ic}
%  J.~D.~Barrow and C.~G.~Tsagas,
  %``Cosmological magnetic field survival,''
  Mon.\ Not.\ Roy.\ Astron.\ Soc.\  {\bf 414}, 512 (2011)
  [arXiv:1101.2390 [astro-ph.CO]];\ 
  %%CITATION = MNRAA,414,512;%%
%
%\cite{Maeda:2009hy}
%\bibitem{Maeda:2009hy} 
  S.~Maeda, S.~Mukohyama and T.~Shiromizu,
  %``Primordial magnetic field from non-inflationary 
  %cosmic expansion in Horava-Lifshitz gravity,''
  Phys.\ Rev.\ D {\bf 80}, 123538 (2009)
  [arXiv:0909.2149 [astro-ph.CO]];\ 
  %%CITATION = ARXIV:0909.2149;%%
  %26 citations counted in INSPIRE as of 12 Oct 2013
%
%\cite{Giovannini:2013oga}
%\bibitem{Giovannini:2013oga} 
  M.~Giovannini,
  %``Anomalous Magnetohydrodynamics,''
  %Phys.\ Rev.\ D {\bf 88}, 063536 (2013)
\textit{ibid}.\ {\bf 88}, 063536 (2013) 
  [arXiv:1307.2454 [hep-th]];\ 
  %%CITATION = ARXIV:1307.2454;%%
  %2 citations counted in INSPIRE as of 20 Dec 2013
%
%\cite{Kouretsis:2013cga}
%\bibitem{Kouretsis:2013cga} 
  A.~P.~Kouretsis,
  %``Cosmic magnetization in curved and Lorentz violating space-times,''
  arXiv:1312.4631 [gr-qc];\ 
  %%CITATION = ARXIV:1312.4631;%%
%
%\cite{Membiela:2013cea}
%\bibitem{Membiela:2013cea} 
  F.~A.~Membiela,
  %``Primordial magnetic fields from a nonsingular bouncing cosmology,''
  arXiv:1312.2162 [astro-ph.CO];\ 
  %%CITATION = ARXIV:1312.2162;%%
%
C.~G.~Tsagas, 
%``On the magnetic evolution in Friedmann universes and the question of 
%cosmic magnetogenesis,''
arXiv:1412.4806 [astro-ph.CO]. 
%%%%%%%%%

\bibitem{ANS-ANSL}
%
%\cite{Agullo:2013tba}
%\bibitem{Agullo:2013tba} 
  I.~Agullo and J.~Navarro-Salas,
  %``Conformal anomaly and primordial magnetic fields,''
  arXiv:1309.3435 [gr-qc];\ 
  %%CITATION = ARXIV:1309.3435;%%
%
%%%%%%%%%%%%%%
I.~Agullo, J.~Navarro-Salas and A.~Landete,
%``Electric-magnetic duality and renormalization 
%in curved spacetimes,''
arXiv:1409.6406 [gr-qc]. 
%%%%%%%%%%%%%%

%\cite{Parker:1968mv}
\bibitem{Parker:1968mv}
  L.~Parker,
  %``Particle creation in expanding universes,''
  Phys.\ Rev.\ Lett.\  {\bf 21}, 562 (1968).
  %%CITATION = PRLTA,21,562;%%

%%%%%
\bibitem{N-G-C}
%
%\cite{Mazzitelli:1995mp}
%\bibitem{Mazzitelli:1995mp} 
  F.~D.~Mazzitelli and F.~M.~Spedalieri,
  %``Scalar electrodynamics and primordial magnetic fields,''
  Phys.\ Rev.\ D {\bf 52}, 6694 (1995)
  [astro-ph/9505140];\ 
  %%CITATION = ASTRO-PH/9505140;%%
%
%\cite{Lambiase:2004zb}
%\bibitem{Lambiase:2004zb} 
  G.~Lambiase and A.~R.~Prasanna,
  %``Gauge invariant wave equations in curved space-times and primordial 
  %magnetic fields,''
%  Phys.\ Rev.\ D {\bf 70}, 063502 (2004)
\textit{ibid}.\  {\bf 70}, 063502 (2004) 
  [gr-qc/0407071];\ 
  %%CITATION = GR-QC/0407071;%%
%
%\cite{Bamba:2008ja}
%\bibitem{Bamba:2008ja} 
  K.~Bamba and S.~D.~Odintsov,
  %``Inflation and late-time cosmic acceleration in non-minimal 
  %Maxwell-$F(R)$ gravity and the generation of large-scale magnetic fields,''
  JCAP {\bf 0804}, 024 (2008)
  [arXiv:0801.0954 [astro-ph]];\ 
  %%CITATION = ARXIV:0801.0954;%%
%\cite{Bamba:2008xa}
%\bibitem{Bamba:2008xa} 
  K.~Bamba, S.~Nojiri and S.~D.~Odintsov,
  %``Inflationary cosmology and the late-time accelerated expansion 
  %of the universe in non-minimal Yang-Mills-F(R) gravity and 
  %non-minimal vector-F(R) gravity,''
  Phys.\ Rev.\ D {\bf 77}, 123532 (2008)
  [arXiv:0803.3384 [hep-th]];\ 
  %%CITATION = ARXIV:0803.3384;%%
%
%\cite{Bamba:2008be}
%\bibitem{Bamba:2008be} 
  K.~Bamba and S.~Nojiri,
  %``Cosmology in non-minimal Yang-Mills/Maxwell theory,''
  arXiv:0811.0150 [hep-th];\ 
  %%CITATION = ARXIV:0811.0150;%%
%
%\cite{Campanelli:2008qp}
%\bibitem{Campanelli:2008qp}
  L.~Campanelli, P.~Cea, G.~L.~Fogli and L.~Tedesco,
  %``Inflation-Produced Magnetic Fields in R^n F^2 and I F^2 models,''
  Phys.\ Rev.\  D {\bf 77}, 123002 (2008)
%\textit{ibid}.\ {\bf 77}, 123002 (2008); 
  [arXiv:0802.2630 [astro-ph]];\ 
  %%CITATION = PHRVA,D77,123002;%%
%
%\cite{Lambiase:2008zz}
%\bibitem{Lambiase:2008zz}
  G.~Lambiase, S.~Mohanty and G.~Scarpetta,
  %``Magnetic field amplification in f(R) theories of gravity,''
  JCAP {\bf 0807}, 019 (2008);\ 
  %%CITATION = JCAPA,0807,019;%%
%
%\cite{Kunze:2009bs}
%\bibitem{Kunze:2009bs}
  K.~E.~Kunze,
  %``Large scale magnetic fields from gravitationally coupled 
  %electrodynamics,''
  Phys.\ Rev.\  D {\bf 81}, 043526 (2010)
%\textit{ibid}.\ {\bf 81}, 043526 (2010). 
  [arXiv:0911.1101 [astro-ph.CO]];\ 
  %%CITATION = PHRVA,D81,043526;%%
%
%\cite{Jimenez:2010uh}
%\bibitem{Jimenez:2010uh}
  J.~B.~Jimenez and A.~L.~Maroto,
  %``Dark energy, non-minimal couplings and the origin 
  %of cosmic magnetic fields,''
  JCAP {\bf 1012}, 025 (2010)
  [arXiv:1010.4513 [astro-ph.CO]];\ 
  %%CITATION = JCAPA,1012,025;%%
%
%\cite{Kunze:2012rq}
%\bibitem{Kunze:2012rq} 
  K.~E.~Kunze,
  %``Completing magnetic field generation 
  %from gravitationally coupled electrodynamics 
  %with the curvaton mechanism,''
  Phys.\ Rev.\ D {\bf 87}, 063505 (2013)
  [arXiv:1210.6899 [astro-ph.CO]];\ 
  %%CITATION = ARXIV:1210.6899;%%
  %1 citations counted in INSPIRE as of 01 Sep 2013
%
%\bibitem{BGL-BGL}
%
%\cite{Bamba:2012mi}
%\bibitem{Bamba:2012mi} 
  K.~Bamba, C.~-Q.~Geng and L.~-W.~Luo,
  %``Generation of large-scale magnetic fields 
  %from inflation in teleparallelism,''
  JCAP {\bf 1210}, 058 (2012)
  [arXiv:1208.0665 [astro-ph.CO]];\ 
  %%CITATION = ARXIV:1208.0665;%%
  %5 citations counted in INSPIRE as of 12 Oct 2013
%
%\cite{Bamba:2013rra}
%\bibitem{Bamba:2013rra} 
%  K.~Bamba, C.~-Q.~Geng and L.~-W.~Luo,
  %``Large-scale magnetic fields from inflation in teleparallel gravity,''
  arXiv:1307.7448 [astro-ph.CO].
  %%CITATION = ARXIV:1307.7448;%%
%
%%%%%

%\cite{Drummond:1979pp}
\bibitem{Drummond:1979pp}
  I.~T.~Drummond and S.~J.~Hathrell,
  %``QED Vacuum Polarization In A Background Gravitational Field 
  %And Its Effect On The Velocity Of Photons,''
  Phys.\ Rev.\  D {\bf 22}, 343 (1980).
  %%CITATION = PHRVA,D22,343;%%

%\cite{Ratra:1991bn}
\bibitem{Ratra:1991bn}
  B.~Ratra,
  %``Cosmological 'seed' magnetic field from inflation,''
  Astrophys.\ J.\  {\bf 391}, L1 (1992).
  %%CITATION = ASJOA,391,L1;%%

\bibitem{MF-Scalar} 
%
%\cite{Lemoine:1995vj}
%\bibitem{Lemoine:1995vj}
  D.~Lemoine and M.~Lemoine,
  %``Primordial magnetic fields in string cosmology,''
  Phys.\ Rev.\  D {\bf 52}, 1955 (1995);\ 
\textit{ibid}.\  {\bf 52}, 1955 (1995);\ 
  %%CITATION = PHRVA,D52,1955;%%
%
%\cite{Gasperini:1995dh}
%\bibitem{Gasperini:1995dh}
  M.~Gasperini, M.~Giovannini and G.~Veneziano,
  %``Primordial magnetic fields from string cosmology,''
  Phys.\ Rev.\ Lett.\  {\bf 75}, 3796 (1995)
  [arXiv:hep-th/9504083];\ 
  %%CITATION = PRLTA,75,3796;%%
%
%PRD 64 (2001) 061301; 
%\cite{Giovannini:2001nh}
%\bibitem{Giovannini:2001nh} 
  M.~Giovannini,
  %``On the variation of the gauge couplings during inflation,''
  Phys.\ Rev.\ D {\bf 64}, 061301 (2001)
  [astro-ph/0104290];\ 
  %%CITATION = ASTRO-PH/0104290;%%
%
%\cite{Martin:2007ue}
%\bibitem{Martin:2007ue}
  J.~Martin and J.~Yokoyama, 
  %``Generation of Large-Scale Magnetic Fields in Single-Field Inflation,''
  JCAP {\bf 0801}, 025 (2008) 
%\textit{ibid}.\ {\bf 0801}, 025 (2008). 
  [arXiv:0711.4307 [astro-ph]];\ 
  %%CITATION = JCAPA,0801,025;%%
%
%\bibitem{M-BS-B}
%
%\cite{Bamba:2006ga}
%\bibitem{Bamba:2006ga} 
  K.~Bamba and M.~Sasaki,
  %``Large-scale magnetic fields in the inflationary universe,''
%  JCAP {\bf 0702}, 030 (2007)
\textit{ibid}.\ {\bf 0702}, 030 (2007)
  [astro-ph/0611701];\ 
  %%CITATION = ASTRO-PH/0611701;%%
%
%\cite{Bamba:2007sx}
%\bibitem{Bamba:2007sx} 
  K.~Bamba,
  %``The interrelation between the generation of large-scale 
  %electric fields and that of large-scale magnetic fields 
  %during inflation,''
%  JCAP {\bf 0710}, 015 (2007)
\textit{ibid}.\  {\bf 0710}, 015 (2007) 
  [arXiv:0710.1906 [astro-ph]];\ 
  %%CITATION = ARXIV:0710.1906;%%
%
%PLB 659 (2008) 661; 
%\cite{Giovannini:2007rh}
%\bibitem{Giovannini:2007rh} 
  M.~Giovannini,
  %``Magnetogenesis, spectator fields and CMB signatures,''
  Phys.\ Lett.\ B {\bf 659}, 661 (2008)
  [arXiv:0711.3273 [astro-ph]];\ 
  %%CITATION = ARXIV:0711.3273;%%
%
%\cite{Bamba:2008my}
%\bibitem{Bamba:2008my} 
  K.~Bamba, N.~Ohta and S.~Tsujikawa,
  %``Generic estimates for magnetic fields generated during inflation 
  %including Dirac-Born-Infeld theories,''
  Phys.\ Rev.\ D {\bf 78}, 043524 (2008)
  [arXiv:0805.3862 [astro-ph]];\ 
  %%CITATION = ARXIV:0805.3862;%%
%
%\cite{Bamba:2008hr}
%\bibitem{Bamba:2008hr} 
  K.~Bamba, C.~Q.~Geng and S.~H.~Ho,
  %``Large-scale magnetic fields from inflation due to Chern-Simons-like 
  %effective interaction,''
  JCAP {\bf 0811}, 013 (2008)
  [arXiv:0806.1856 [astro-ph]];\ 
  %%CITATION = ARXIV:0806.1856;%%
%
%JCAP 1004 (2010) 003
%\cite{Giovannini:2009xa}
%\bibitem{Giovannini:2009xa} 
  M.~Giovannini,
  %``Electric-magnetic duality and the conditions of inflationary 
  %magnetogenesis,''
  %JCAP {\bf 1004}, 003 (2010)
\textit{ibid}.\  {\bf 1004}, 003 (2010)
  [arXiv:0911.0896 [astro-ph.CO]];\ 
  %%CITATION = ARXIV:0911.0896;%%
%%%%%%%%
%\bibitem{M-BGHK-HKBG}
%
%\cite{Bamba:2011si}
%\bibitem{Bamba:2011si} 
  K.~Bamba, C.~Q.~Geng, S.~H.~Ho and W.~F.~Kao,
  %``Large-scale magnetic fields from inflation due to a $CPT$-even 
  %Chern-Simons-like term with Kalb-Ramond and scalar fields,''
  Eur.\ Phys.\ J.\ C {\bf 72}, 1978 (2012)
  [arXiv:1108.0151 [astro-ph.CO]];\ 
  %%CITATION = ARXIV:1108.0151;%%
%
%\cite{Ho:2010aq}
%\bibitem{Ho:2010aq} 
  S.~H.~Ho, W.~F.~Kao, K.~Bamba and C.~Q.~Geng,
  %``Cosmological birefringence due to CPT-even Chern-Simons-like 
  %term with Kalb-Ramond and scalar fields,''
  arXiv:1008.0486 [hep-ph];\ 
  %%CITATION = ARXIV:1008.0486;%%
%
%\cite{Giovannini:2013rza}
%\bibitem{Giovannini:2013rza} 
  M.~Giovannini,
  %``Inflationary susceptibilities, duality and 
  %large-scale magnetic fields generation,''
  Phys.\ Rev.\ D {\bf 88}, 083533 (2013)
  [arXiv:1310.1802 [hep-th]];\ 
  %%CITATION = ARXIV:1310.1802;%%
  %1 citations counted in INSPIRE as of 19 Dec 2013
%
%\cite{Giovannini:2013qga}
%\bibitem{Giovannini:2013qga} 
  M.~Giovannini,
  %``No-hair conjectures, primordial shear and protoinflationary initial 
  %conditions,''
%  Phys.\ Rev.\ D {\bf 89}, 063512 (2014)
\textit{ibid}.\  {\bf 89}, 063512 (2014)
  [arXiv:1312.4832 [hep-th]];\ 
  %%CITATION = ARXIV:1312.4832;%%
%
%\cite{Kastor:2013nha}
%\bibitem{Kastor:2013nha} 
  D.~Kastor and J.~Traschen,
  %``Magnetic Fields in an Expanding Universe,''
  Class.\ Quant.\ Grav.\  {\bf 31}, 075023 (2014)
  [arXiv:1312.4923 [hep-th]];\ 
  %%CITATION = ARXIV:1312.4923;%%
  %1 citations counted in INSPIRE as of 23 Apr 2014
%
%\cite{Atmjeet:2013yta}
%\bibitem{Atmjeet:2013yta} 
  K.~Atmjeet, I.~Pahwa, T.~R.~Seshadri and K.~Subramanian,
  %``Cosmological Magnetogenesis From Extra-dimensional Gauss Bonnet Gravity,''
  arXiv:1312.5815 [astro-ph.CO].
  %%CITATION = ARXIV:1312.5815;%%
%
%%%%%%%%

\bibitem{B-Y}
%
%\cite{Bamba:2003av}
%\bibitem{Bamba:2003av} 
  K.~Bamba and J.~Yokoyama,
  %``Large scale magnetic fields from inflation in dilaton electromagnetism,''
  Phys.\ Rev.\ D {\bf 69}, 043507 (2004)
%\textit{ibid}.\  {\bf 69}, 043507 (2004) 
  [astro-ph/0310824];\ 
  %%CITATION = ASTRO-PH/0310824;%%
%
%\cite{Bamba:2004cu}
%\bibitem{Bamba:2004cu} 
%  K.~Bamba and J.~Yokoyama,
  %``Large-scale magnetic fields from dilaton inflation 
  %in noncommutative spacetime,''
%  Phys.\ Rev.\ D {\bf 70}, 083508 (2004)
%\textit{ibid}.\  
{\bf 70}, 083508 (2004) 
  [hep-ph/0409237].
  %%CITATION = HEP-PH/0409237;%%
%

%\cite{Garretson:1992vt}
\bibitem{Garretson:1992vt} 
  W.~D.~Garretson, G.~B.~Field and S.~M.~Carroll,
  %``Primordial magnetic fields from pseudoGoldstone bosons,''
  Phys.\ Rev.\ D {\bf 46}, 5346 (1992)
  [hep-ph/9209238].
  %%CITATION = HEP-PH/9209238;%%
%

%\cite{Dolgov:1993vg}
\bibitem{Dolgov:1993vg} 
  A.~D.~Dolgov,
  %``Breaking of conformal invariance and electromagnetic field 
  %generation in the universe,''
  Phys.\ Rev.\ D {\bf 48}, 2499 (1993)
  [hep-ph/9301280].
  %%CITATION = HEP-PH/9301280;%%

%\cite{BlancoPillado:2004ns}
\bibitem{BlancoPillado:2004ns} 
  J.~J.~Blanco-Pillado, C.~P.~Burgess, J.~M.~Cline, C.~Escoda, M.~Gomez-Reino, R.~Kallosh, A.~D.~Linde and F.~Quevedo,
  %``Racetrack inflation,''
  JHEP {\bf 0411}, 063 (2004)
  [hep-th/0406230].
  %%CITATION = HEP-TH/0406230;%%

%\cite{Kachru:2003aw}
\bibitem{Kachru:2003aw} 
  S.~Kachru, R.~Kallosh, A.~D.~Linde and S.~P.~Trivedi,
  %``De Sitter vacua in string theory,''
  Phys.\ Rev.\ D {\bf 68}, 046005 (2003)
  [hep-th/0301240].
  %%CITATION = HEP-TH/0301240;%%

%\cite{Barnaby:2011qe}
\bibitem{Barnaby:2011qe} 
  N.~Barnaby, E.~Pajer and M.~Peloso,
  %``Gauge Field Production in Axion Inflation: Consequences for Monodromy, 
  %non-Gaussianity in the CMB, and Gravitational Waves at Interferometers,''
  Phys.\ Rev.\ D {\bf 85}, 023525 (2012)
  [arXiv:1110.3327 [astro-ph.CO]].
  %%CITATION = ARXIV:1110.3327;%%

%\cite{Barnaby:2012xt}
\bibitem{Barnaby:2012xt} 
  N.~Barnaby, J.~Moxon, R.~Namba, M.~Peloso, G.~Shiu and P.~Zhou,
  %``Gravity waves and non-Gaussian features from particle production 
  %in a sector gravitationally coupled to the inflaton,''
  Phys.\ Rev.\ D {\bf 86}, 103508 (2012)
  [arXiv:1206.6117 [astro-ph.CO]].
  %%CITATION = ARXIV:1206.6117;%%
  %12 citations counted in INSPIRE as of 03 Jun 2013
%%%%%%

%%%%%%%%%%%%%%%%%%%%%%%%%%%%%%%%%%%
%\cite{Ferreira:2014zia}
\bibitem{Ferreira:2014zia} 
  R.~Z.~Ferreira and M.~S.~Sloth,
  %``Universal Constraints on Axions from Inflation,''
  arXiv:1409.5799 [hep-ph].
  %%CITATION = ARXIV:1409.5799;%%
  %4 citations counted in INSPIRE as of 19 Nov 2014
%%%%%%%%%%%%%%%%%%%%%%%%%%%%%%%%%%%

%\cite{Cheng:2014kga}
\bibitem{Cheng:2014kga} 
  S.~L.~Cheng, W.~Lee and K.~W.~Ng,
  %``Inflationary dilaton-axion magnetogenesis,''
  arXiv:1409.2656 [astro-ph.CO].
  %%CITATION = ARXIV:1409.2656;%%
  %2 citations counted in INSPIRE as of 18 Nov 2014

%\bibitem{A-S}
%
%\cite{Anber:2006xt}
\bibitem{Anber:2006xt} 
  M.~M.~Anber and L.~Sorbo,
  %``N-flationary magnetic fields,''
  JCAP {\bf 0610}, 018 (2006)
  [astro-ph/0606534].
  %%CITATION = ASTRO-PH/0606534;%%
%

%\cite{Anber:2009ua}
\bibitem{Anber:2009ua} 
  M.~M.~Anber and L.~Sorbo,
  %``Naturally inflating on steep potentials through 
  %electromagnetic dissipation,''
  Phys.\ Rev.\ D {\bf 81}, 043534 (2010)
  [arXiv:0908.4089 [hep-th]].
  %%CITATION = ARXIV:0908.4089;%%
%

%\cite{Bugaev:2013fya}
\bibitem{Bugaev:2013fya} 
  E.~Bugaev and P.~Klimai,
  %``Axion inflation with gauge field production 
  %and primordial black holes,''
  arXiv:1312.7435 [astro-ph.CO].
  %%CITATION = ARXIV:1312.7435;%%

%\cite{Meerburg:2012id}
\bibitem{Meerburg:2012id} 
  P.~D.~Meerburg and E.~Pajer,
  %``Observational Constraints on Gauge Field Production in Axion Inflation,''
  JCAP {\bf 1302}, 017 (2013)
  [arXiv:1203.6076 [astro-ph.CO]].
  %%CITATION = ARXIV:1203.6076;%%
  %7 citations counted in INSPIRE as of 01 Sep 2013
%%%%%

%\cite{Pajer:2013fsa}
\bibitem{Pajer:2013fsa} 
  E.~Pajer and M.~Peloso,
  %``A review of Axion Inflation in the era of Planck,''
  Class.\ Quant.\ Grav.\  {\bf 30}, 214002 (2013)
  [arXiv:1305.3557 [hep-th]].
  %%CITATION = ARXIV:1305.3557;%%
  %4 citations counted in INSPIRE as of 20 Dec 2013

%\cite{Barnaby:2010vf}
\bibitem{Barnaby:2010vf} 
  N.~Barnaby and M.~Peloso,
  %``Large Nongaussianity in Axion Inflation,''
  Phys.\ Rev.\ Lett.\  {\bf 106}, 181301 (2011)
  [arXiv:1011.1500 [hep-ph]].
  %%CITATION = ARXIV:1011.1500;%%

%%%%%%
%\cite{Barnaby:2011vw}
\bibitem{Barnaby:2011vw} 
  N.~Barnaby, R.~Namba and M.~Peloso,
  %``Phenomenology of a Pseudo-Scalar Inflaton: 
  %Naturally Large Nongaussianity,''
  JCAP {\bf 1104}, 009 (2011)
  [arXiv:1102.4333 [astro-ph.CO]].
  %%CITATION = ARXIV:1102.4333;%%
  %29 citations counted in INSPIRE as of 03 Jun 2013

%\cite{Komatsu:2001rj}
\bibitem{Komatsu:2001rj} 
  E.~Komatsu and D.~N.~Spergel,
  %``Acoustic signatures in the primary microwave background bispectrum,''
  Phys.\ Rev.\ D {\bf 63}, 063002 (2001)
  [astro-ph/0005036].
  %%CITATION = ASTRO-PH/0005036;%%

%\cite{Tanaka:2010km}
\bibitem{Tanaka:2010km} 
  T.~Tanaka, T.~Suyama and S.~Yokoyama,
  %``Use of delta N formalism - Difficulties 
　%in generating large local-type non-Gaussianity during inflation -,''
  Class.\ Quant.\ Grav.\  {\bf 27}, 124003 (2010)
  [arXiv:1003.5057 [astro-ph.CO]].
  %%CITATION = ARXIV:1003.5057;%%
  %23 citations counted in INSPIRE as of 13 Oct 2013

%\cite{Linde:2012bt}
\bibitem{Linde:2012bt} 
  A.~Linde, S.~Mooij and E.~Pajer,
  %``Gauge field production in SUGRA inflation: 
  %local non-Gaussianity and primordial black holes,''
  Phys.\ Rev.\ D {\bf 87}, 103506 (2013)
  [arXiv:1212.1693 [hep-th]].
  %%CITATION = ARXIV:1212.1693;%%
  %9 citations counted in INSPIRE as of 01 Sep 2013

%\cite{Brown:2005kr}
\bibitem{Brown:2005kr} 
  I.~Brown and R.~Crittenden,
  %``Non-Gaussianity from cosmic magnetic fields,''
  Phys.\ Rev.\ D {\bf 72}, 063002 (2005)
  [astro-ph/0506570].
  %%CITATION = ASTRO-PH/0506570;%%
  %46 citations counted in INSPIRE as of 06 Jun 2013

%%%%% Detect %%%%%

%\bibitem{Planck-1}
% See 
%http://www.sciops.esa.int/index.php?project=PLANCK.

%\bibitem{Planck-2}
%http://www.rssd.esa.int/SA/PLANCK/docs/Bluebook-ESA-SCI(2005)1\_V2.pdf.

%\cite{Ade:2013ydc}
\bibitem{Ade:2013ydc} 
  P.~A.~R.~Ade {\it et al.}  [Planck Collaboration],
  %``Planck 2013 Results. XXIV. Constraints on primordial non-Gaussianity,''
  arXiv:1303.5084 [astro-ph.CO].
  %%CITATION = ARXIV:1303.5084;%%
  %49 citations counted in INSPIRE as of 29 May 2013

%\cite{Ade:2014xna}
\bibitem{Ade:2014xna} 
  P.~A.~R.~Ade {\it et al.}  [BICEP2 Collaboration],
  %``BICEP2 I: Detection Of B-mode Polarization 
  %at Degree Angular Scales,''
  arXiv:1403.3985 [astro-ph.CO].
  %%CITATION = ARXIV:1403.3985;%%

\bibitem{K-K} 
%
M.~J.~Duff, B.~E.~W.~Nilsson and C.~N.~Pope,
%``Kaluza-Klein Supergravity,''
Phys.\ Rept.\  {\bf 130}, 1 (1986);\ 
T.~Appelquist, A.~Chodos and P.~G.~O.~Freund, 
\textit{Modern Kaluza-Klein Theories} 
(Addison-Wesley, Reading, 1987);\ 
J.~M.~Overduin and P.~S.~Wesson,
%``Kaluza-Klein gravity,''
Phys.\ Rept.\  {\bf 283}, 303 (1997) 
[gr-qc/9805018];\  
Y.~Fujii and K.~Maeda,
\textit{The Scalar-Tensor Theory of Gravitation}
(Cambridge University Press, Cambridge, United Kingdom, 2003);\ 
C.~N.~Pope, 
\textit{Lectures on Kaluza-Klein theory} (2000), 
http://people.physics.tamu.edu/pope/ihplec.pdf
.
%
%%%%%%%%%%%%%%%%%%%

%\cite{Bamba:2006km}
\bibitem{Bamba:2006km} 
  K.~Bamba,
  %``Baryon asymmetry from hypermagnetic helicity 
  %in dilaton hypercharge electromagnetism,''
  Phys.\ Rev.\ D {\bf 74}, 123504 (2006)
  [hep-ph/0611152].
  %%CITATION = HEP-PH/0611152;%%

%\cite{Bamba:2007hf}
\bibitem{Bamba:2007hf} 
  K.~Bamba, C.~Q.~Geng and S.~H.~Ho,
  %``Hypermagnetic Baryogenesis,''
  Phys.\ Lett.\ B {\bf 664}, 154 (2008)
  [arXiv:0712.1523 [hep-ph]].
  %%CITATION = ARXIV:0712.1523;%%

%%%%%

\bibitem{N-A-I}
%
%\cite{Freese:1990rb}
%\bibitem{Freese:1990rb} 
  K.~Freese, J.~A.~Frieman and A.~V.~Olinto,
  %``Natural inflation with pseudo - Nambu-Goldstone bosons,''
  Phys.\ Rev.\ Lett.\  {\bf 65}, 3233 (1990);\ 
  %%CITATION = PRLTA,65,3233;%%
  %360 citations counted in INSPIRE as of 08 Jun 2013
%
%\bibitem{Prof-Mazumdar}
%
%\cite{Liddle:1998jc}
%\bibitem{Liddle:1998jc} 
  A.~R.~Liddle, A.~Mazumdar and F.~E.~Schunck,
  %``Assisted inflation,''
  Phys.\ Rev.\ D {\bf 58}, 061301 (1998)
  [astro-ph/9804177];\ 
  %%CITATION = ASTRO-PH/9804177;%%
  %334 citations counted in INSPIRE as of 28 Nov 2014
%
%\cite{Copeland:1999cs}
%\bibitem{Copeland:1999cs} 
  E.~J.~Copeland, A.~Mazumdar and N.~J.~Nunes,
  %``Generalized assisted inflation,''
%  Phys.\ Rev.\ D {\bf 60}, 083506 (1999)
\textit{ibid}. {\bf 60}, 083506 (1999)
  [astro-ph/9904309];\ 
  %%CITATION = ASTRO-PH/9904309;%%
  %131 citations counted in INSPIRE as of 28 Nov 2014
%
%\cite{Mazumdar:2001mm}
%\bibitem{Mazumdar:2001mm} 
  A.~Mazumdar, S.~Panda and A.~Perez-Lorenzana,
  %``Assisted inflation via tachyon condensation,''
  Nucl.\ Phys.\ B {\bf 614}, 101 (2001)
  [hep-ph/0107058];\ 
  %%CITATION = HEP-PH/0107058;%%
  %209 citations counted in INSPIRE as of 28 Nov 2014
%
%\cite{Kim:2004rp}
%\bibitem{Kim:2004rp} 
  J.~E.~Kim, H.~P.~Nilles and M.~Peloso,
  %``Completing natural inflation,''
  JCAP {\bf 0501}, 005 (2005)
  [hep-ph/0409138];\ 
  %%CITATION = HEP-PH/0409138;%%
  %63 citations counted in INSPIRE as of 08 Jun 2013
%
%\cite{Dimopoulos:2005ac}
%\bibitem{Dimopoulos:2005ac} 
  S.~Dimopoulos, S.~Kachru, J.~McGreevy and J.~G.~Wacker,
  %``N-flation,''
%  JCAP {\bf 0808}, 003 (2008)
\textit{ibid}.\  {\bf 0808}, 003 (2008) 
  [hep-th/0507205];\ 
  %%CITATION = HEP-TH/0507205;%%
  %230 citations counted in INSPIRE as of 08 Jun 2013
%
%\cite{Easther:2005zr}
%\bibitem{Easther:2005zr} 
  R.~Easther and L.~McAllister,
  %``Random matrices and the spectrum of N-flation,''
%  JCAP {\bf 0605}, 018 (2006)
\textit{ibid}.\  {\bf 0605}, 018 (2006)
  [hep-th/0512102];\ 
  %%CITATION = HEP-TH/0512102;%%
  %96 citations counted in INSPIRE as of 08 Jun 2013
%
%\cite{McAllister:2008hb}
%\bibitem{McAllister:2008hb} 
  L.~McAllister, E.~Silverstein and A.~Westphal,
  %``Gravity Waves and Linear Inflation from Axion Monodromy,''
  Phys.\ Rev.\ D {\bf 82}, 046003 (2010)
  [arXiv:0808.0706 [hep-th]];\ 
  %%CITATION = ARXIV:0808.0706;%%
  %141 citations counted in INSPIRE as of 08 Jun 2013
%
%\cite{Flauger:2009ab}
%\bibitem{Flauger:2009ab} 
  R.~Flauger, L.~McAllister, E.~Pajer, A.~Westphal and G.~Xu,
  %``Oscillations in the CMB from Axion Monodromy Inflation,''
  JCAP {\bf 1006}, 009 (2010)
  [arXiv:0907.2916 [hep-th]];\ 
  %%CITATION = ARXIV:0907.2916;%%
  %90 citations counted in INSPIRE as of 08 Jun 2013
%
%\cite{Kaloper:2008fb}
%\bibitem{Kaloper:2008fb} 
  N.~Kaloper and L.~Sorbo,
  %``A Natural Framework for Chaotic Inflation,''
  Phys.\ Rev.\ Lett.\  {\bf 102}, 121301 (2009)
  [arXiv:0811.1989 [hep-th]].
  %%CITATION = ARXIV:0811.1989;%%
  %30 citations counted in INSPIRE as of 08 Jun 2013
%

\bibitem{CN-inflation}
%
%\cite{Adshead:2012kp}
%\bibitem{Adshead:2012kp} 
  P.~Adshead and M.~Wyman,
  %``Chromo-Natural Inflation: Natural inflation on a steep potential 
  %with classical non-Abelian gauge fields,''
  Phys.\ Rev.\ Lett.\  {\bf 108}, 261302 (2012)
  [arXiv:1202.2366 [hep-th]];\ 
  %%CITATION = ARXIV:1202.2366;%%
  %29 citations counted in INSPIRE as of 03 Oct 2013
%
%\cite{Adshead:2012qe}
%\bibitem{Adshead:2012qe} 
%  P.~Adshead and M.~Wyman,
  %``Gauge-flation trajectories in Chromo-Natural Inflation,''
  Phys.\ Rev.\ D {\bf 86}, 043530 (2012)
  [arXiv:1203.2264 [hep-th]];\ 
  %%CITATION = ARXIV:1203.2264;%%
  %17 citations counted in INSPIRE as of 03 Oct 2013
%
%\cite{Martinec:2012bv}
%\bibitem{Martinec:2012bv} 
  E.~Martinec, P.~Adshead and M.~Wyman,
  %``Chern-Simons EM-flation,''
  JHEP {\bf 1302}, 027 (2013)
  [arXiv:1206.2889 [hep-th]].
  %%CITATION = ARXIV:1206.2889;%%
  %10 citations counted in INSPIRE as of 03 Oct 2013
%

%
\bibitem{B-D}
N.~D.~Birrell and P.~C.~W.~Davies,
\textit{Quantum fields in curved space}
(Cambridge University Press, New York, 1982);\ 
%
V.~F.~Mukhanov and S.~Winirzki, 
\textit{Introduction to Quantum Effects in Gravity}
(Cambridge University Press, New York, 2007).
%

%\cite{Smoot:1992td}
\bibitem{Smoot:1992td} 
  G.~F.~Smoot, C.~L.~Bennett, A.~Kogut, E.~L.~Wright, J.~Aymon, N.~W.~Boggess, E.~S.~Cheng and G.~De Amici {\it et al.},
  %``Structure in the COBE differential microwave radiometer 
  %first year maps,''
  Astrophys.\ J.\  {\bf 396}, L1 (1992).
  %%CITATION = ASJOA,396,L1;%%
  %1726 citations counted in INSPIRE as of 09 Nov 2013

%%%%%%%%
%\cite{Ade:2013uln}
\bibitem{Ade:2013uln} 
  P.~A.~R.~Ade {\it et al.}  [Planck Collaboration],
  %``Planck 2013 results. XXII. Constraints on inflation,''
  arXiv:1303.5082 [astro-ph.CO].
  %%CITATION = ARXIV:1303.5082;%%
  %70 citations counted in INSPIRE as of 03 Jun 2013
%%%%%%%%

%\cite{Ade:2013lta}
\bibitem{Ade:2013lta} 
  P.~A.~R.~Ade {\it et al.}  [Planck Collaboration],
  %``Planck 2013 results. XVI. Cosmological parameters,''
  arXiv:1303.5076 [astro-ph.CO].
  %%CITATION = ARXIV:1303.5076;%%
  %21 citations counted in INSPIRE as of 02 Apr 2013
%

%%%%%%%%%%%%%%

\bibitem{Kolb and Turner}
E.~W.~Kolb and M.~S.~Turner,
\textit{The Early Universe}
(Addison-Wesley, Redwood City, California, 1990).

%\cite{Bassett:2000aw}
\bibitem{Bassett:2000aw} 
  B.~A.~Bassett, G.~Pollifrone, S.~Tsujikawa and F.~Viniegra,
  %``Preheating as cosmic magnetic dynamo,''
  Phys.\ Rev.\ D {\bf 63}, 103515 (2001)
  [astro-ph/0010628].
  %%CITATION = ASTRO-PH/0010628;%%
  %65 citations counted in INSPIRE as of 17 Dec 2014

%%%%%%%%%%%%%%
%%% G-part %%%
%%%%%%%%%%%%%%

\bibitem{BHKP-B} 
%
%\cite{Barnaby:2009mc}
%\bibitem{Barnaby:2009mc} 
  N.~Barnaby, Z.~Huang, L.~Kofman and D.~Pogosyan,
  %``Cosmological Fluctuations from Infra-Red Cascading During Inflation,''
  Phys.\ Rev.\ D {\bf 80}, 043501 (2009)
  [arXiv:0902.0615 [hep-th]];\ 
  %%CITATION = ARXIV:0902.0615;%%
  %57 citations counted in INSPIRE as of 29 Sep 2013
%
%\cite{Barnaby:2010ke}
%\bibitem{Barnaby:2010ke} 
  N.~Barnaby,
  %``On Features and Nongaussianity from Inflationary Particle Production,''
%  Phys.\ Rev.\ D {\bf 82}, 106009 (2010)
\textit{ibid}.\  {\bf 82}, 106009 (2010)
  [arXiv:1006.4615 [astro-ph.CO]].
  %%CITATION = ARXIV:1006.4615;%%
  %45 citations counted in INSPIRE as of 29 Sep 2013

%\cite{Hinshaw:2012aka}
\bibitem{Hinshaw:2012aka} 
  G.~Hinshaw {\it et al.}  [WMAP Collaboration],
  %``Nine-Year Wilkinson Microwave Anisotropy Probe (WMAP) 
  %Observations: Cosmological Parameter Results,''
  Astrophys.\ J.\ Suppl.\  {\bf 208}, 19 (2013)
  [arXiv:1212.5226 [astro-ph.CO]].
  %%CITATION = ARXIV:1212.5226;%%
  %616 citations counted in INSPIRE as of 20 Dec 2013

%%%%%%%%%%%%%%% \delta N %%%%%%%%%%%%%%%

%\cite{Sasaki:1995aw}
\bibitem{Sasaki:1995aw} 
  M.~Sasaki and E.~D.~Stewart,
  %``A General analytic formula for the spectral index 
  %of the density perturbations produced during inflation,''
  Prog.\ Theor.\ Phys.\  {\bf 95}, 71 (1996)
  [astro-ph/9507001].
  %%CITATION = ASTRO-PH/9507001;%%
  %533 citations counted in INSPIRE as of 26 Sep 2013

%\cite{Starobinsky:1986fxa}
\bibitem{Starobinsky:1986fxa} 
  A.~A.~Starobinsky,
  %``Multicomponent de Sitter (Inflationary) Stages 
  %and the Generation of Perturbations,''
  JETP Lett.\  {\bf 42}, 152 (1985)
  [Pisma Zh.\ Eksp.\ Teor.\ Fiz.\  {\bf 42}, 124 (1985)].
  %%CITATION = JTPLA,42,152;%%
  %375 citations counted in INSPIRE as of 26 Sep 2013

%\cite{Lyth:2005fi}
\bibitem{Lyth:2005fi} 
  D.~H.~Lyth and Y.~Rodriguez,
  %``The Inflationary prediction for primordial non-Gaussianity,''
  Phys.\ Rev.\ Lett.\  {\bf 95}, 121302 (2005)
  [astro-ph/0504045].
  %%CITATION = ASTRO-PH/0504045;%%
  %364 citations counted in INSPIRE as of 26 Sep 2013

%%%%%%%%%%%%%%%%%%%%%%%%%%%%%%%%%%%%%%%

%\cite{Komatsu:2010fb}
\bibitem{Komatsu:2010fb}
  E.~Komatsu {\it et al.}  [WMAP Collaboration],
  %``Seven-Year Wilkinson Microwave Anisotropy Probe (WMAP) Observations:
  %Cosmological Interpretation,''
  Astrophys.\ J.\ Suppl.\  {\bf 192}, 18 (2011)
  [arXiv:1001.4538 [astro-ph.CO]].
  %%CITATION = APJSA,192,18;%%

\bibitem{POLARBEAR}
http://mountainpolarbear.blogspot.jp/.

\bibitem{LiteBIRD}
http://cmbpol.kek.jp/litebird/.

%%%%%%%%%%%%%%%%%%%
\bibitem{A-A}
%
%\cite{Ade:2014gna}
%\bibitem{Ade:2014gna} 
  P.~A.~R.~Ade {\it et al.}  [Planck Collaboration],
  %``Planck intermediate results. XIX. An overview of the polarized thermal 
  %emission from Galactic dust,''
  arXiv:1405.0871 [astro-ph.GA];\ 
  %%CITATION = ARXIV:1405.0871;%%
  %4 citations counted in INSPIRE as of 27 May 2014
%
%\cite{Ade:2014zja}
%\bibitem{Ade:2014zja} 
%  P.~A.~R.~Ade {\it et al.}  [Planck Collaboration],
  %``Planck intermediate results. XXII. Frequency dependence of 
  %thermal emission from Galactic dust in intensity and polarization,''
  arXiv:1405.0874 [astro-ph.GA];\
  %%CITATION = ARXIV:1405.0874;%%
  %3 citations counted in INSPIRE as of 27 May 2014
%
%\cite{Adam:2014oea}
%\bibitem{Adam:2014oea} 
  R.~Adam {\it et al.}  [Planck Collaboration],
  %``Planck intermediate results. XXX. The angular power spectrum 
  %of polarized dust emission at intermediate and 
  %high Galactic latitudes,''
  arXiv:1409.5738 [astro-ph.CO].
  %%CITATION = ARXIV:1409.5738;%%
  %3 citations counted in INSPIRE as of 25 Sep 2014
%

\bibitem{MS-KK}
%
%\cite{Mortonson:2014bja}
%\bibitem{Mortonson:2014bja} 
  M.~J.~Mortonson and U.~Seljak,
  %``A joint analysis of Planck and 
  %BICEP2 B modes including dust polarization uncertainty,''
  arXiv:1405.5857 [astro-ph.CO];\ 
  %%CITATION = ARXIV:1405.5857;%%
%
%\bibitem{Kamionkowski:2014}
  M.~Kamionkowski and E.~D.~Kovetz, 
  %``Statistical diagnostics to identify Galactic foregrounds 
  %in B-mode maps,''
  arXiv:1408.4125 [astro-ph.CO].
%
%%%%%%%%%%%%%%%%%%%

%%%%%
\bibitem{KS}
%
%\cite{Kobayashi:2014jga}
%\bibitem{Kobayashi:2014jga}
  T.~Kobayashi and O.~Seto,
  %``Polynomial inflation models after BICEP2,''
  Phys.\ Rev.\ D {\bf 89}, 103524 (2014)
  [arXiv:1403.5055 [astro-ph.CO]];\ 
  %%CITATION = ARXIV:1403.5055;%%
  %20 citations counted in INSPIRE as of 05 Jun 2014
%
%\cite{Kobayashi:2014rla}
%\bibitem{Kobayashi:2014rla} 
%  T.~Kobayashi and O.~Seto,
  %``Beginning of Universe through large field hybrid inflation,''
  arXiv:1404.3102 [hep-ph].
  %%CITATION = ARXIV:1404.3102;%%
  %2 citations counted in INSPIRE as of 23 Apr 2014
%

\bibitem{S-B}
%
%\cite{Joergensen:2014rya}
%\bibitem{Joergensen:2014rya} 
  J.~Joergensen, F.~Sannino and O.~Svendsen,
  %``Primordial tensor modes from quantum corrected inflation,''
  Phys.\ Rev.\ D {\bf 90}, 043509 (2014)
  [arXiv:1403.3289 [hep-ph]];\ 
  %%CITATION = ARXIV:1403.3289;%%
  %14 citations counted in INSPIRE as of 02 Oct 2014
%
%\cite{Gong:2014cqa}
%\bibitem{Gong:2014cqa}
  Q.~Gao and Y.~Gong,
  %``The challenge for single field inflation with BICEP2 result,''
  Phys.\ Lett.\ B {\bf 734}, 41 (2014)
  [arXiv:1403.5716 [gr-qc]];\ 
  %%CITATION = ARXIV:1403.5716;%%
  %21 citations counted in INSPIRE as of 01 Jun 2014
%
%\cite{Ashoorioon:2014nta}
%\bibitem{Ashoorioon:2014nta}
  A.~Ashoorioon, K.~Dimopoulos, M.~M.~Sheikh-Jabbari and G.~Shiu,
  %``Non-Bunch-Davis Initial State Reconciles Chaotic Models 
  %with BICEP and Planck,''
  arXiv:1403.6099 [hep-th];\ 
  %%CITATION = ARXIV:1403.6099;%%
  %41 citations counted in INSPIRE as of 05 Jul 2014
%
%\cite{Sloth:2014sga}
%\bibitem{Sloth:2014sga}
  M.~S.~Sloth,
  %``Chaotic inflation with curvaton induced running,''
  arXiv:1403.8051 [hep-ph];\ 
  %%CITATION = ARXIV:1403.8051;%%
  %8 citations counted in INSPIRE as of 13 Jun 2014
%
%\cite{Cheng:2014cja}
%\bibitem{Cheng:2014cja}
  C.~Cheng and Q.~-G.~Huang,
  %``Constraint on inflation model from BICEP2 and WMAP 9-year data,''
  arXiv:1404.1230 [astro-ph.CO];\ 
  %%CITATION = ARXIV:1404.1230;%%
  %10 citations counted in INSPIRE as of 05 Jun 2014
%
%\cite{Cheng:2014bta}
%\bibitem{Cheng:2014bta}
  C.~Cheng, Q.~-G.~Huang and W.~Zhao,
  %``Constraints on the extensions to the base $\Lambda$CDM 
  %model from BICEP2, Planck and WMAP,''
  arXiv:1404.3467 [astro-ph.CO];\  
  %%CITATION = ARXIV:1404.3467;%%
  %8 citations counted in INSPIRE as of 05 Jun 2014
%
%\cite{Hu:2014aua}
%\bibitem{Hu:2014aua}
  B.~Hu, J.~-W.~Hu, Z.~-K.~Guo and R.~-G.~Cai,
  %``Reconstruction of the primordial power spectra with Planck and BICEP2,''
  arXiv:1404.3690 [astro-ph.CO];\ 
  %%CITATION = ARXIV:1404.3690;%%
  %8 citations counted in INSPIRE as of 11 Jun 2014
%
%\cite{Hamada:2014xka}
%\bibitem{Hamada:2014xka}
  Y.~Hamada, H.~Kawai and K.~-y.~Oda,
  %``Predictions on mass of Higgs portal scalar dark matter from 
  %Higgs inflation and flat potential,''
  arXiv:1404.6141 [hep-ph];\ 
  %%CITATION = ARXIV:1404.6141;%%
  %6 citations counted in INSPIRE as of 05 Jun 2014
%
%\cite{Gao:2014pca}
%\bibitem{Gao:2014pca} 
  Q.~Gao, Y.~Gong and T.~Li,
  %``The Modified Lyth Bound and Implications of BICEP2 Results,''
  arXiv:1405.6451 [gr-qc];\ 
  %%CITATION = ARXIV:1405.6451;%%
%
%\cite{Barranco:2014ira}
%\bibitem{Barranco:2014ira} 
  L.~Barranco, L.~Boubekeur and O.~Mena,
  %``A model-independent fit to Planck and BICEP2 data,''
  arXiv:1405.7188 [astro-ph.CO];\ 
  %%CITATION = ARXIV:1405.7188;%%
%
%\cite{Martin:2014lra}
%\bibitem{Martin:2014lra} 
  J.~Martin, C.~Ringeval, R.~Trotta and V.~Vennin,
  %``Compatibility of Planck and BICEP2 in the Light of Inflation,''
  arXiv:1405.7272 [astro-ph.CO];\ 
  %%CITATION = ARXIV:1405.7272;%%
%
%\cite{Garcia-Bellido:2014eva}
%\bibitem{Garcia-Bellido:2014eva} 
  J.~Garcia-Bellido, D.~Roest, M.~Scalisi and I.~Zavala,
  %``Can CMB data constrain the inflationary field range?,''
  arXiv:1405.7399 [hep-th];\ 
  %%CITATION = ARXIV:1405.7399;%%
%
%\cite{Hossain:2014ova}
%\bibitem{Hossain:2014ova} 
  M.~Wail Hossain, R.~Myrzakulov, M.~Sami and E.~N.~Saridakis,
  %``Evading Lyth bound in models of quintessential inflation,''
  arXiv:1405.7491 [gr-qc];\ 
  %%CITATION = ARXIV:1405.7491;%%
%
%\cite{Bamba:2014daa}
%\bibitem{Bamba:2014daa} 
  K.~Bamba, S.~Nojiri and S.~D.~Odintsov,
  %``Reconstruction of scalar field theories realizing inflation consistent 
  %with the Planck and BICEP2 results,''
  Phys.\ Lett.\ B {\bf 737}, 374 (2014)
  [arXiv:1406.2417 [hep-th]];\ 
  %%CITATION = ARXIV:1406.2417;%%
  %5 citations counted in INSPIRE as of 02 Oct 2014
%
%\cite{Inagaki:2014wva}
%\bibitem{Inagaki:2014wva} 
  T.~Inagaki, R.~Nakanishi and S.~D.~Odintsov,
  %``Inflationary Parameters in Renormalization Group 
  %Improved $\phi^4$ Theory,''
  arXiv:1408.1270 [gr-qc];\ 
  %%CITATION = ARXIV:1408.1270;%%
%
%\cite{Elizalde:2014xva}
%\bibitem{Elizalde:2014xva} 
  E.~Elizalde, S.~D.~Odintsov, E.~O.~Pozdeeva and S.~Y.~Vernov,
  %``Renormalization-group inflationary scalar electrodynamics 
  %and SU(5) scenarios confronted with Planck2013 and BICEP2 results,''
  arXiv:1408.1285 [hep-th];\ 
  %%CITATION = ARXIV:1408.1285;%%
%
%\cite{Dine:2014gba}
%\bibitem{Dine:2014gba} 
  M.~Dine and L.~Stephenson-Haskins,
  %``Hybrid Inflation with Planck Scale Fields,''
  arXiv:1408.0046 [hep-ph];\ 
  %%CITATION = ARXIV:1408.0046;%%
%
%\cite{Hamada:2014wna}
%\bibitem{Hamada:2014wna} 
  Y.~Hamada, H.~Kawai, K.~y.~Oda and S.~C.~Park,
  %``Higgs inflation from Standard Model criticality,''
  arXiv:1408.4864 [hep-ph].
  %%CITATION = ARXIV:1408.4864;%%
  %2 citations counted in INSPIRE as of 02 Oct 2014
%

\bibitem{KSY-HKSY}
%
%\cite{Kobayashi:2014ooa}
%\bibitem{Kobayashi:2014ooa} 
  T.~Kobayashi, O.~Seto and Y.~Yamaguchi,
  %``Axion monodromy inflation with sinusoidal corrections,''
  arXiv:1404.5518 [hep-ph];\ 
  %%CITATION = ARXIV:1404.5518;%%
  %6 citations counted in INSPIRE as of 21 May 2014
%
%\cite{Higaki:2014sja}
%\bibitem{Higaki:2014sja} 
  T.~Higaki, T.~Kobayashi, O.~Seto and Y.~Yamaguchi,
  %``Axion monodromy inflation with multi-natural modulations,''
  arXiv:1405.0775 [hep-ph].
  %%CITATION = ARXIV:1405.0775;%%
  %4 citations counted in INSPIRE as of 21 May 2014
%

%%%%%%%%%%%%%%%%%%%%%%%%%%
%%% Refs. for Appendix A %%%
%%%%%%%%%%%%%%%%%%%%%%%%%%%%%%%%%%%%%%%%%%%%%%%%%%%%%%%%%%%%%%%%%%%%%%%%%%%

%\cite{Demozzi:2009fu}
\bibitem{Demozzi:2009fu}
  V.~Demozzi, V.~Mukhanov and H.~Rubinstein,
  %``Magnetic fields from inflation?,''
  JCAP {\bf 0908}, 025 (2009) 
  [arXiv:0907.1030 [astro-ph.CO]]. 
  %%CITATION = JCAPA,0908,025;%%

%\bibitem{Backreaction}
%
%\cite{Kanno:2009ei}
\bibitem{Kanno:2009ei}
  S.~Kanno, J.~Soda and M.~a.~Watanabe,
  %``Cosmological Magnetic Fields from Inflation and Backreaction,''
  JCAP {\bf 0912}, 009 (2009)
%\textit{ibid}. {\bf 0912}, 009 (2009). 
  [arXiv:0908.3509 [astro-ph.CO]].
  %%CITATION = JCAPA,0912,009;%%
%

%\cite{Suyama:2012wh}
\bibitem{Suyama:2012wh} 
  T.~Suyama and J.~Yokoyama,
  %``Metric perturbation from inflationary magnetic field 
  %and generic bound on inflation models,''
  Phys.\ Rev.\ D {\bf 86}, 023512 (2012)
  [arXiv:1204.3976 [astro-ph.CO]].
  %%CITATION = ARXIV:1204.3976;%%

%\cite{Fujita:2012rb}
\bibitem{Fujita:2012rb} 
  T.~Fujita and S.~Mukohyama,
  %``Universal upper limit on inflation energy scale 
  %from cosmic magnetic field,''
  JCAP {\bf 1210}, 034 (2012)
  [arXiv:1205.5031 [astro-ph.CO]].
  %%CITATION = ARXIV:1205.5031;%%

\bibitem{DMR-C}
%
%\cite{Durrer:2013xla}
%\bibitem{Durrer:2013xla} 
  R.~Durrer, G.~Marozzi and M.~Rinaldi,
  %``Comment on "Origin of cosmic magnetic fields",''
  Phys.\ Rev.\ Lett.\  {\bf 111}, 229001 (2013)
%\textit{ibid}.\  {\bf 111}, 229001 (2013)
  [arXiv:1305.3192 [astro-ph.CO]];\ 
  %%CITATION = ARXIV:1305.3192;%%
  %2 citations counted in INSPIRE as of 20 Dec 2013
%
L.~Campanelli,
%``Reply to "Comments on Origin of cosmic magnetic fields",''
%Phys. Rev. Lett.  {\bf 111}, 229002 (2013)
\textit{ibid}.\  {\bf 111}, 229002 (2013)
[arXiv:1305.7062 [astro-ph.CO]].


%\cite{Maleknejad:2012fw}
\bibitem{Maleknejad:2012fw} 
  A.~Maleknejad, M.~M.~Sheikh-Jabbari and J.~Soda,
  %``Gauge Fields and Inflation,''
  Phys.\ Rept.\  {\bf 528}, 161 (2013)
  [arXiv:1212.2921 [hep-th]].
  %%CITATION = ARXIV:1212.2921;%%
  %14 citations counted in INSPIRE as of 19 Aug 2013

%\cite{Cembranos:2012ng}
\bibitem{Cembranos:2012ng} 
  J.~A.~R.~Cembranos, A.~L.~Maroto and S.~J.~N\'{u}\~{n}ez.~Jare\~{n}o,
  %``Isotropy theorem for cosmological Yang-Mills theories,''
  Phys.\ Rev.\ D {\bf 87}, 043523 (2013)
  [arXiv:1212.3201 [astro-ph.CO]].
  %%CITATION = ARXIV:1212.3201;%%
  %8 citations counted in INSPIRE as of 19 Aug 2013

%\cite{Bartolo:2013msa}
\bibitem{Bartolo:2013msa} 
  N.~Bartolo, S.~Matarrese, M.~Peloso and A.~Ricciardone,
  %``Anisotropy in solid inflation,''
  JCAP {\bf 1308}, 022 (2013)
  [arXiv:1306.4160 [astro-ph.CO]].
  %%CITATION = ARXIV:1306.4160;%%
  %3 citations counted in INSPIRE as of 01 Sep 2013

%\cite{Namba:2013kia}
\bibitem{Namba:2013kia} 
  R.~Namba, E.~Dimastrogiovanni and M.~Peloso,
  %``Gauge-flation confronted with Planck,''
  JCAP {\bf 1311}, 045 (2013)
  [arXiv:1308.1366 [astro-ph.CO]].
  %%CITATION = ARXIV:1308.1366;%%
  %1 citations counted in INSPIRE as of 20 Dec 2013

%\cite{Nurmi:2013gpa}
\bibitem{Nurmi:2013gpa} 
  S.~Nurmi and M.~S.~Sloth,
  %``Constraints on Gauge Field Production during Inflation,''
  arXiv:1312.4946 [astro-ph.CO].
  %%CITATION = ARXIV:1312.4946;%%

%\cite{Fujita:2014sna}
\bibitem{Fujita:2014sna} 
  T.~Fujita and S.~Yokoyama,
  %``Critical constraint on inflationary magnetogenesis,''
  JCAP {\bf 1403}, 013 (2014) 
  [arXiv:1402.0596 [astro-ph.CO]].
  %%CITATION = ARXIV:1402.0596;%%

%\cite{Ferreira:2013sqa}
\bibitem{Ferreira:2013sqa} 
  R.~J.~Z.~Ferreira, R.~K.~Jain and M.~S.~Sloth,
  %``Inflationary magnetogenesis without the strong coupling problem,''
  JCAP {\bf 1310}, 004 (2013)
  [arXiv:1305.7151 [astro-ph.CO]].
  %%CITATION = ARXIV:1305.7151;%%
  %11 citations counted in INSPIRE as of 20 Dec 2013

%\cite{Ferreira:2014hma}
\bibitem{Ferreira:2014hma} 
  R.~J.~Z.~Ferreira, R.~K.~Jain and M.~S.~Sloth,
  %``Inflationary Magnetogenesis without the Strong Coupling 
  %Problem II: Constraints from CMB anisotropies and B-modes,''
  arXiv:1403.5516 [astro-ph.CO].
  %%CITATION = ARXIV:1403.5516;%%

%%%%%%%%%%%%%%%%%%%%%%%%%%%%%%%%%%%%%%%%%%%%%%%%%%%%%
%\cite{Caprini:2014mja}
\bibitem{Caprini:2014mja} 
  C.~Caprini and L.~Sorbo,
  %``Adding helicity to inflationary magnetogenesis,''
  JCAP {\bf 1410}, 056 (2014)
  [arXiv:1407.2809 [astro-ph.CO]].
  %%CITATION = ARXIV:1407.2809;%%
  %7 citations counted in INSPIRE as of 21 Nov 2014
%%%%%%%%%%%%%%%%%%%%%%%%%%%%%%%%%%%%%%%%%%%%%%%%%%%%%

\bibitem{AHK-AHKS}
%
%\cite{Abe:2005rx}
%\bibitem{Abe:2005rx} 
  H.~Abe, T.~Higaki and T.~Kobayashi,
  %``KKLT type models with moduli-mixing superpotential,''
  Phys.\ Rev.\ D {\bf 73}, 046005 (2006)
  [hep-th/0511160];\ 
  %%CITATION = HEP-TH/0511160;%%
%\cite{Abe:2008xu}
%\bibitem{Abe:2008xu} 
  H.~Abe, T.~Higaki, T.~Kobayashi and O.~Seto,
  %``Non-perturbative moduli superpotential with positive exponents,''
%  Phys.\ Rev.\ D {\bf 78}, 025007 (2008)
\textit{ibid}.\  {\bf 78}, 025007 (2008)
  [arXiv:0804.3229 [hep-th]].
  %%CITATION = ARXIV:0804.3229;%%
%

%%%%%%%%%%%%%%%%%%%%%%%%%%%%%%%%%%%%%%%%%%%%%%%%%%%%%%%%%%%%%%%%%%%%%%%%%%%

%%%%%%%%%%%%%%%%%%%%%%%%%%
%%% Refs. for Appendix B %%%
%%%%%%%%%%%%%%%%%%%%%%%%%%%%%%%%%%%%%%%%%%%%%%%%%%%%%%%%%%%%%%%%%%%%%%%%%%%

\bibitem{CMB-Limit}
%
%\bibitem{Subramanian}
%\cite{Subramanian:1998fn}
%\bibitem{Subramanian:1998fn}
  K.~Subramanian and J.~D.~Barrow,
  %``Microwave Background Signals from Tangled Magnetic Fields,''
  Phys.\ Rev.\ Lett.\  {\bf 81}, 3575 (1998)
  [arXiv:astro-ph/9803261];\ 
  %%CITATION = PRLTA,81,3575;%%
%
%\cite{Seshadri:2000ky}
%\bibitem{Seshadri:2000ky}
  T.~R.~Seshadri and K.~Subramanian,
  %``CMBR Polarization Signals from Tangled Magnetic Fields,''
%  Phys.\ Rev.\ Lett.\  {\bf 87}, 101301 (2001)
\textit{ibid}. {\bf 87}, 101301 (2001) 
  [arXiv:astro-ph/0012056];\ 
  %%CITATION = PRLTA,87,101301;%%
%
%\cite{Subramanian:2002nh}
%\bibitem{Subramanian:2002nh}
  K.~Subramanian and J.~D.~Barrow,
  %``Small-scale microwave background anisotropies due to tangled primordial
  %magnetic fields,''
  Mon.\ Not.\ Roy.\ Astron.\ Soc.\  {\bf 335}, L57 (2002)
  [arXiv:astro-ph/0205312];\ 
  %%CITATION = MNRAA,335,L57;%%
%
%\cite{Subramanian:2003sh}
%\bibitem{Subramanian:2003sh}
  K.~Subramanian, T.~R.~Seshadri and J.~D.~Barrow,
  %``Small-scale CMB polarization anisotropies due to tangled primordial
  %magnetic fields,''
%  Mon.\ Not.\ Roy.\ Astron.\ Soc.\  {\bf 344}, L31 (2003)
\textit{ibid}. {\bf 344}, L31 (2003)
  [arXiv:astro-ph/0303014];\ 
  %%CITATION = MNRAA,344,L31;%%
%
%\cite{Tashiro:2005hc}
%\bibitem{Tashiro:2005hc}
  H.~Tashiro, N.~Sugiyama and R.~Banerjee,
  %``Nonlinear Evolution of Cosmic Magnetic Fields and Cosmic Microwave
  %Background Anisotropies,''
  Phys.\ Rev.\  D {\bf 73}, 023002 (2006)
  [arXiv:astro-ph/0509220];\ 
  %%CITATION = PHRVA,D73,023002;%%
%
%\cite{Yamazaki:2008gr}
%\bibitem{Yamazaki:2008gr}
  D.~G.~Yamazaki, K.~Ichiki, T.~Kajino and G.~J.~Mathews,
  %``Effects of a Primordial Magnetic Field on Low 
  %and High Multipoles of the CMB,''
%  Phys.\ Rev.\  D {\bf 77}, 043005 (2008)
\textit{ibid}. {\bf 77}, 043005 (2008) 
  [arXiv:0801.2572 [astro-ph]];\ 
  %%CITATION = PHRVA,D77,043005;%%
%
%\cite{Kahniashvili:2008hx}
%\bibitem{Kahniashvili:2008hx}
  T.~Kahniashvili, Y.~Maravin and A.~Kosowsky,
  %``Faraday Rotation Limits On A Primordial Magnetic Field From Wilkinson
  %Microwave Anisotropy Probe Five-Year Data,''
%  Phys.\ Rev.\  D {\bf 80}, 023009 (2009)
\textit{ibid}. {\bf 80}, 023009 (2009)
  [arXiv:0806.1876 [astro-ph]];\ 
  %%CITATION = PHRVA,D80,023009;%%
%
%\cite{Yamazaki:2010nf}
%\bibitem{Yamazaki:2010nf}
  D.~G.~Yamazaki, K.~Ichiki, T.~Kajino and G.~J.~Mathews,
  %``New Constraints on the Primordial Magnetic Field,''
%  Phys.\ Rev.\  D {\bf 81}, 023008 (2010)
\textit{ibid}. {\bf 81}, 023008 (2010) 
  [arXiv:1001.2012 [astro-ph.CO]];\ 
  %%CITATION = PHRVA,D81,023008;%%
%
%\cite{Shaw:2010ea}
%\bibitem{Shaw:2010ea} 
  J.~R.~Shaw and A.~Lewis,
  %``Constraining Primordial Magnetism,''
%  Phys.\ Rev.\ D {\bf 86}, 043510 (2012)
\textit{ibid}. {\bf 86}, 043510 (2012) 
  [arXiv:1006.4242 [astro-ph.CO]];\ 
  %%CITATION = ARXIV:1006.4242;%%
  %34 citations counted in INSPIRE as of 01 Sep 2013
%
%\cite{Trivedi:2010gi}
%\bibitem{Trivedi:2010gi}
  P.~Trivedi, K.~Subramanian and T.~R.~Seshadri,
  %``Primordial Magnetic Field Limits from Cosmic Microwave Background
  %Bispectrum of Magnetic Passive Scalar Modes,''
%  Phys.\ Rev.\  D {\bf 82}, 123006 (2010)
\textit{ibid}. {\bf 82}, 123006 (2010) 
  [arXiv:1009.2724 [astro-ph.CO]];\ 
  %%CITATION = PHRVA,D82,123006;%%
%
%\cite{Shiraishi:2010yk}
%\bibitem{Shiraishi:2010yk} 
  M.~Shiraishi, D.~Nitta, S.~Yokoyama, K.~Ichiki and K.~Takahashi,
  %``Cosmic microwave background bispectrum of vector modes induced 
  %from primordial magnetic fields,''
%  Phys.\ Rev.\ D {\bf 82}, 121302 (2010)
\textit{ibid}.\  {\bf 82}, 121302 (2010) 
  [Erratum-ibid.\ D {\bf 83}, 029901 (2011)] 
  [arXiv:1009.3632 [astro-ph.CO]];\ 
  %%CITATION = ARXIV:1009.3632;%%
  %18 citations counted in INSPIRE as of 01 Sep 2013
%
%\cite{Yamazaki:2014xna}
%\bibitem{Yamazaki:2014xna} 
  D.~G.~Yamazaki,
  %``CMB with the background primordial magnetic field,''
%  Phys.\ Rev.\ D {\bf 89}, 083528 (2014)
\textit{ibid}.\  {\bf 89}, 083528 (2014)
  [arXiv:1404.5310 [astro-ph.CO]].
  %%CITATION = ARXIV:1404.5310;%%
%

%\cite{Giovannini:2008df}
\bibitem{Giovannini:2008df}
  M.~Giovannini and K.~E.~Kunze,
  %``CMB polarization induced by stochastic magnetic fields,''
  arXiv:0804.2238 [astro-ph].
  %%CITATION = ARXIV:0804.2238;%%

%\cite{Barrow:1997mj}
\bibitem{Barrow:1997mj}
  J.~D.~Barrow, P.~G.~Ferreira and J.~Silk,
  %``Constraints on a Primordial Magnetic Field,''
  Phys.\ Rev.\ Lett.\  {\bf 78}, 3610 (1997) 
  [arXiv:astro-ph/9701063].
  %%CITATION = PRLTA,78,3610;%%

%%%
%\cite{Pogosian:2013dya}
\bibitem{Pogosian:2013dya} 
  L.~Pogosian,
  %``Searching for primordial magnetism with multi-frequency CMB experiments,''
  %Monthly Notices of the Royal Astronomical Society, Volume 438,
  %Issue 3, p.2508-2512, 2014
  Mon.\ Not.\ Roy.\ Astron.\ Soc.\  {\bf 438}, 2508 (2014)
  [arXiv:1311.2926 [astro-ph.CO]].
  %%CITATION = ARXIV:1311.2926;%%
  %1 citations counted in INSPIRE as of 24 Apr 2014

%\cite{Andre:2013afa}
\bibitem{Andre:2013afa} 
  P.~Andre {\it et al.}  [PRISM Collaboration],
  %``PRISM (Polarized Radiation Imaging and Spectroscopy Mission): 
  %A White Paper on the Ultimate Polarimetric Spectro-Imaging of 
  %the Microwave and Far-Infrared Sky,''
  arXiv:1306.2259 [astro-ph.CO].
  %%CITATION = ARXIV:1306.2259;%%
  %10 citations counted in INSPIRE as of 14 Nov 2013
%%%

%\cite{Tashiro:2006uv}
\bibitem{Tashiro:2006uv} 
  H.~Tashiro and N.~Sugiyama,
  %``Probing Primordial Magnetic Fields with the 21cm Fluctuations,''
  Mon.\ Not.\ Roy.\ Astron.\ Soc.\  {\bf 372}, 1060 (2006)
  [astro-ph/0607169].
  %%CITATION = ASTRO-PH/0607169;%%

\bibitem{YIKM}
%
%\cite{Yamazaki:2008jh}
%\bibitem{Yamazaki:2008jh}
  D.~G.~Yamazaki, K.~Ichiki, T.~Kajino and G.~J.~Mathews,
  %``Constraints on the Primordial Magnetic Field from $\sigma_8$,''
  Phys.\ Rev.\  D {\bf 78}, 123001 (2008)
  [arXiv:0811.2221 [astro-ph]];\ 
  %%CITATION = PHRVA,D78,123001;%%
%
%\cite{Yamazaki:2010jw}
%\bibitem{Yamazaki:2010jw}
%  D.~G.~Yamazaki, K.~Ichiki, T.~Kajino and G.~J.~Mathews,
  %``Constraints on the neutrino mass and the primordial magnetic field 
  %from the matter density fluctuation parameter $\sigma_8$,''
%  Phys.\ Rev.\  D {\bf 81}, 103519 (2010)
\textit{ibid}. {\bf 81}, 103519 (2010)
  [arXiv:1005.1638 [astro-ph.CO]].
  %%CITATION = PHRVA,D81,103519;%%
%
%%%%%

%\cite{Ganc:2014wia}
\bibitem{Ganc:2014wia} 
  J.~Ganc and M.~S.~Sloth,
  %``Probing correlations of early magnetic fields using $\mu$-distortion,''
  arXiv:1404.5957 [astro-ph.CO].
  %%CITATION = ARXIV:1404.5957;%%

%%%%%
%\cite{Wang:2008vp}
\bibitem{Wang:2008vp}
  S.~Wang,
  %``New primordial-magnetic-field limit from the latest LIGO S-5 data,''
  Phys.\ Rev.\  D {\bf 81}, 023002 (2010)
  [arXiv:0810.5620 [astro-ph]].
  %%CITATION = PHRVA,D81,023002;%%
%%%%%

%%%%%
\bibitem{TSLS-TS-TTI}
%\bibitem{Tashiro:2010st}
%
%\cite{Tashiro:2008fc}
%\bibitem{Tashiro:2008fc} 
  H.~Tashiro, J.~Silk, M.~Langer and N.~Sugiyama,
  %``The Sunyaev-Zel'dovich effect and 
  %Faraday rotation contributions of galaxy groups 
  %to the CMB angular power spectrum,''
  arXiv:0807.3888 [astro-ph];\ 
  %%CITATION = ARXIV:0807.3888;%%
%
%\cite{Tashiro:2009hx}
%\bibitem{Tashiro:2009hx} 
  H.~Tashiro and N.~Sugiyama,
  %``S-Z power spectrum produced by primordial magnetic fields,''
  arXiv:0908.0113 [astro-ph.CO];\ 
  %%CITATION = ARXIV:0908.0113;%%
  %3 citations counted in INSPIRE as of 12 Oct 2013
%
%\cite{Tashiro:2010st}
%\bibitem{Tashiro:2010st}
  H.~Tashiro, K.~Takahashi and K.~Ichiki,
  %``Primordial magnetic fields with X-ray and S-Z cluster survey,''
  arXiv:1010.4407 [astro-ph.CO].
  %%CITATION = ARXIV:1010.4407;%%
%
%%%%%

%\cite{Kuroyanagi:2009ez}
\bibitem{Kuroyanagi:2009ez} 
  S.~Kuroyanagi, H.~Tashiro and N.~Sugiyama,
  %``A numerical study of primordial magnetic field amplification 
  %by inflation-produced gravitational waves,''
  Phys.\ Rev.\ D {\bf 81}, 023510 (2010)
  [arXiv:0909.0907 [astro-ph.CO]].
  %%CITATION = ARXIV:0909.0907;%%
  %1 citations counted in INSPIRE as of 12 Oct 2013

%%%%%
%\cite{Bamba:2007hm}
\bibitem{Bamba:2007hm}
  K.~Bamba,
  %``Property of the spectrum of large-scale magnetic fields 
  %from inflation,''
  Phys.\ Rev.\  D {\bf 75}, 083516 (2007) 
  [arXiv:astro-ph/0703647].
  %%CITATION = PHRVA,D75,083516;%%

%\cite{Bonvin:2013tba}
\bibitem{Bonvin:2013tba} 
  C.~Bonvin, C.~Caprini and R.~Durrer,
  %``Magnetic fields from inflation: The CMB 
  %temperature anisotropies,''
  Phys.\  Rev.\  D88, {\bf 083515} (2013)
  [arXiv:1308.3348 [astro-ph.CO]].
  %%CITATION = ARXIV:1308.3348;%%
  %2 citations counted in INSPIRE as of 20 Dec 2013
%%%%%

%\cite{Takahashi:2013uoa}
\bibitem{Takahashi:2013uoa} 
  K.~Takahashi, M.~Mori, K.~Ichiki, S.~Inoue and H.~Takami,
  %``Lower Bounds on Magnetic Fields in Intergalactic 
  %Voids from Long-term GeV-TeV Light Curves of 
  %the Blazar Mrk 421,''
  Astrophys.\ J.\  {\bf 771}, L42 (2013)
  arXiv:1303.3069 [astro-ph.CO].
  %%CITATION = ASJOA,771,L42;%%
  %2 citations counted in INSPIRE as of 12 Oct 2013

\bibitem{BBN}
%
%\cite{Grasso:1996kk}
%\bibitem{Grasso:1996kk}
  D.~Grasso and H.~R.~Rubinstein,
  %``Revisiting Nucleosynthesis Constraints on Primordial Magnetic Fields,''
  Phys.\ Lett.\  B {\bf 379}, 73 (1996) 
  [arXiv:astro-ph/9602055];\ 
  %%CITATION = PHLTA,B379,73;%%
%
%\cite{Cheng:1996yi}
%\bibitem{Cheng:1996yi}
  B.~Cheng, A.~V.~Olinto, D.~N.~Schramm and J.~W.~Truran,
  %``Constraints on the strength of primordial magnetic fields from big bang
  %nucleosynthesis revisited,''
  Phys.\ Rev.\  D {\bf 54}, 4714 (1996)
  [arXiv:astro-ph/9606163]. 
  %%CITATION = PHRVA,D54,4714;%%

%\cite{Nikiel-Wroczynski:2013zwa}
\bibitem{Nikiel-Wroczynski:2013zwa} 
  B.~Nikiel-Wroczy\'{n}ski, M.~Soida, M.~Urbanik, R.~Beck and D.~J.~Bomans,
  %``Intergalactic magnetic fields in Stephan's Quintet,''
  Mon.\ Not.\ Roy.\ Astron.\ Soc.\ {\bf 435}, 149 (2013)
  [arXiv:1307.3447 [astro-ph.CO]].
  %%CITATION = ARXIV:1307.3447;%%

%\cite{Tashiro:2013bxa}
\bibitem{Tashiro:2013bxa} 
  H.~Tashiro and T.~Vachaspati,
  %``Cosmological magnetic field correlators from blazar induced cascade,''
  Phys.\ Rev.\ D {\bf 87}, 123527 (2013)
  [arXiv:1305.0181 [astro-ph.CO]].
  %%CITATION = ARXIV:1305.0181;%%

%\cite{Miyamoto:2013oua}
\bibitem{Miyamoto:2013oua} 
  K.~Miyamoto, T.~Sekiguchi, H.~Tashiro and S.~Yokoyama,
  %``CMB distortion anisotropies due to the decay of 
  %primordial magnetic fields,''
  arXiv:1310.3886 [astro-ph.CO].
  %%CITATION = ARXIV:1310.3886;%%

%\cite{Kunze:2013iwa}
\bibitem{Kunze:2013iwa} 
  K.~E.~Kunze,
  %``Secondary CMB anisotropies from bulk motions 
  %in the presence of stochastic magnetic fields,''
  arXiv:1312.5630 [astro-ph.CO].
  %%CITATION = ARXIV:1312.5630;%%

%%%
%\cite{Chen:2013gva}
\bibitem{Chen:2013gva} 
  P.~Chen and T.~Suyama,
  %``Constraining Primordial Magnetic Fields 
  %by CMB Photon-Graviton Conversion,''
  Phys.\ Rev.\ D {\bf 88}, 123521 (2013)
  [arXiv:1309.0537 [astro-ph.CO]].
  %%CITATION = ARXIV:1309.0537;%%
  %1 citations counted in INSPIRE as of 24 Apr 2014
%%%

%\cite{Tashiro:2013yea}
\bibitem{Tashiro:2013yea} 
  H.~Tashiro, J.~Silk and D.~J.~E.~Marsh,
  %``Constraints on primordial magnetic fields from 
  %CMB distortions in the axiverse,''
  Phys.\ Rev.\ D {\bf 88}, 125024 (2013)
  [arXiv:1308.0314 [astro-ph.CO]].
  %%CITATION = ARXIV:1308.0314;%%
  %6 citations counted in INSPIRE as of 24 Apr 2014

\bibitem{Axiverse}
%
%\cite{Arvanitaki:2009fg}
%\bibitem{Arvanitaki:2009fg} 
  A.~Arvanitaki, S.~Dimopoulos, S.~Dubovsky, N.~Kaloper and J.~March-Russell,
  %``String Axiverse,''
  Phys.\ Rev.\ D {\bf 81}, 123530 (2010)
  [arXiv:0905.4720 [hep-th]];\ 
  %%CITATION = ARXIV:0905.4720;%%
  %136 citations counted in INSPIRE as of 30 Sep 2013
%
%\cite{Yoshino:2012kn}
%\bibitem{Yoshino:2012kn} 
  H.~Yoshino and H.~Kodama,
  %``Bosenova collapse of axion cloud around a rotating black hole,''
  Prog.\ Theor.\ Phys.\  {\bf 128}, 153 (2012)
  [arXiv:1203.5070 [gr-qc]];\ 
  %%CITATION = ARXIV:1203.5070;%%
  %14 citations counted in INSPIRE as of 30 Sep 2013
%
%\cite{Yoshino:2013ofa}
%\bibitem{Yoshino:2013ofa} 
%  H.~Yoshino and H.~Kodama,
  %``Gravitational radiation from an axion cloud around 
  %a black hole: Superradiant phase,''
  arXiv:1312.2326 [gr-qc];\ 
  %%CITATION = ARXIV:1312.2326;%%
  %1 citations counted in INSPIRE as of 20 Dec 2013
%
%\cite{Kodama:2011zc}
%\bibitem{Kodama:2011zc} 
%  H.~Kodama and H.~Yoshino,
  %``Axiverse and Black Hole,''
  Int.\ J.\ Mod.\ Phys.\ Conf.\ Ser.\  {\bf 7}, 84 (2012)
  [arXiv:1108.1365 [hep-th]].
  %%CITATION = ARXIV:1108.1365;%%
  %16 citations counted in INSPIRE as of 30 Sep 2013
%

%\cite{Trivedi:2013wqa}
\bibitem{Trivedi:2013wqa} 
  P.~Trivedi, K.~Subramanian and T.~R.~Seshadri,
  %``Primordial Magnetic Field Limits from CMB Trispectrum - 
  %Scalar Modes and Planck Constraints,''
  Phys.\ Rev.\ D {\bf 89}, 043523 (2014)
  [arXiv:1312.5308 [astro-ph.CO]].
  %%CITATION = ARXIV:1312.5308;%%
  %2 citations counted in INSPIRE as of 24 Apr 2014

\bibitem{BS-HF}
%
%\cite{Brandenburg:2014zia}
%\bibitem{Brandenburg:2014zia} 
  A.~Brandenburg and R.~Stepanov,
  %``Faraday signature of magnetic helicity from 
  %reduced depolarization,''
  Astrophys.\ J.\  {\bf 786}, 91 (2014)
  [arXiv:1401.4102 [astro-ph.CO]];\ 
  %%CITATION = ARXIV:1401.4102;%%
  %1 citations counted in INSPIRE as of 24 Apr 2014
%
%\cite{Horellou:2014dja}
%\bibitem{Horellou:2014dja} 
  C.~Horellou and A.~Fletcher,
  %``Magnetic field tomography and differential Faraday rotation,''
  arXiv:1401.4152 [astro-ph.CO].
  %%CITATION = ARXIV:1401.4152;%%
%

%\cite{Caprini:2003vc}
\bibitem{Caprini:2003vc}
  C.~Caprini, R.~Durrer and T.~Kahniashvili,
  %``The Cosmic Microwave Background and Helical 
  %Magnetic Fields: the tensor mode,''
  Phys.\ Rev.\  D {\bf 69}, 063006 (2004) 
  [arXiv:astro-ph/0304556]. 
  %%CITATION = PHRVA,D69,063006;%%

%\cite{Tashiro:2012mf}
\bibitem{Tashiro:2012mf} 
  H.~Tashiro, T.~Vachaspati and A.~Vilenkin,
  %``Chiral Effects and Cosmic Magnetic Fields,''
  Phys.\ Rev.\ D {\bf 86}, 105033 (2012)
  [arXiv:1206.5549 [astro-ph.CO]].
  %%CITATION = ARXIV:1206.5549;%%
  %3 citations counted in INSPIRE as of 12 Oct 2013

%\cite{Joyce:1997uy}
\bibitem{Joyce:1997uy} 
  M.~Joyce and M.~E.~Shaposhnikov,
  %``Primordial magnetic fields, right-handed electrons, 
  %and the Abelian anomaly,''
  Phys.\ Rev.\ Lett.\  {\bf 79}, 1193 (1997)
  [astro-ph/9703005].
  %%CITATION = ASTRO-PH/9703005;%%
  %126 citations counted in INSPIRE as of 12 Oct 2013

\bibitem{GS-GS}
%
%\cite{Giovannini:1997gp}
%\bibitem{Giovannini:1997gp} 
  M.~Giovannini and M.~E.~Shaposhnikov,
  %``Primordial magnetic fields, anomalous isocurvature fluctuations 
  %and big bang nucleosynthesis,''
  Phys.\ Rev.\ Lett.\  {\bf 80}, 22 (1998)
  [hep-ph/9708303];\ 
  %%CITATION = HEP-PH/9708303;%%
  %82 citations counted in INSPIRE as of 12 Oct 2013
%
%\cite{Giovannini:1997eg}
%\bibitem{Giovannini:1997eg} 
%  M.~Giovannini and M.~E.~Shaposhnikov,
  %``Primordial hypermagnetic fields and triangle anomaly,''
  Phys.\ Rev.\ D {\bf 57}, 2186 (1998)
  [hep-ph/9710234].
  %%CITATION = HEP-PH/9710234;%%
  %160 citations counted in INSPIRE as of 12 Oct 2013
%

%%%%%%%%

%\cite{Long:2013tha}
\bibitem{Long:2013tha} 
  A.~J.~Long, E.~Sabancilar and T.~Vachaspati,
  %``Leptogenesis and Primordial Magnetic Fields,''
  JCAP {\bf 1402}, 036 (2014)
  [arXiv:1309.2315 [astro-ph.CO]].
  %%CITATION = ARXIV:1309.2315;%%
  %6 citations counted in INSPIRE as of 24 Apr 2014

%\cite{Montiel:2014dia}
\bibitem{Montiel:2014dia} 
  A.~Montiel, N.~Breton and V.~Salzano,
  %``Parameter estimation of a nonlinear magnetic universe from observations,''
  arXiv:1403.6493 [astro-ph.CO].
  %%CITATION = ARXIV:1403.6493;%%

%%%%%%%%
\bibitem{QUIET-1}
 See http://quiet.uchicago.edu/index.php.

%\cite{Samtleben:2008rb}
\bibitem{Samtleben:2008rb}
  D.~Samtleben and f.~t.~Q.~Collaboration,
  %``Measuring the Cosmic Microwave Background Radiation (CMBR) polarization
  %with QUIET,''
  Nuovo Cim.\  {\bf 122B}, 1353 (2007)
  [arXiv:0802.2657 [astro-ph]].
  %%CITATION = NUCIA,122B,1353;%%

%\cite{Araujo:2012yh}
\bibitem{Araujo:2012yh} 
  D.~Araujo {\it et al.}  [QUIET Collaboration],
  %``Second Season QUIET Observations: 
  %Measurements of the CMB Polarization Power Spectrum at 95 GHz,''
  Astrophys.\ J.\  {\bf 760}, 145 (2012)
  [arXiv:1207.5034 [astro-ph.CO]].
  %%CITATION = ARXIV:1207.5034;%%
  %12 citations counted in INSPIRE as of 30 Sep 2013

\bibitem{B-Pol}
 See http://www.b-pol.org/index.php.

%\bibitem{LiteBIRD}
% See http://cmbpol.kek.jp/litebird/.

%\bibitem{POLARBEAR}
% See http://mountainpolarbear.blogspot.jp/.

%%%%

\bibitem{Test}
%
%\cite{Kahniashvili:2005xe}
%\bibitem{Kahniashvili:2005xe}
  T.~Kahniashvili and B.~Ratra,
  %``Effects of Cosmological Magnetic Helicity 
  %on the Cosmic Microwave Background,''
  Phys.\ Rev.\  D {\bf 71}, 103006 (2005)
%\textit{ibid}. {\bf 71}, 103006 (2005) 
  [arXiv:astro-ph/0503709];\ 
  %%CITATION = PHRVA,D71,103006;%%
%
%\cite{Kahniashvili:2006zs}
%\bibitem{Kahniashvili:2006zs}
  T.~Kahniashvili,
  %``Effects of primordial helicity on CMB,''
  New Astron.\ Rev.\  {\bf 50}, 1015 (2006)
  [arXiv:astro-ph/0605440];\ 
  %%CITATION = ASTRE,50,1015;%%
%
%\cite{Kristiansen:2008tx}
%\bibitem{Kristiansen:2008tx}
  J.~R.~Kristiansen and P.~G.~Ferreira,
  %``Constraining primordial magnetic fields with CMB polarization
  %experiments,''
  Phys.\ Rev.\  D {\bf 77}, 123004 (2008)
  [arXiv:0803.3210 [astro-ph]].
  %%CITATION = PHRVA,D77,123004;%%
%

%\cite{Tashiro:2013ita}
\bibitem{Tashiro:2013ita} 
  H.~Tashiro, W.~Chen, F.~Ferrer and T.~Vachaspati,
  %``Search for CP Violation in the Gamma Ray Sky,''
  arXiv:1310.4826 [astro-ph.CO].
  %%CITATION = ARXIV:1310.4826;%%

%\cite{Giovannini:2014oia}
\bibitem{Giovannini:2014oia} 
  M.~Giovannini,
  %``Scaling laws and sum rules for the B-mode polarization,''
  Phys.\ Rev.\ D {\bf 89}, 061301 (2014)
  [arXiv:1402.0394 [astro-ph.CO]].
  %%CITATION = ARXIV:1402.0394;%%
 
%\cite{Calabrese:2013lga}
\bibitem{Calabrese:2013lga} 
  E.~Calabrese, M.~Martinelli, S.~Pandolfi, V.~F.~Cardone, C.~J.~A.~P.~Martins, S.~Spiro and P.~E.~Vielzeuf,
  %``Dark Energy coupling with electromagnetism as seen 
  %from future low-medium redshift probes,''
  arXiv:1311.5841 [astro-ph.CO].
  %%CITATION = ARXIV:1311.5841;%%

%\cite{Camera:2013fva}
\bibitem{Camera:2013fva} 
  S.~Camera, C.~Fedeli and L.~Moscardini,
  %``Magnification bias as a novel probe for 
  %primordial magnetic fields,''
  JCAP {\bf 1403}, 027 (2014)
  [arXiv:1311.6383 [astro-ph.CO]].
  %%CITATION = ARXIV:1311.6383;%%

\end{thebibliography}
\end{document}